\documentclass[a4paper, 11pt, times]{elsarticle}

\biboptions{sort&compress}

\usepackage{graphicx}
\usepackage{amsmath}
\usepackage{subfig}
\usepackage{float}
\usepackage{algorithm}
\usepackage{setspace}
\usepackage{bm}
\usepackage{enumitem}
\usepackage{comment}
\usepackage{amssymb}
\usepackage{bbm, bbold}
\usepackage{xcolor}
\usepackage{multirow}
\usepackage{url}
\usepackage{appendix}
\usepackage{lineno}

\let\Algorithm\algorithm
\renewcommand\algorithm[1][]{\Algorithm[#1]\setstretch{1.3}}

\newcommand{\pderiv}[2]{\frac{\partial #1}{\partial #2}}

\newcommand{\IBi}{\int_{\mathcal{B}_{0}}}
\newcommand{\IBt}{\int_{\mathcal{B}_{t}}}

\newcommand{\T}{\text{T}}
\newcommand{\dV}{\, \text{d}V}
\newcommand{\dv}{\, \text{d}v}
\newcommand{\dA}{\, \text{d}A}

\newcommand{\magn}{\text{magn}}

\newcommand{\PsiVol}{\Psi^{\text{vol}}}

\newcommand{\PsiMagn}{\Psi^{\text{magn}}}

\newcommand{\thickbar}[1]{\bm \bar{#1}}

\newcommand{\gram}{\mathrm{g}}

\newcommand{\millimeter}{\mathrm{mm}}
\newcommand{\second}{\mathrm{s}}

\newcommand{\Ampere}{\mathrm{A}}

\newcommand{\milliTesla}{\mathrm{mT}}

\RequirePackage[margin=20mm]{geometry}

\begin{document}
\fontfamily{ptm}\selectfont

\begin{frontmatter}

\title{A unified numerical approach for soft to hard magneto-viscoelastically coupled polymers}

\author[add1]{Chennakesava Kadapa \corref{cor1}}
\ead{c.kadapa@bolton.ac.uk}

\author[add2]{Mokarram Hossain\corref{cor1} }
\ead{mokarram.hossain@swansea.ac.uk}

\cortext[cor1]{Corresponding authors}

\address[add1]{School of Engineering, University of Bolton, Bolton BL3 5AB, United Kingdom}
\address[add2]{Zienkiewicz Centre for Computational Engineering (ZCCE), Swansea University, Swansea SA1 8EN, United Kingdom}

\begin{abstract}
The last decade has witnessed the emergence of magneto-active polymers (MAPs) as one of the most advanced multi-functional soft composites. Depending on the magnetisation mechanisms and responsive behaviour, MAPs are mainly classified into two groups: i) hard magnetic MAPs in which a large residual magnetic flux density sustains even after the removal of the external magnetic field, and ii) soft magnetic MAPs where the magnetisation of the filler particles disappear upon the removal of the external magnetic field. Polymeric materials are widely treated as fully incompressible solids that require special numerical treatment to solve the associated boundary value problem. Furthermore, both soft and hard magnetic particles-filled soft polymers are inherently viscoelastic. Therefore, the aim of this paper is to devise a unified finite element method-based numerical framework for magneto-mechanically coupled systems that can work for compressible and fully incompressible materials and from hard to soft MAPs, including the effects of the time-dependent viscoelastic behaviour of the underlying matrix. First, variational formulations for the uncoupled problem for hard MAPs and the coupled problem for soft MAPs are derived. The weak forms are then discretised with higher-order B\'ezier elements while the evolution equation for internal variables in viscoelastic models is solved using the generalised-alpha time integration scheme, which is implicit and second-order accurate. Finally, a series of experimentally-driven boundary value problems consisting of the beam and robotic gripper models are solved in magneto-mechanically coupled settings, demonstrating the versatility of the proposed numerical framework. The effect of viscoelastic material parameters on the response characteristics of MAPs under coupled magneto-mechanical loading is also studied.
\end{abstract}


\end{frontmatter}


\graphicspath{{./figures/}}

\section{Introduction}

\noindent   Materials are integral parts of the construction of machines. Machines are designed to perform specific tasks such as lifting or moving objects. For decades, scientists have been thinking of designing multi-functional and multi-responsive materials that can not only be used for building machines, but also themselves act like machines \cite{McCrakenAdvMat2020}. The mechanical and rheological properties of such responsive materials can be tuned to make them  move, deform, bend, twist, crawl and jump upon the application of external stimuli such as electric field \cite{Pelrine2000Science, Mehnert2021JMPS1}, magnetic field \cite{bell02, stepanov1}, temperature \cite{LendleinACIE2002, LengMRS2009}, light \cite{LendleinNature2005, LiNatureComm2020}, and pH \cite{XiaoSoftMatter2015,li2019}. Among other multi-functional materials, magneto-active polymers (MAPs) gained unprecedented attention in recent years due to their remote and contactless large actuation mechanisms.  A magneto-active polymeric composite consists of a soft elastomer embedded with magnetisable particles. While the bulk polymers could be any elastomers such as acrylics, polyurethanes and silicones, the magnetisable particles are two types: soft magnetics and hard magnetics. In soft MAPs, the particles can be magnetised by an external field but will demagnetise as soon as the external field disappears. In the case of hard MAPs, particles will retain their magnetisations even after the removal of the applied field. Some promising applications of MAPs include  remote-controlled soft robotics, precision and controlled drug deliver, smart vibration absorbers, morphing structures, base isolation in seismic devices, tunable stiffness actuators, soft and flexible electronics, automotive suspension bushing, sensing devices, to mention a few \cite{breger2015,bosnjak1,kim1,lu1, hu1,ren1,bose2,bedn99,bell02,bica12,bocz09,boczkowska1,ginder1,ginder2,ginder3,varga1}. \\

\noindent  In order to understand various mechanical and rheological properties of both hard and soft magnetic materials, experimental characterisations are essential. Hence, over the years, an extensive amount of experimental studies have been conducted \cite{danas1,gordaninejad1,kallio1,schubert1, burhannuddin1}. These experiments include tensile and compression tests under a wide range of deformation modes such as uniaxial, bi-axial (both equal and unequal), pure shear etc.  Such tests are conducted under pure mechanical loads as well as magneto-mechanical fields to differentiate how an externally applied field changes the mechanical properties of MAPs. For soft MAPs, Stepanov et al. \cite{stepanov1} imposed a magneto-mechanical load by stretching samples up to 100\% strains. However, as expected, for a high mechanical stretching, soft magnetisable particles are separated largely, which reduce the so-called \emph{magneto-restrictive} (change of sample sizes) or \emph{magneto-rheological} (change of various moduli) effects. In contrast to tensile tests, Kallio \cite{kallio1}, Gordaninejad et al. \cite{gordaninejad1}, and Moreno et al. \cite{moreno1} performed compressive tests for a wide range of mechanical strains under a magneto-mechanical load to quantify the magneto-rheological effects on soft MAPs. In addition to uniaxial tensile and compression experiments,  Schubert et al.  \cite{schubert1,schubert2} conducted a series of shear and equi-biaxial tensile tests both on isotropic and anisotropic soft MAPs. For a review on experimental characterisations of soft MAPs, the paper of Bastola and Hossain \cite{Bastola2020} could be a starting point. In contrast to the soft MAPs,  the experimental characterisations for hard MAPs are scarce in the literature. One of the earliest experiments on hard MAPs was due to Stepanov and co-workers, e.g., \cite{stepanov2012, kramarenko2015, borin2020}.   More recent works on  experimental characterisations of hard MAPs can be found in  Antonel et al. \cite{antonel2011}, Kramarenko et al. \cite{kramarenko2015}, Koo et al. \cite{koo2012}, and Lee et al. \cite{lee2018}.  For a comprehensive overview of syntheses, experimental characterisations, computational modelling, and potential applications of hard MAPs, the reader can consult recent reviews by Wu et al. \cite{WuMultiMat2020},  Lucarini et al. \cite{lucarini1}. \\

\noindent  Based on the seminal theoretical works of Pao and Nemat-Nasser \cite{pao1}, Eringen and Maugin \cite{eringen1}, Maugin \cite{maugin1}; Dorfmann and Ogden \cite{brigadnov1, dorfmann4,dorfmann5} and Bustamante and co-workers \cite{bustamante1, BustamanteMMS2008, shariff5} developed constitutive frameworks for magneto-mechanics taking into account the underlying isotropy and transverse isotropy for  MAPs.  Furthermore,  Saxena et al. \cite{saxena1,saxena2}, Nedjar \cite{nedjar1}, Haldar et al. \cite{haldar1, haldar2}, Mukherjee et al. \cite{MukherjeeIJNLM2020} developed constitutive models that take into account the time-dependent behaviour of polymeric matrices. In contrast to phenomenologically-motivated ideas that are formulated in terms of strain or stretch invariants \cite{shariff6}, few constitutive models have been proposed over the years that consider the micro-mechanical information of the MAPs' underlying microstructures. For instance, exploiting the pioneering lattice model of Jolly and co-workers \cite{jolly1},  Ivaneyko et al. \cite{Ivaneyko2011,Ivaneyko2012,Ivaneyko2014} extended it to a wider framework. Recently, Khanouki et al. \cite{khanouki1}, Gao and Wang \cite{gao1} performed experimental studies on soft MAPs and fitted with the lattice model proposed by Ivaneyko and co-workers. Furthermore, Garcia-Gonzalez and Hossain \cite{garcia-gonzalez1, garcia-gonzalez2021b} extended the Ivaneyko model to finite strains by incorporating time-dependent viscoelastic behaviours of soft and hard MAPs.  In contrast to soft MAPs, the computational modelling of hard MAPs is receiving worthy attention only after the seminal work of Kim et al. \cite{kim1}. Therein, they systematically demonstrated that the effects of the externally applied magnetic field are permanent when hard particles in the MAP composites are magnetically fully saturated. That means the residual magnetisation of saturated particles can be considered as a user-given parameter of a certain magnitude. Recently, Garcia-Gonzalez \cite{garcia-gonzalez2019} devised a computational framework for modelling the time-dependent behaviour of hard magneto-materials. As mentioned earlier, hard magnetic particles distributed inside MAPs produce significant forces and torques when internally placed magnetic dipoles react with the externally applied field. This unique feature facilitates in production of slender/strip rods, beams, shell-like responsive structures, termed as hard-magnetic elastica by Wang and co-workers \cite{wang2020},  for making flexible and active machine components. Wang et al. \cite{wang2020} proposed a finite strain theory for such largely deformable elastic materials in predicting precisely the amount of deflections produced by an externally applied magnetic field. Following the initial formulations of Wang et al. \cite{wang2020}, Chen et al. \cite{chenwei1,chenwei2} proposed large strain beam theories for simple hard magnetic materials and functionally graded hard MAPs. \\

\noindent  Parallel to the mechanical characterisation and nonlinear constitutive models for MAPs, computer simulation techniques for MAPs are also being explored towards facilitating the design and optimisation of smart devices based on magneto-responsive soft polymers. The deformation behaviour of MAPs is nonlinear, truly incompressible, and time/strain rate-dependent \cite{KankanalaJMPS2004,BustamanteQJMAM2006,BustamanteMMS2008,VuMMS2010}, which makes their numerical simulation extremely challenging.   In the literature, Haldar et al. \cite{haldar1}, Ethiraj and Miehe \cite{EthirajIJES2016}, Garcia-Gonzalez \cite{garcia-gonzalez2019}, and  Liu et al. \cite{LiuIJAM2020}   extended the standard displacement formulation to develop a displacement-potential formulation for soft MAPs. However, the incompressible nature of deformations in the finite strain regime means that conventional finite element methods based on the pure displacement formulations for the mechanics problem prove to be inadequate for the simulation of large-scale problems  \cite{KadapaCMAME2016elast,KadapaIJNME2019mixed}. To model the incompressible deformations in an approximate manner by choosing large values of the bulk modulus, Pelteret et al. \cite{PelteretIJNME2016} adapted the Hu-Washizu mixed displacement-pressure-Jacobian formulation of Simo et al. \cite{SimoCMAME1985}. Zhao et al. \cite{ZhaoJMPS2019}  used the so-called F-bar formulation of de Souza Neto et al. \cite{NetoIJSS1996} to model incompressible deformations in hard MAPs. Nevertheless, in spite of the success of mixed displacement-pressure-Jacobian and F-bar formulations in modelling incompressible deformations in which a large value for the bulk modulus is chosen, they are not free of shortcomings. As demonstrated comprehensively in our recent paper on computational electromechanics \cite{KadapaCMAME2020}, the F-bar formulation is computationally expensive, and the mixed displacement-pressure-Jacobian formulation is sensitive to the bulk modulus. Moreover, despite the availability of a few computational frameworks for the simulation of nearly incompressible MAPs, a unified framework for modelling both soft and hard MAPs by considering compressible and truly incompressible deformation behaviour as well as the time-dependent material response into account is still lacking. Hence, this contribution aims to address the gap. \\

\noindent In this work, we propose a novel unified finite element framework that can work for both compressible and truly incompressible hard MAPs (an uncoupled problem) as well as soft MAPs (a coupled problem). The present work is an extension of the mixed displacement-pressure formulation recently proposed by the authors \cite{KadapaMAMS2020,KadapaJMPS2021}. The advantage of this formulation lies in its ability to account for compressible and truly incompressible hyperelastic material models and their extension in a single framework. For hard MAPs, which are solved as uncoupled problems, the finite element formulation is the mixed displacement-pressure formulation of \cite{KadapaMAMS2020} with some minor modifications accounting for the contribution from the applied magnetic fields to the strain energy functions. For the soft MAPs for which the governing equations need to be solved together, a mixed displacement-pressure-potential formulation is developed. Thanks to the tremendous success of B\'ezier elements in our previous works \cite{KadapaIJNME2019bbar,KadapaIJNME2019mixed,KadapaCMAME2020,KadapaJMPS2021,KadapaAMech2021}, the spatial discretisation in the present work is performed using B\'ezier elements. In particular, quadratic B\'ezier hexahedron elements (BQ2) are considered for the displacement and magnetic potential, while linear hexahedron elements (BQ1) are used for discretising the pressure field. Additionally, time-dependent viscoelastic constitutive models \cite{HossainCMS2012,saxena1} are considered for both hard and soft MAPs. The evolution equation for the internal variables are numerically integrated using the generalised-alpha scheme \cite{JansenCMAME2000,KadapaCS2017}, which is a second-order accurate implicit time integration scheme. \\

\noindent The rest of the manuscript is organised as follows. In Section \ref{sec-gov-eqns}, basic equations governing magneto-mechanics at finite strains are introduced. Strain energy functions and finite element formulations are discussed in Sections \ref{sec-fem-hm} and \ref{sec-fem-sm}, respectively, for hard and soft magnetic materials.  The evolution equation to track the internal variables in viscoelastic models is also elaborated in these sections. Proposed finite element formulations are validated against several experimentally-driven numerical examples in Section \ref{sec-fem-examples}. Finally, the paper is concluded in Section \ref{sec-conclusion} with a summary of observations, including an outlook for future works.

\section{Governing equations for finite strain magnetomechanics}  \label{sec-gov-eqns}

\subsection{Kinematics and stress measures}
Let us consider an arbitrary solid body whose original (or reference) configuration is $\mathcal{B}_{0}$. Under the influence of external loads, the body change into a new configuration $\mathcal{B}_{t}$, which can be identified from the original configuration using the displacement field defined as
\begin{equation} \label{eqn-displ-definition}
\bm{u}(\bm{X}) := \bm{x} - \bm{X}
\end{equation}
where $\bm{X}$ and $\bm{x}$ are positions of a point in the reference and current configurations, respectively. The components of displacement in X-, Y- and Z-coordinate directions are denoted as $u_x$, $u_y$ and $u_z$, respectively. Following the definition of the displacement field in Eq. (\ref{eqn-displ-definition}), the deformation gradient is defined as
\begin{align}
\bm{F} := \pderiv{\bm{x}}{\bm{X}} = \bm{I} + \pderiv{\bm{u}}{\bm{X}},
\end{align}
where $\bm{I}$ is the second-order identity tensor. MAPs are widely known to undergo incompressible deformations. That is, the deformation of the solid is such that the total volume change at any time instant is zero. This condition acts as a constraint, known as the \textit{incompressibility constraint}, on the deformation field. This constraint in the finite strain regime is represented mathematically as,
\begin{align} \label{eqn-constraint-fic}
J = 1,
\end{align}
where $J := \det{\bm{F}}$. To model the incompressible deformation behaviour in the finite strain regime, $\bm{F}$ is decomposed into deviatoric and volumetric components as
\begin{equation}
\bm{F} = \bm{F}_{\mathrm{vol}} \, \bm{F}_{\mathrm{dev}},
\end{equation}
with
\begin{equation}
\bm{F}_{\mathrm{vol}} := J^{1/3} \, \bm{I}, \quad \mathrm{and} \quad \bm{F}_{\mathrm{dev}} := J^{-1/3} \, \bm{F}.
\end{equation}

Using the multiplicative decomposition of the deformation gradient, the modified deformation gradient $\thickbar{\bm{F}}$ and a modified right Cauchy-Green deformation tensor $\thickbar{\bm{C}}$ are defined as
\begin{align}
\thickbar{\bm{F}} & := J^{-1/3} \, \bm{F}, \\
\thickbar{\bm{C}} & := \thickbar{\bm{F}}^{\T} \, \thickbar{\bm{F}}.
\end{align}
Strain energy density functions for incompressible materials are devised in terms of the first and the second invariants, $\thickbar{I}_1$ and $\thickbar{I}_2$, of $\thickbar{\bm{C}}$, respectively,  see \cite{SteinmannAAM2012, HossainJMBM2013, MihaiJRSCI2015} for comprehensive details on the topic.

\subsection{Magnetostatics - Ampere's and Gauss's laws}
Ampere's law and Gauss's law are the two laws that govern the magnetostatics \cite{book-Dorfmann-EM}. Ampere's law can be expressed in the reference and current configurations as,
\begin{align}
\text{Curl } \mathbbm{H} &= \nabla_{\bm{X}} \times \mathbbm{H} = \bm{0}, \\
\text{curl } \mathbbm{h} &= \nabla_{\bm{x}} \times \mathbbm{h} = 0,
\end{align}
where $\mathbbm{H}$ and $\nabla_{\bm{X}}$, respectively, are the magnetic field vector (or magnetic displacement vector) and gradient operator in the reference configuration, and $\mathbbm{h}$ and $\nabla_{\bm{x}}$, are the counterparts of $\mathbbm{H}$ and $\nabla_{\bm{X}}$ in the current configuration. Ampere's law implies the existence of a scalar magnetic potential $\phi$ such that
\begin{align}
\mathbbm{H} &= - \, \text{Grad } \phi = - \, \nabla_{\bm{X}} \phi, \\
\mathbbm{h} &= - \, \text{grad } \phi = - \, \nabla_{\bm{x}} \phi.
\end{align}

Gauss's law in the reference and current configurations can be written as
\begin{align}
\text{Div } \mathbbm{B} &= \nabla_{\bm{X}} \cdot \mathbbm{B} = \bm{0}, \\
\text{div } \mathbbm{b} &= \nabla_{\bm{x}} \cdot \mathbbm{b} = 0,
\end{align}
where $\mathbbm{B}$ and $\mathbbm{b}$ are the magnetic inductors vectors, respectively, in the reference and current configurations.

Various quantities for magnetostatics introduced so far are related between the reference configuration to the current configuration by following relations,
\begin{align}
\mathbbm{h}  &= \bm{F}^{-\T} \, \mathbbm{H}, \\
\mathbbm{b}  &= \frac{1}{J} \, \bm{F} \, \mathbbm{B}, \\
\mathbbm{b}  &= \mu_0 \, \left[ \mathbbm{h} + \mathbbm{m} \right], \\
\mathbbm{B}  &= \mu_0 \, J \, \bm{C}^{-1} \, \left[ \mathbbm{H} + \mathbbm{M} \right],
\end{align}
where $\mu_0$ is the free space magnetic permeability constant and $\mathbbm{m}$ is the magnetic moment density vector in the current configuration, and it is related to its counterpart in the reference configuration via the relation
\begin{align}
\mathbbm{M} = \bm{F}^{\T} \, \mathbbm{m}.
\end{align}

For a given energy function $\Psi$, the magnetic displacement vector in the reference configuration ($\mathbbm{B}$) is defined as
\begin{align}
\mathbbm{B} := -\pderiv{\Psi}{\mathbbm{H}}.
\end{align}

\subsection{Equilibrium equations} 
By ignoring the effects of acceleration, the complete set of governing equations for the coupled magneto-statics problem in the current configuration can be written as
\begin{subequations} \label{govern-eqns-curconfig}
\begin{align}
- \, \nabla_{\bm{x}} \cdot \widehat{\bm{\sigma}}(\bm{x},t) &= \bm{f}(\bm{x},t) && \forall \, \bm{x} \in \mathcal{B}_{t}, \; t \in [0, t_f], \\
J(\bm{x},t) &= 1,  && \forall \, \bm{x} \in \mathcal{B}_{t}, \; t \in [0, t_f] \\
\nabla_{\bm{x}} \cdot \mathbbm{b}(\bm{x},t) &= 0  &&  \forall \, \bm{x} \in \mathcal{B}_{t}, \; t \in [0, t_f], \\
\bm{u}(\bm{x},t) &= \overline{\bm{u}}(\bm{x},t)  &&  \forall \, \bm{x} \in \partial \mathcal{B}_{t}^{\text{mech,D}}, \; t \in [0, t_f], \\
\bm{\sigma}(\bm{x},t) \cdot \bm{n} &= \overline{\bm{t}}(\bm{x},t) &&   \forall \, \bm{x} \in \partial \mathcal{B}_{t}^{\text{mech,N}}, \; t \in [0, t_f], \\
\phi(\bm{x},t) &= \overline{\phi}(\bm{x},t)  &&   \forall \, \bm{x} \in \partial \mathcal{B}_{t}^{\text{magn,D}}, \; t \in [0, t_f], \\
- \, \mathbbm{b} \cdot \bm{n} &= \overline{\omega} &&   \forall \, \bm{x} \in \partial \mathcal{B}_{t}^{\text{magn,N}}, \; t \in [0, t_f],
\end{align}
\end{subequations}
where, $\widehat{\bm{\sigma}}$ is the total Cauchy stress tensor, $t$ is the pseudo time, $t_f$ is the final time, $\bm{f}$ is the body force per unit deformed volume, $\bm{n}$ is the unit outward normal on the boundary $\partial \mathcal{B}_{t}$, $\overline{\bm{u}}$ is the prescribed value of displacement on the Dirichlet boundary $\partial \mathcal{B}_{t}^{\text{mech,D}}$, $\overline{\bm{t}}$ is the specified traction force per unit deformed area on the Neumann boundary $\partial \mathcal{B}_{t}^{\text{mech,N}}$, $\overline{\phi}$ is the prescribed value of magnetic potential on the Dirichlet boundary $\partial \mathcal{B}_{t}^{\text{magn,D}}$, and $\overline{\omega}$ is the specified magnetic surface charge density per unit deformed area on the Neumann boundary $\partial \mathcal{B}_{t}^{\text{magn,N}}$.


\section{Hard magneto-active polymers} \label{sec-fem-hm}
\noindent  Kim et al. \cite{kim1} systematically demonstrated that the effects of the externally applied magnetic field are fixed when hard particles in the MAP composites are magnetically fully saturated. That means the residual magnetisation of saturated particles can be considered as a user given internal parameters of a certain magnitude. Based on their initial work, Zhao et al. \cite{ZhaoJMPS2019} put all the details of mathematical modelling of hard MAPs meeting experimental results with numerical ones side by side. For the case of hard magnetic materials, the residual magnetic field, $\mathbbm{B}^{r}$, is assumed to exist throughout or part of the domain. Then the material deforms under the action of an applied magnetic field $\mathbbm{B}^{a}$.

\subsection{Strain energy functions}
In the case of hard MAPs, the effect of residual and applied magnetic fields is accounted for by adding their energy contribution to the total strain energy density function. Accordingly, the strain energy density function for the hard MAPs is given by
\begin{align} \label{eqn-enfun-hm}
\Psi =\Psi^{\text{vol}}(J) +    \Psi^{\text{mech}}_{\infty}(\thickbar{I}_1, \thickbar{I}_2, J) +  \Psi^{\text{mech}}_{v}(\thickbar{\bm{C}}, \bm{A}) + \Psi^{\text{magn}}_{\text{hm}},
\end{align}
where $\Psi^{\text{mech}}_{\infty}(\thickbar{I}_1, \thickbar{I}_2, J)$ is the deviatoric hyperelastic energy function, $\Psi^{\text{vol}}(J)$, is the volumetric energy function,
$\Psi^{\text{mech}}_{v}(\thickbar{\bm{C}}, \bm{A})$ is the energy function contribution from the time-dependent viscoelastic behaviour, and $\Psi^{\text{magn}}_{\text{hm}}$ is the energy function due to the applied magnetic field. Here, $\bm{A}$ is the strain-like internal state variable, which is a second order tensor. The deviatoric hyperelastic and volumetric parts can take any compatible energy function available in the literature, see \cite{SteinmannAAM2012, HossainJMBM2013, MihaiJRSCI2015}. Note that the volumetric energy function $\Psi^{\text{vol}}(J)$ \textit{vanishes for the truly incompressible case}, i.e., when $J=1$. In this study, we use Neo-Hookean and Gent models for the deviatoric part which  are given as
\begin{itemize}
\item \textbf{Neo-Hookean model}
\begin{align}
\Psi^{\text{mech}}_{\infty} = \frac{\mu}{2} \, \left[ \thickbar{I}_1 - 3 \right]
\end{align}
\item \textbf{Gent model}
\begin{align}
\Psi^{\text{mech}}_{\infty} = - \, \frac{\mu \, I_m}{2} \, \ln\left(1-\frac{\thickbar{I}_1 - 3}{I_m}\right)
\end{align}
where $\mu$ is the shear modulus and  $I_m$ is a material parameter representing the upper limit of $\left[ \thickbar{I}_1 - 3 \right]$.
\end{itemize}

The energy function for the viscoelastic part is assumed to depend only on the elastic deformations. This, however, is not due to any limitations of the proposed numerical framework but only for the sake of demonstration using numerical examples in the present work. The expression for the energy function is given \cite{Linder2011} as
\begin{align} \label{eqn-enerfun-visco}
\Psi^{\text{mech}}_{v}(\thickbar{\bm{C}}, \bm{A}) = \sum_{k=1}^{s} \, \frac{\mu_{v}^{(k)}}{2} \, \left[ \bm{A}^{(k)} : \thickbar{\bm{C}} - 3 - \ln(\det(\bm{A}^{(k)})) \right],
\end{align}
while the evolution equation to track  the internal variable $\bm{A}$ is 
\begin{align} \label{eqn-iv-evol}
\dot{\bm{A}}^{(k)} = \frac{1}{\tau^{(k)}} \, \left[ \thickbar{\bm{C}}^{-1} - \bm{A}^{(k)} \right],
\end{align}
where $k$ is the summation counter, $s$ is the number of Maxwell elements, and $\mu_{v}^{(k)}$, $\bm{A}^{(k)}$ and $\tau^{(k)}$, respectively, are the shear modulus, internal variable and relaxation time for the $k$th member of the Maxwell model. See \ref{section-appndx-evol}, for the details on the solution of the evolution equation (\ref{eqn-iv-evol}).

The energy function for the magnetic part is considered as proposed by Zhao et al. \cite{ZhaoJMPS2019}, which they derived based on their experimental work. The free energy function for the magnetic part is given as
\begin{align} \label{eqn-Psi-magn-hm}
\Psi^{\text{magn}}_{\text{hm}} = - \, \frac{1}{\mu_0} \, \mathbbm{B}^{a} \cdot \bm{F} \, \mathbbm{B}^{r}
 = - \, \frac{1}{\mu_0} \, \mathbbm{B}^{a}_{s} \, F_{sM} \, \mathbbm{B}^r_{M}.
\end{align}

For $\Psi^{\text{magn}}_{\text{hm}}$ given in Eq. (\ref{eqn-Psi-magn-hm}), the first Piola-Kirchhoff stress tensor ($\bm{P}^{\magn}$), Cauchy stress tensor ($\bm{\sigma}^{\magn}$) and the elasticity tensor ($\mathsf{e}^{\magn}$) become
\begin{align}
P^{\magn}_{iJ} &= \pderiv{\Psi^{\text{magn}}_{\text{hm}}}{F_{iJ}} = - \, \frac{1}{\mu_0} \, \mathbbm{B}_{i} \, \mathbbm{B}^r_{J} \\
\sigma^{\magn}_{ij}
&= \frac{1}{J} \, P^{\magn}_{iJ} \, F_{jJ} = - \, \frac{1}{J} \, \frac{1}{\mu_0} \, \mathbbm{B}_{i} \, \mathbbm{B}^r_{J} \, F_{jJ} \\
\mathsf{e}^{\magn}_{ijkl} &= \frac{1}{J} \, F_{jJ} \, F_{lL} \, \pderiv{P^{\magn}_{iJ}}{F_{kL}} = 0
\end{align}

\subsection{Finite element formulation for hard magnetic polymers}
Since the magnetic field is a user-defined input quantity for hard magnetic materials, only the quantities related to the mechanical deformation are independent solution variables in the finite element formulation. Therefore, for the hard MAPs, only minor modifications are required for the standard finite element formulation for hyperelasticity. In this work, the mixed displacement-pressure formulation recently proposed by the authors in \cite{KadapaMAMS2020} is adapted. As demonstrated in \cite{KadapaMAMS2020,KadapaCMAME2020,KadapaJMPS2021}, this formulation is proven to be robust and computationally efficient for modelling large-deformation and large-strain behaviour of components made of materials whose stress-strain response is described using hyperelastic constitutive models and their extension to soft multi-functional composites, e.g., electro-active polymers \cite{KadapaCMAME2020}.

Following Kadapa and Hossain \cite{KadapaMAMS2020}, the energy functional for the displacement-pressure formulation is given by
\begin{align} \label{eqn-Pi-hm}
\Pi(\bm{u},p) = \IBi \, \left[ \Psi(\thickbar{\bm{C}}, p) + \Psi_{p} \right] \dV - \Pi_{\text{ext}}
\end{align}
where $\Pi_{\text{ext}}$ is the energy contribution due to the external forces, given as
\begin{align}
\Pi_{\text{ext}} &= \IBi  \, \bm{u}^{\T} \, \bm{f}_{0} \, \dV + \int_{\partial \mathcal{B}^{\text{mech,N}}_{0}} \bm{u}^{\T} \, \bm{t}_{0} \, \dA + \int_{\partial \mathcal{B}^{\text{magn,N}}_{0}} \phi \, \omega_{0} \, \dA.
\end{align}
In the aforementioned equation, $\boldsymbol{f}_{0}$ is the body force per unit original volume, $\boldsymbol{t}_{0}$ is the traction force per unit undeformed area, $\omega_{0}$ is the magnetic surface density per unit undeformed area, and $\dA$ is the elemental area in the reference configuration.

$\Psi_{p}$ in Eq. (\ref{eqn-Pi-hm}) is the energy function associated with the imposition of the incompressibility constraint in the case of truly incompressible deformations and, for the compressible and nearly incompressible cases, it is related to the volumetric energy function ($\PsiVol(J)$) as \cite{KadapaMAMS2020}
\begin{align} \label{eqn-enerfun-PsiP}
\Psi_{p} = p \, \left[ J - \widehat{J} - \frac{\widehat{\vartheta} \, p}{2} \right],
\end{align}
where, $\widehat{J}$ and $\widehat{\vartheta}$ are constants which depend only on the displacement field from the previously converged load step ($t_n$). They are evaluated using the expressions
\begin{align} \label{eqn-Psi-linear3}
\widehat{J} = J_{n} - \frac{\pderiv{\PsiVol}{J}\biggr\rvert_{J_{n}}}{\pderiv{^2\PsiVol}{J^2} \biggr\rvert_{J_{n}}}; \qquad
\widehat{\vartheta} = \frac{1}{\pderiv{^2\PsiVol}{J^2} \biggr\rvert_{J_{n}}},
\end{align}
where, the subscript $n$ denotes the previously converged load step. Note that the truly incompressible case is easily  recovered by setting $\widehat{J}=1$ and $\widehat{\vartheta}=0$ in (\ref{eqn-enerfun-PsiP}). We refer the reader to Kadapa and Hossain \cite{KadapaMAMS2020} for the comprehensive details and the advantages of this approach.

The first ($\delta$) and the second ($d$) variations for the energy functional (\ref{eqn-Pi-hm}) become
\begin{align} \label{eqn-hm-variation1}
\delta \Pi
&= \IBi \, \left[ \delta F_{iJ} \, P_{iJ} + \delta J \, p \right] \dV + \IBi \, \delta p \, \left[ J - \widehat{J} - \widehat{\vartheta} \, p \right] \dV - \delta \, \Pi_{\mathrm{ext}}
\end{align}
and
\begin{align} \label{eqn-hm-variation2}
d(\delta \Pi)
= \IBi \, \left[ \delta F_{iJ} \, \pderiv{P_{iJ}}{F_{kL}} \, dF_{kL} + p \, d(\delta J)  \right] \dV + \IBi \, \left[ \delta p \, dJ + \delta J \, dp - \widehat{\vartheta} \, \delta p \, dp \right] \dV - d(\delta \, \Pi_{\mathrm{ext}}),
\end{align}
where $P_{iJ}$ is the first Piola-Kirchhoff stress tensor in index notation, and it is computed for a given energy function $\Psi$ by the relation
\begin{align}
P_{iJ} = \pderiv{\Psi}{F_{iJ}}.
\end{align}

Using the relations,
\begin{align} \label{eqn-relations-1}
\delta F_{iJ} &= \delta u_{i,j} \, F_{jJ} \\
\delta J &= J \, \delta u_{i,j} \, \delta_{ij}
\end{align}
we can rewrite the first variation (\ref{eqn-hm-variation1}) and second variation (\ref{eqn-hm-variation2}) of the energy functional as
\begin{align}
\delta \Pi = \IBt \, \delta u_{i,j} \, \widehat{\sigma}_{ij} \, \dv + \IBi \, \delta p \, \left[ J - \widehat{J} - \widehat{\vartheta} \, p \right] \dV - \delta \, \Pi_{\mathrm{ext}}
\end{align}

\begin{align}
d(\delta \Pi)
&= \IBt \, \left[ \delta u_{i,j} \, \mathsf{e}_{ijkl} \, du_{k,l} + \delta u_{i,i} \, dp + \delta p \, du_{i,i}  \right] \dv
- \IBi \, \widehat{\vartheta} \, \delta p \, dp \dV - d(\delta \, \Pi_{\mathrm{ext}})
\end{align}
where $\widehat{\sigma}_{ij}$ and $\mathsf{e}_{ijkl}$, respectively, are the effective Cauchy stress tensor and the material tangent tensor (of order four). They are computed using the relations,
\begin{align}
\widehat{\sigma}_{ij} &= \frac{1}{J} \, F_{jJ} \, P_{iJ} + p \, \delta_{ij}, \\
\mathsf{e}_{ijkl} &= \frac{1}{J} \, F_{jJ} \, \pderiv{P_{iJ}}{F_{kL}} \, F_{lL} + p \, [\delta_{ij} \, \delta_{lk}-\delta_{il} \, \delta_{jk}].
\end{align}

We consider the approximations for the displacement and pressure, and their variations as
\begin{subequations}  \label{eqn-fe-approxs1}
\begin{align}
 \bm{u} &= \mathbf{N}_{\bm{u}} \, \mathbf{u}, \qquad
\delta \bm{u} = \mathbf{N}_{\bm{u}} \, \delta \mathbf{u}, \qquad
 d \bm{u} = \mathbf{N}_{\bm{u}} \, d \mathbf{u}, \\
  p &= \mathbf{N}_p \, \mathbf{p}, \qquad
\delta    p = \mathbf{N}_p \, \delta \mathbf{p}, \qquad
 d    p = \mathbf{N}_p \, d \mathbf{p},
\end{align}
\end{subequations}
where $\mathbf{N}_{\bm{u}}$ and $\mathbf{N}_{p}$ are the matrices of basis functions for displacement and pressure, respectively; and $\mathbf{u}$ and $\mathbf{p}$ are the vectors of displacement and pressure degrees of freedom, respectively.

Using the finite element approximations given in (\ref{eqn-fe-approxs1}), and by employing the Newton-Raphson scheme for solving the resulting nonlinear problem in an iterative manner, the coupled matrix system for the mixed displacement-pressure formulation can be written as
\begin{equation}  \label{matrixsystem-proposed}
 \begin{bmatrix}
  \mathbf{K}_{\bm{u}\bm{u}}  &  \mathbf{K}_{\bm{u}p} \\
  \mathbf{K}_{p\bm{u}}  &  \mathbf{K}_{pp}
 \end{bmatrix}
 \begin{Bmatrix}  \Delta \mathbf{u} \\ \Delta \mathbf{p} \end{Bmatrix}
=
- \begin{Bmatrix}  \mathbf{R}_{\bm{u}} \\ \mathbf{R}_{p} \end{Bmatrix}
\end{equation}
where,
\begin{align}
\mathbf{K}_{\bm{u}\bm{u}} &= \IBt \mathbf{G}^{\T}_{\bm{u}} \, \bm{\mathsf{e}}(\bm{u}^k, p^k) \, \mathbf{G}_{\bm{u}} \, \dv, \label{eqn-Kuu-hm} \\
\mathbf{K}_{\bm{u}p}  &= \IBt \mathbf{D}^{\T}_{\bm{u}} \, \mathbf{N}_{p} \, \dv = \mathbf{K}_{p\bm{u}}^{\T} \\
\mathbf{K}_{pp}  &= - \, \IBi \widehat{\vartheta} \, \mathbf{N}_{p}^{\T} \mathbf{N}_{p} \, \dV, \\
\mathbf{R}_{\bm{u}} &= \IBt \mathbf{G}^{\T}_{\bm{u}} \, \widehat{\bm{\sigma}}(\bm{u}^k, p^k) \, \dv - \mathbf{F}^{\text{ext}}_{\bm{u}}, \\
\mathbf{R}_p     &= \IBi \mathbf{N}_p^\mathrm{T} \, \left[ J(\bm{u}^k) - \widehat{J} - \widehat{\vartheta} \, p^k \right] \, \dV.
\end{align}
In the aforementioned equations,  $\mathbf{F}^{\text{ext}}_{\bm{u}}$ is the external force vector the elasticity problem, and $\mathbf{G}_{\bm{u}}$ and $\mathbf{D}_{\bm{u}}$, respectively, are the gradient-displacement and divergence-displacement matrices. For a single basis function, they are given as
\renewcommand{\arraystretch}{1.2}
\begin{equation}
\mathbf{G}_{\bm{u}} =
\begin{bmatrix}
\pderiv{N_{\bm{u}}}{x}  &               0            &             0        &
\pderiv{N_{\bm{u}}}{y}  &               0            &             0        &
\pderiv{N_{\bm{u}}}{z}  &               0            &             0        \\
           0            &   \pderiv{N_{\bm{u}}}{x}   &             0        &
           0            &   \pderiv{N_{\bm{u}}}{y}   &             0        &
           0            &   \pderiv{N_{\bm{u}}}{z}   &             0        \\
           0            &               0            &  \pderiv{N_{\bm{u}}}{x} &
           0            &               0            &  \pderiv{N_{\bm{u}}}{y} &
           0            &               0            &  \pderiv{N_{\bm{u}}}{z} 
\end{bmatrix}^{\T},
\end{equation}
\renewcommand{\arraystretch}{1.0}
\begin{equation}
\mathbf{D}_{\bm{u}}
=
\begin{bmatrix}
\pderiv{N_{\bm{u}}}{x}  &  \pderiv{N_{\bm{u}}}{y}  &  \pderiv{N_{\bm{u}}}{z}
\end{bmatrix}.
\end{equation}

In Eq. (\ref{eqn-Kuu-hm}), $\bm{\mathsf{e}}$ is the matrix representation of the material tangent tensor $\mathsf{e}_{ijkl}$, and it is of size $9 \times 9$. Note that the Cauchy stress tensor ($\widehat{\bm{\sigma}}$) and elasticity tensor ($\bm{\mathsf{e}}$) also contain contributions from the viscoelastic part as well. The evolution equation (\ref{eqn-iv-evol}) for the internal variable in the viscoelastic part of the free energy function is solved at each integration point using the generalised-alpha scheme \cite{JansenCMAME2000,KadapaCS2017} which is second-order accurate, unlike the backward-Euler scheme which is only first-order accurate. Solution of (\ref{eqn-iv-evol}) using the generalised-alpha scheme is presented in \ref{section-appndx-evol}.

\section{Soft magneto-active polymers}  \label{sec-fem-sm}
\noindent In contrast to hard magneto-active polymers, soft MAPs operate in different manners as filler particles are gradually being magnetised and lose their magnetisations once the external magnetic field is completely removed. Therefore, the applied magnetic field not only changes the particles' magnetisations but also induces deformations which motivate us to consider the problem as a magneto-mechanically coupled problem. In the contribution, we will discretise all governing equations  (\ref{govern-eqns-curconfig})  for the magneto-mechanical coupled problem in which the displacement, as well as potential fields, are considered as independent variables.

\subsection{Strain energy functions}
Strain energy functions for soft magnetic materials consist of additional contributions due to the magnetic fields and interactions between mechanical and magnetic fields. The total strain energy density function for the soft magnetic materials is assumed as
\begin{align} \label{eqn-enfun-sm}
\Psi =    \Psi^{\text{vol}}(J) +     \Psi^{\text{mech}}_{\infty}(\thickbar{I}_1, \thickbar{I}_2, J) +  \Psi^{\text{mech}}_{v}(\thickbar{\bm{C}}, \bm{A}) + \Psi^{\text{free}}(\mathbbm{H},\bm{C},J) + \PsiMagn_{\text{sm}}(\mathbbm{H}, \thickbar{\bm{C}},J)
\end{align}
where $\Psi^{\text{free}}$ is the energy function due to the free space and $\PsiMagn_{\text{sm}}$ is the free energy function that accounts for the magneto-mechanical coupling while the rest of the energy functions are the same as the ones used in the case of hard MAPs. The free-space strain energy function is given as,
\begin{equation}
\Psi^{\text{free}}(\mathbbm{H},\bm{C},J) =   - \, \frac{1}{2} \, \mu_{0} \, J \, \bm{C}^{-1} :[ \mathbbm{H} \otimes \mathbbm{H}] =  - \, \frac{1}{2} \, \mu_{0} \, J \, \mathbbm{H}_{M} \, C_{MN}^{-1} \, \mathbbm{H}_{N},
\end{equation}
and the strain energy function $\PsiMagn_{\text{sm}}$ adapted from Pelteret et al.  \cite{PelteretIJNME2016} is
\begin{align} \label{eqn-Psi-magn-sm}
\PsiMagn_{\text{sm}} = \alpha \, \mu_{0} \, \Bigg[ \bm{I} : \left[ \mathbbm{H} \otimes \mathbbm{H} \right] \Bigg] + \beta \, \mu_{0} \, \Bigg[ \thickbar{\bm{C}} : \left[ \mathbbm{H} \otimes \mathbbm{H} \right] \Bigg] + \eta \, \mu_{0} \, \Bigg[ \thickbar{\bm{C}}^{-1} : \left[ \mathbbm{H} \otimes \mathbbm{H} \right] \Bigg],
\end{align}
where the constants $\alpha$, $\beta$ and $\eta$ are assumed as $\alpha = - \, 0.5$, $\beta = - \, 4.0$ and $\eta = - \, 0.5$.

\subsection{Finite element formulation for soft  MAPs}
For soft magnetic materials, the magnetic potential ($\phi$) is also solved as an independent variable in addition to displacement and pressure in the mixed formulation for the hard magnetic case. Therefore, the energy functional for the displacement-pressure-potential ($\bm{u}-p-\phi$) formulation is given by
\begin{align}  \label{eqn-Pi-sm}
\Pi(\bm{u},p,\phi) = \IBi \, \left[ \Psi(\thickbar{\bm{C}}, p, \phi) + \Psi_{p} \right] \dV - \Pi_{\mathrm{ext}}.
\end{align}

The first variation of $\Pi$ can be written as
\begin{align} \label{eqn-sm-variation1}
\delta \Pi = \IBi \, \Bigg[ \delta F_{iJ} \, P_{iJ} - \delta \mathbbm{H}_{I} \, \mathbbm{B}_{I} + \delta J \, p + \delta p \, \left[ J - \widehat{J} - \widehat{\vartheta} \, p \right] \Bigg] \dV - \delta \, \Pi_{\mathrm{ext}},
\end{align}
where
\begin{align}
\mathbbm{B}_{I}     &= -\pderiv{\Psi}{\mathbbm{H}_{I}}.
\end{align}

Using the relations in Eq. (\ref{eqn-relations-1}) and the following relations for the magnetostatic variables,
\begin{align}
\delta \mathbbm{H}_{I} &= F_{pI} \, \delta \mathbbm{h}_{p} \\
\mathbbm{B}_{I} &= J \, F^{-1}_{Iq} \, \mathbbm{b}_{q} \\
\delta \mathbbm{h}_{q} &= - \, \delta \phi_{,q}
\end{align}
where $(\bullet)_{,q}$ denotes the derivative of $(\bullet)$ with respect to the $q$th component, we can rewrite the first variation (\ref{eqn-sm-variation1}) as
\begin{align}
\delta \Pi = \IBt \, \Bigg[ \delta u_{i,j} \, \widehat{\sigma}_{ij} + \delta \phi_{,i} \, \mathbbm{b}_{i} \Bigg] \dv + \IBi \, \delta p \, \left[ J - \widehat{J} - \widehat{\vartheta} \, p \right] \dV - \delta \, \Pi_{\mathrm{ext}}.
\end{align}

Similarly, the second variation (\ref{eqn-Pi-sm}) can be written as,
\begin{align}
d(\delta \Pi)
&= \IBt \, \left[ \delta u_{i,j} \, \mathsf{e}_{ijkl} \, du_{k,l} + \delta u_{i,j} \, \mathsf{p}_{ijk} \, d \phi_{,k} + \delta \phi_{,i} \, \widehat{\mathsf{p}}_{ijk} \, du_{j,k} + \delta \phi_{,i} \, \mathsf{d}_{ij} \, d\phi_{,j} \right] \dv \\
&+ \IBt \, \left[ \delta u_{i,i} \, dp + \delta p \, du_{i,i} \right] \dv - \IBi \, \widehat{\vartheta} \, \delta p \, dp \dV - d(\delta \, \Pi_{\mathrm{ext}}),
\end{align}
where $\mathsf{p}_{ijk}$ and $\widehat{\mathsf{p}}_{ijk}$ are third-order coupling tensors; and $\mathsf{d}_{ij}$ is the material permittivity tensor of order two, and they are computed using the relations
\begin{align}
\mathsf{p}_{ijk}  &= - \; \frac{1}{J} \, F_{jJ} \, \pderiv{P_{iJ}}{\mathbbm{H}_{K}} \, F_{kK}, \\
\widehat{\mathsf{p}}_{ijk}  &= \hspace*{4mm} \frac{1}{J} \, F_{iI} \, \pderiv{\mathbbm{B}_{I}}{F_{jK}} \, F_{kK},  \\
\mathsf{d}_{ij}   &= - \; \frac{1}{J} \, F_{iI} \, \pderiv{\mathbbm{B}_{I}}{\mathbbm{H}_{J}} \, F_{jJ}.
\end{align}

The finite element approximations for the magnetic potential are taken as
\begin{subequations}  \label{eqn-fe-approxs2}
\begin{align}
\phi &= \mathbf{N}_{\phi} \, \bm{\phi}, \qquad
\delta  \phi = \mathbf{N}_{\phi} \, \delta \bm{\phi}, \qquad
 d  \phi = \mathbf{N}_{\phi} \, d \bm{\phi},
\end{align}
\end{subequations}
where $\mathbf{N}_{\phi}$ is the matrix of basis functions for the magnetic potential, and $\bm{\phi}$ is the vector of magnetic potential degrees of freedom.

By using the approximations (\ref{eqn-fe-approxs1}) and (\ref{eqn-fe-approxs2}), and adapting the Newton-Raphson scheme, the discrete matrix system for the incremental displacements, $\Delta \mathbf{u}$, incremental pressure, $\Delta \mathbf{p}$, and incremental magnetic potential, $\Delta \bm{\phi}$, at iteration $k+1$ can be written as
\begin{equation}  \label{eqn-impl-matrix-full}
 \begin{bmatrix}
  \mathbf{K}_{\bm{u}\bm{u}} &  \mathbf{K}_{\bm{u}p}  &  \mathbf{K}_{\bm{u}\phi} \\
  \mathbf{K}_{p\bm{u}}      &  \mathbf{K}_{pp}       &  \bm{0} \\
  \mathbf{K}_{\phi \bm{u}}  &  \bm{0}                &  \mathbf{K}_{\phi\phi}
 \end{bmatrix}
 \begin{Bmatrix}  \Delta \mathbf{u} \\ \Delta \mathbf{p} \\ \Delta \bm{\phi} \end{Bmatrix}
=
- \begin{Bmatrix}  \mathbf{R}_{\bm{u}} \\ \mathbf{R}_{p} \\ \mathbf{R}_{\phi} \end{Bmatrix}.
\end{equation}

Here, $\mathbf{K}_{\bm{u}p}$, $\mathbf{K}_{p\bm{u}}$, $\mathbf{K}_{pp}$ and $\mathbf{R}_{p}$ are the same as the ones given for the hard magnetic case in Section \ref{sec-fem-hm}. The rest of the sub-matrices and vectors are given as
\begin{align}
\mathbf{K}_{\bm{u}\bm{u}}  &= \IBt \, \mathbf{G}_{\bm{u}}^{\mathrm{T}} \, \bm{\mathsf{e}}(\bm{u}^k, p^k, \phi^k) \, \mathbf{G}_{\bm{u}} \, \dv,  \label{eqn-Kuu-sm} \\
\mathbf{K}_{\bm{u} \phi}  &= \IBt \, \mathbf{G}_{\bm{u}}^\mathrm{T} \, \bm{\mathsf{p}}(\bm{u}^k,\phi^k) \, \mathbf{G}_{\phi} \, \dv = \mathbf{K}_{\phi \bm{u}}^{\mathrm{T}}, \label{eqn-Kuf} \\
\mathbf{K}_{\phi\phi}  &= \IBt \, \mathbf{G}_{\phi}^{\mathrm{T}} \, \bm{\mathsf{d}}(\bm{u}^k,\phi^k) \, \mathbf{G}_{\phi} \, \dv \label{eqn-Kff}, \\
\mathbf{R}_{\bm{u}} &= \IBt \mathbf{G}^{\T}_{\bm{u}} \, \widehat{\bm{\sigma}}(\bm{u}^k, p^k, \phi^k) \, \dv - \mathbf{F}^{\text{ext}}_{\bm{u}}, \\
\mathbf{R}_{\phi} &= \IBt \mathbf{G}^{\T}_{\phi} \, \mathbbm{b}_{i}(\bm{u}^k, \phi^k) \, \dv - \mathbf{F}^{\text{ext}}_{\phi},
\end{align}
where $\mathbf{F}^{\text{ext}}_{\phi}$ is the external force vector for the magnetic problem; $\bm{\mathsf{p}}$ and $\bm{\mathsf{d}}$, respectively, are the matrix versions of the tensors $\mathsf{p}_{ijk}$, $\mathsf{d}_{ij}$, and they are of size $9\times3$ and $3\times3$, respectively; and $\mathbf{G}_{\phi}$ is the gradient-displacement matrix for the magnetic potential field, and is given as
\renewcommand{\arraystretch}{1.2}
\begin{equation}
\mathbf{G}_{\phi} =
\begin{bmatrix}
\pderiv{N_{\phi}}{x}  &
\pderiv{N_{\phi}}{y}  &
\pderiv{N_{\phi}}{z}  \\
\end{bmatrix}^{\T}.
\end{equation}


\section{Numerical examples} \label{sec-fem-examples}
In this section, we will demonstrate the accuracy and computational advantages of the proposed simulation framework for the hard and soft magneto-active polymers by studying several numerical examples. The proposed formulations are implemented in an-house code written in C++. Following our recent work on electro-active polymers (EAPs) \cite{KadapaCMAME2020}, and morphoelasticity \cite{KadapaJMPS2021},  BQ2/BQ1 and BQ2/BQ1/BQ2 elements are considered, respectively, for the hard magnetic and soft magnetic cases. Here, BQ1 and BQ2 denote linear (8-noded) and quadratic (27-noded) B\'ezier hexahedron elements. Therefore, the displacement and magnetic potential fields are approximated using quadratic basis functions, and the pressure field is approximated using linear basis functions. Comparison is made with the results obtained with the Q1/Q0 element to demonstrate the advantage of the higher-order elements. Note that the basis functions for the BQ1 and Q1 elements are the same.

The units adapted for mass, length, time, magnetic potential, and magnetic induction are grams, millimetre, second, Ampere and milliTesla (mT), respectively. Unless stated otherwise explicitly, the free-space magnetic permeability constant is $\mu_0 = 1.2566 \, \frac{\gram \, \millimeter}{\second^2 \, \Ampere^2}$. Note also that since a single Maxwell element is used in the viscoelastic model, superscripts denoting the counter ($k$) in the viscous shear modulus ($\mu_{v}^{(k)}$) and relaxation time ($\tau^{(k)}$) are dropped henceforth.


\subsection{Numerical examples - hard magnetic case}

\subsubsection{Uniaxial actuation of a cube}
\begin{figure}[H]
 \centering
 \subfloat[]{\includegraphics[trim=10mm 0mm 10mm 0mm, clip, scale=0.6]{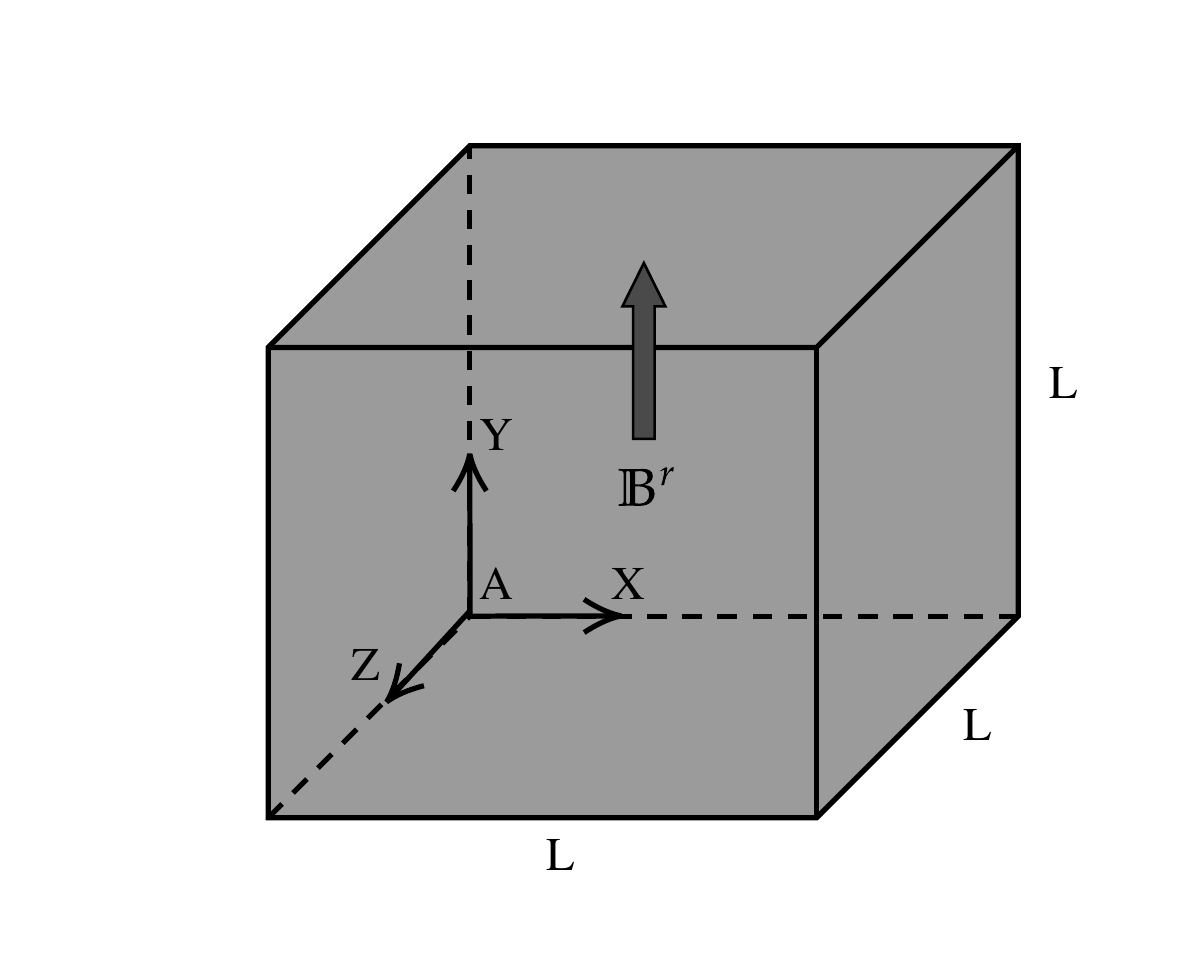} \label{fig-hm-cube-geom}}
 \subfloat[]{\includegraphics[clip, scale=0.4]{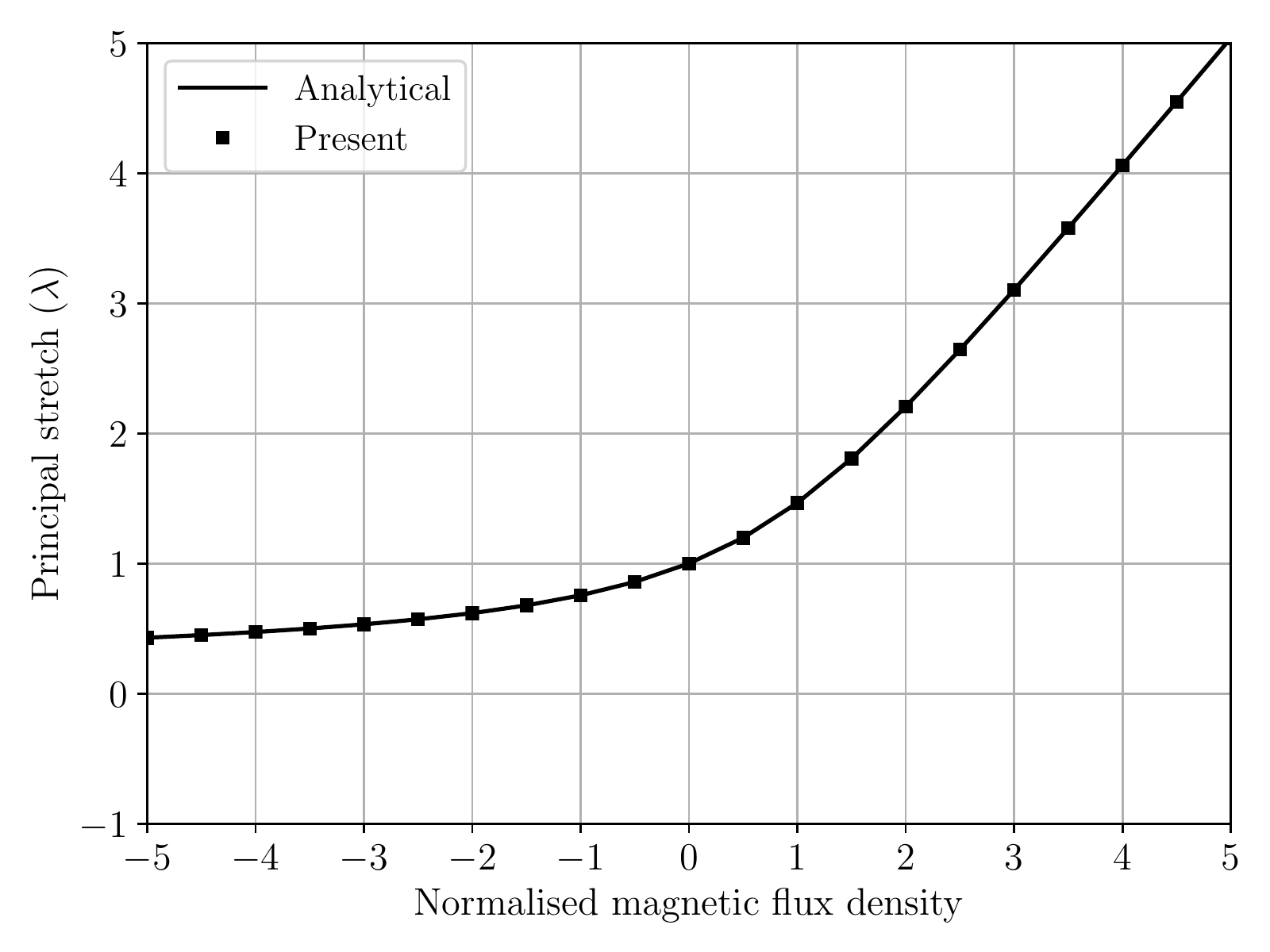} \label{fig-hm-cube-graph}}
 \subfloat[]{\includegraphics[clip, scale=0.25]{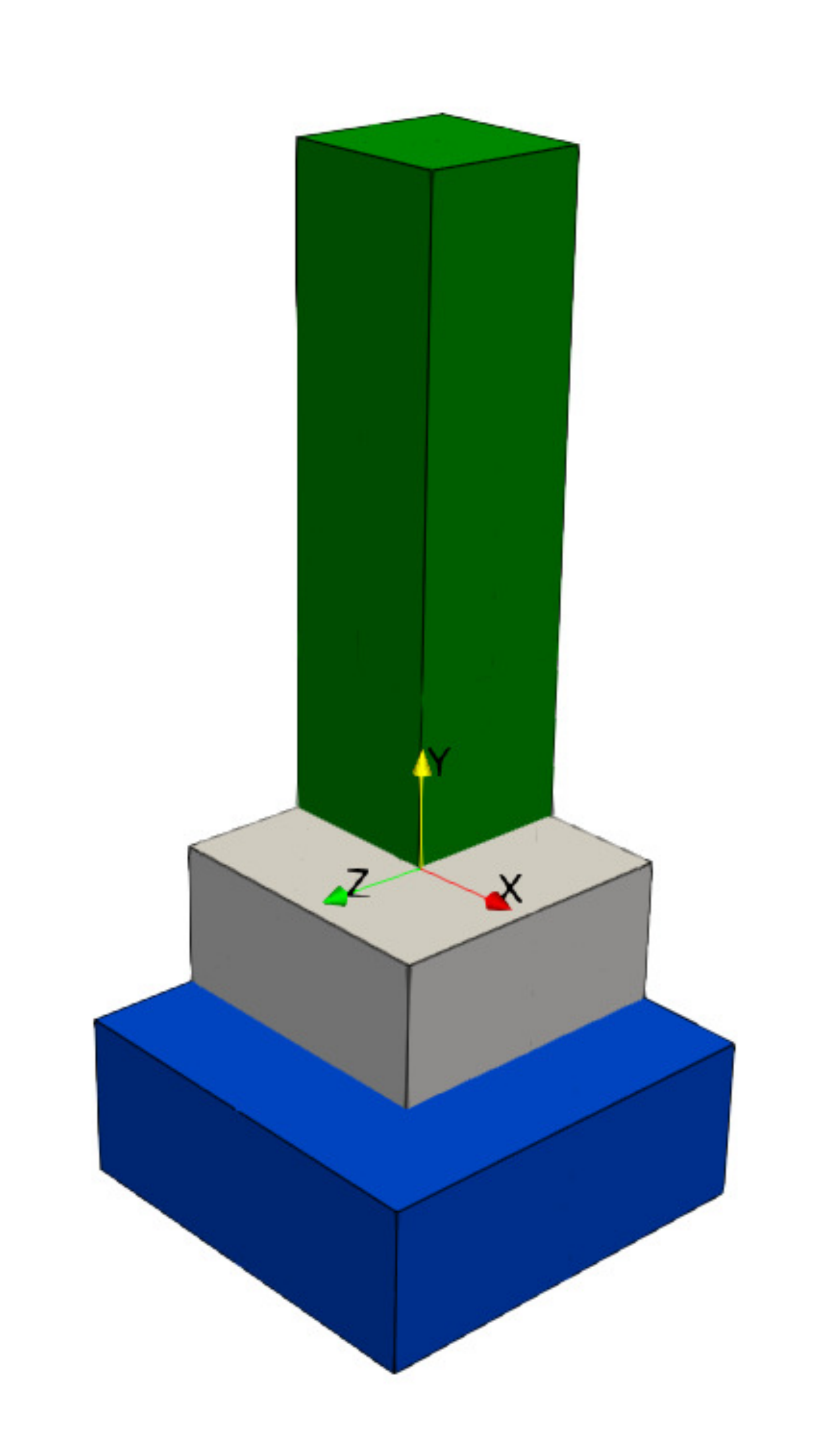}}
 \caption{A cubic block in 3D: (a) problem setup, (b) graph of normalised magnetic flux density versus principal stretch and (c) deformed configurations for $\mathbbm{B}^r \cdot \mathbbm{B}^{a}/(\mu \, \mu_{0})=3$ (green) and $\mathbbm{B}^r \cdot \mathbbm{B}^{a}/(\mu \, \mu_{0})=-3$ (blue) along with the original configuration (gray).}
\end{figure}

In the first example, we consider the actuation of a cube studied previously in Zhao et al. \cite{ZhaoJMPS2019}. The edges of the cube are initially aligned with the coordinate axes, as shown in Fig. \ref{fig-hm-cube-geom}. The hyperelastic material model for the elastic part of the strain energy function is assumed to be fully incompressible Neo-Hookean with a shear modulus of $\mu=1000$, and the permeability is taken as $\mu_0=0.001$. The residual and applied magnetic field are assumed to be oriented parallel to the Y-axis. Simulations are performed using one BQ2/BQ1 element. The deformation of the cube is computed for 20 different values of the dimensionless parameter $\mathbbm{B}^r \cdot \mathbbm{B}^{a}/(\mu \, \mu_{0})$ in the range $[-5,5]$. The values of the principal stretch ($\lambda$) obtained in the present work, as shown in Fig. \ref{fig-hm-cube-graph}, are in good agreement with the analytical solution given the cubic equation \cite{ZhaoJMPS2019}
\begin{align}
\lambda^3 - \frac{\mathbbm{B}^r \cdot \mathbbm{B}^{a}}{\mu \, \mu_{0}} \, \lambda^2 - 1 = 0.
\end{align}

\subsubsection{Large deflection of a cantilever beam}
\begin{figure}[H]
 \centering
 \includegraphics[trim=10mm 20mm 10mm 20mm, clip, scale=1.0]{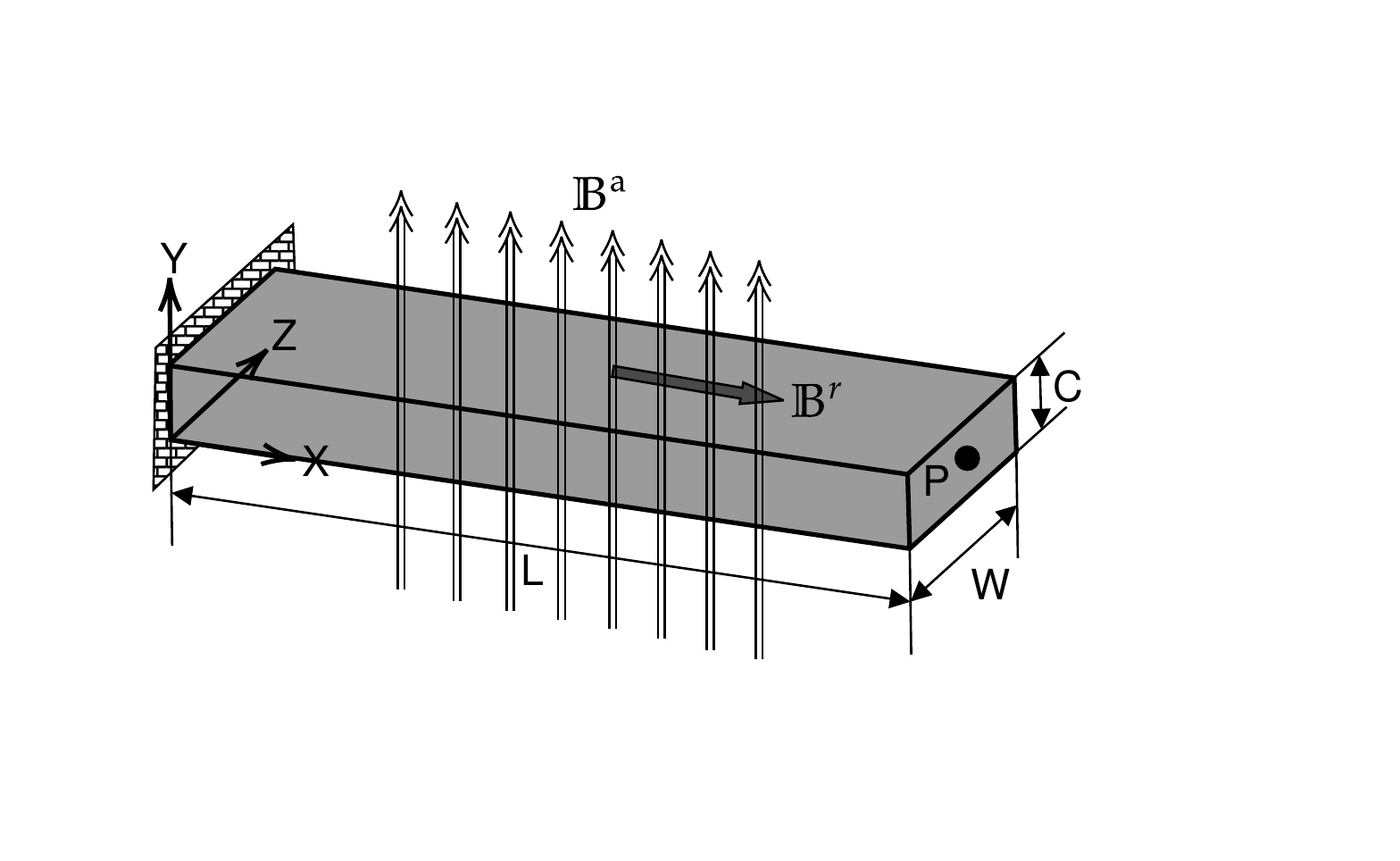}
 \caption{Cantilever beam: setup of the problem.}
 \label{fig-hm-beam-geom}
\end{figure}

The cantilever magnetic beam problem studied in Zhao et al. \cite{ZhaoJMPS2019} is considered in this example. As shown in the setup of the problem in Fig. \ref{fig-hm-beam-geom}, the residual magnetic field, $\mathbbm{B}^r$, is oriented along the length of the beam towards the free end, and the applied magnetic field, $\mathbbm{B}^{a}$, is perpendicular to $\mathbbm{B}^r$. The length of the beam is $L=17.2$ mm, and its width is $W=5$ mm. Two different values of thickness, 0.84 mm ($L/C \approx 20.5$) and 0.42 mm ($L/C \approx 41$) are considered for the demonstration. A truly incompressible Neo-Hookean model with a shear modulus of $\mu=303$ kPa is considered while the magnitude of the residual magnetic field is, $\vert \mathbbm{B}^r \vert = 143 \; \milliTesla$ and the maximum value of applied magnetic field is $\vert \mathbbm{B}^{a} \vert = 50 \; \milliTesla$.

\begin{figure}[H]
 \centering
 \subfloat[$L/C=20.5$]{\includegraphics[clip, scale=0.5]{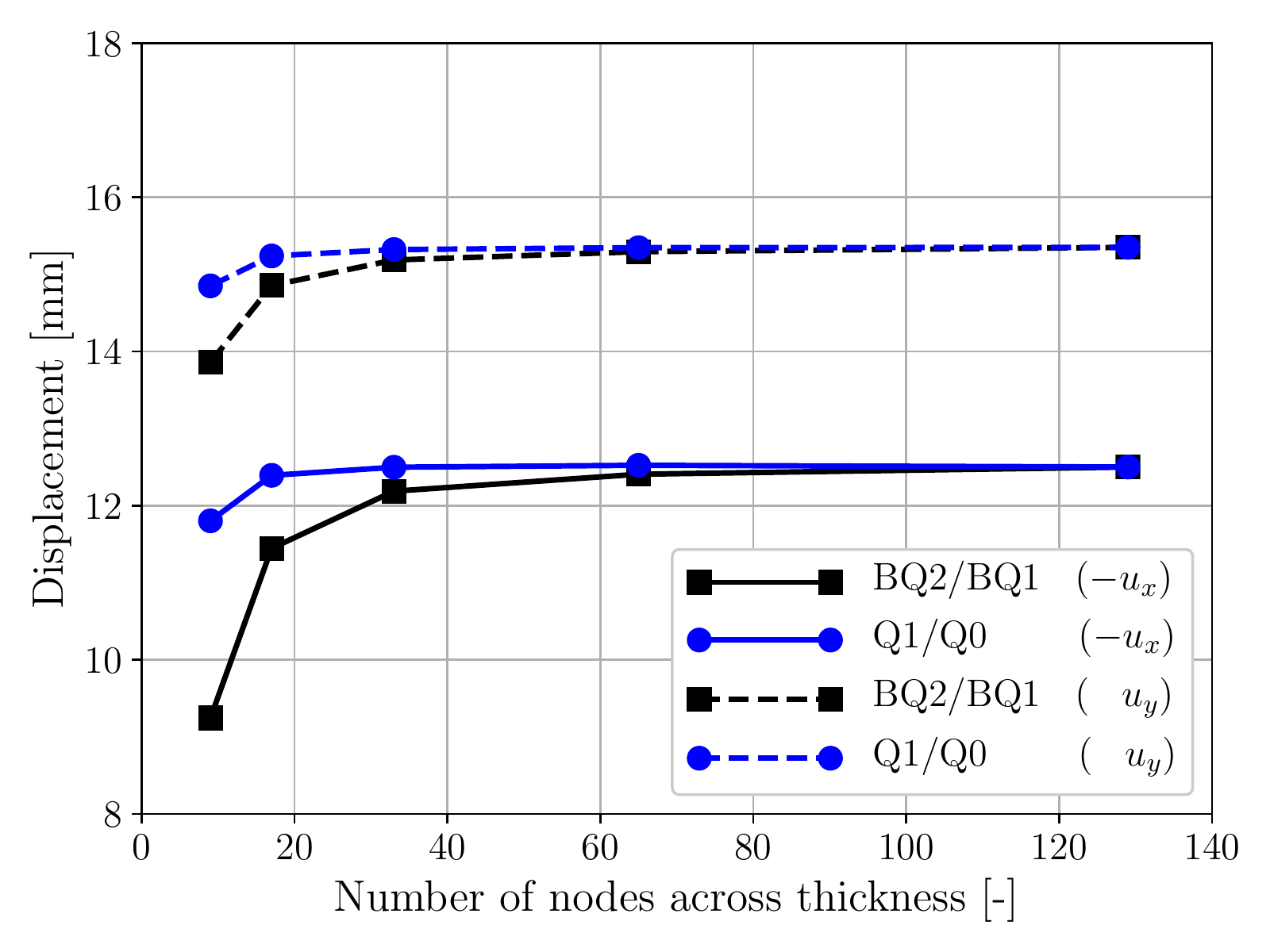}}
 \subfloat[$L/C=41$]{\includegraphics[clip, scale=0.5]{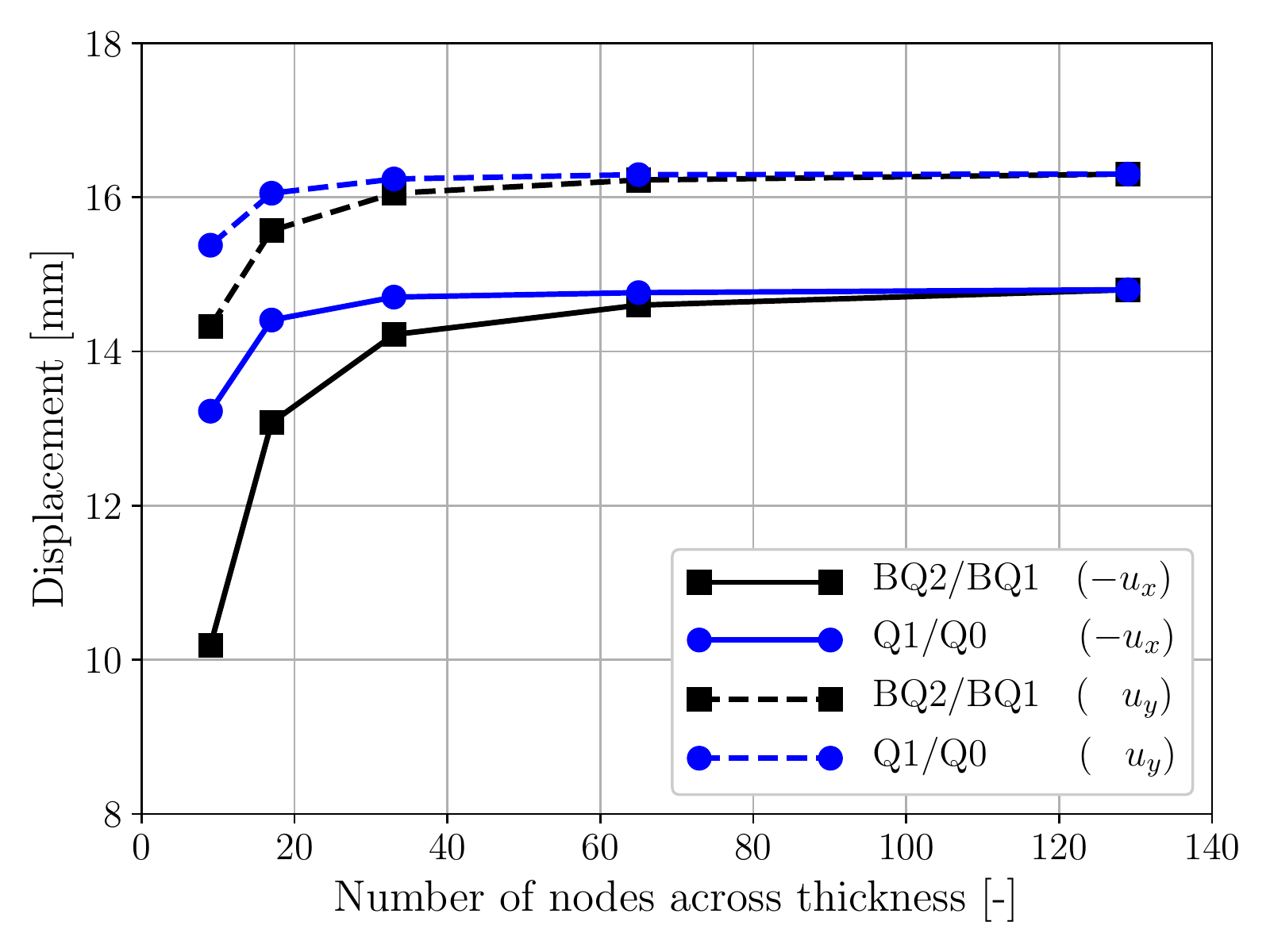}}
 \caption{Cantilever beam: convergence graphs of the displacement of point P with respect to the mesh refinement.}
 \label{fig-hm-beam-graphs-conv}
\end{figure}

Since the deformation in Z-direction is absent, the number of elements in that direction is taken as four and eight, respectively, for BQ2/BQ1 and Q1/Q0 elements. Numerical results are obtained for five successively refined meshes starting with $8 \times 1 \times 4$ and $16 \times 2 \times 8$ meshes, respectively, for BQ2/BQ1 and Q1/Q0 elements. For such meshes, one-to-one comparisons can be provided based on the number of nodes across the thickness of the beam. Meshes with 9, 17, 33, 65 and 129 nodes across the thickness of the beam are denoted as M1, M2, M3, M4 and M5, respectively. The graphs of X-displacement ($u_x$ and Y-displacement ($u_y$) of the midpoint of free end of the beam (point P in Fig. \ref{fig-hm-beam-geom}) as shown in Fig. \ref{fig-hm-beam-graphs-conv} demonstrate clear convergence with respect to the mesh refinement. It can be observed from the convergence graphs that, when compared with the Q1/Q0 element, qualitatively superior results can be obtained with the BQ2/BQ1 element using coarse meshes; the difference in the results obtained with the two different element types is more pronounced for the thinner beam.

\begin{figure}[H]
 \centering
 \includegraphics[clip, scale=0.8]{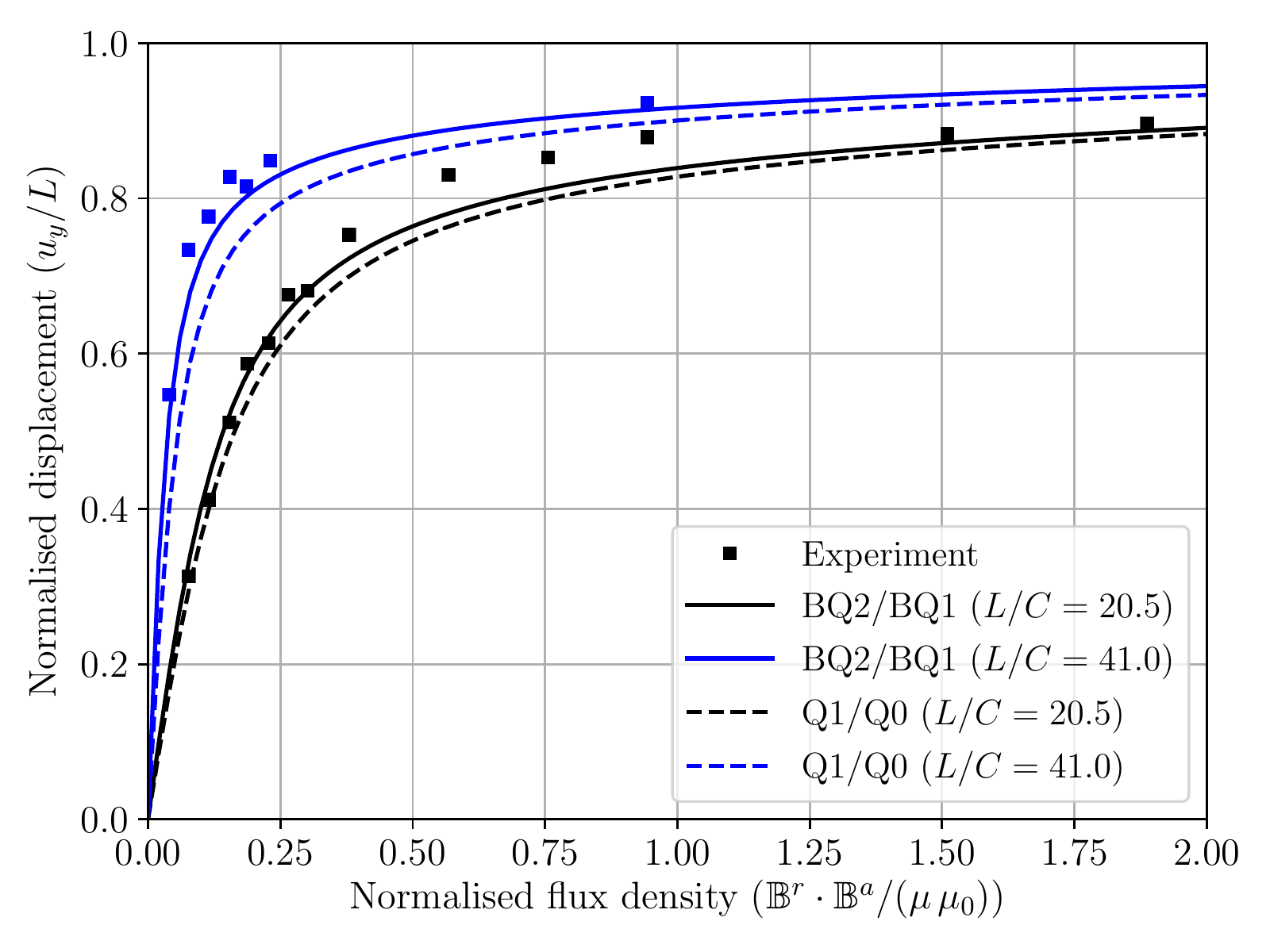}
 \caption{Beam in 3D: comparison of simulation results obtained with M3 mesh against the experimental values from Zhao et al. \cite{ZhaoJMPS2019}.}
 \label{fig-hm-beam-graphs-comp}
\end{figure}

\begin{figure}[H]
 \centering
 \subfloat[$\vert \mathbbm{B}^{a} \vert = 5 \; \milliTesla$]{\includegraphics[trim=10mm 10mm 10mm 10mm, clip, scale=0.35]{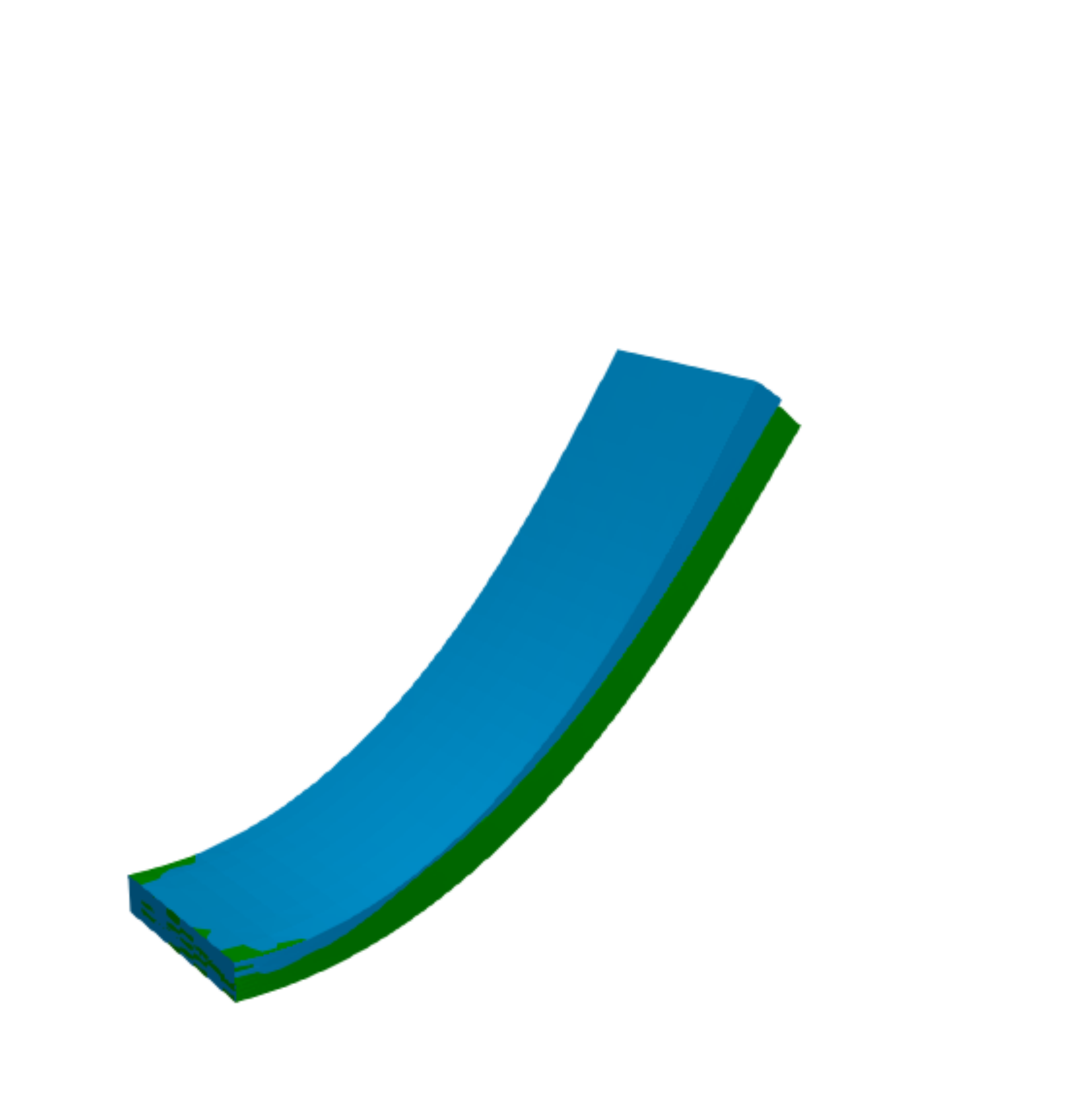}}
 \subfloat[$\vert \mathbbm{B}^{a} \vert = 10 \; \milliTesla$]{\includegraphics[trim=10mm 10mm 10mm 10mm, clip, scale=0.35]{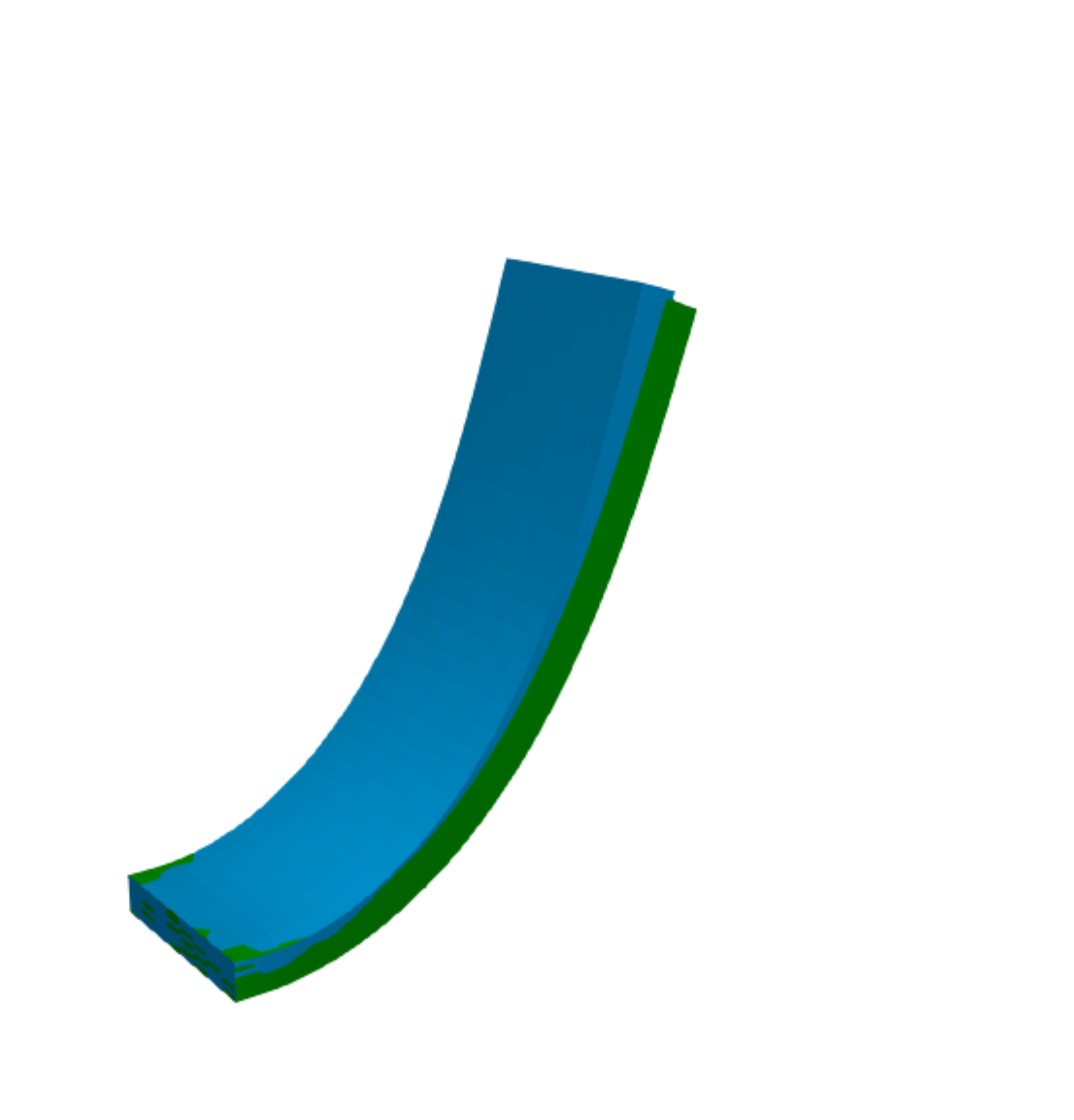}}
 \subfloat[$\vert \mathbbm{B}^{a} \vert = 50 \; \milliTesla$]{\includegraphics[trim=10mm 10mm 40mm 10mm, clip, scale=0.35]{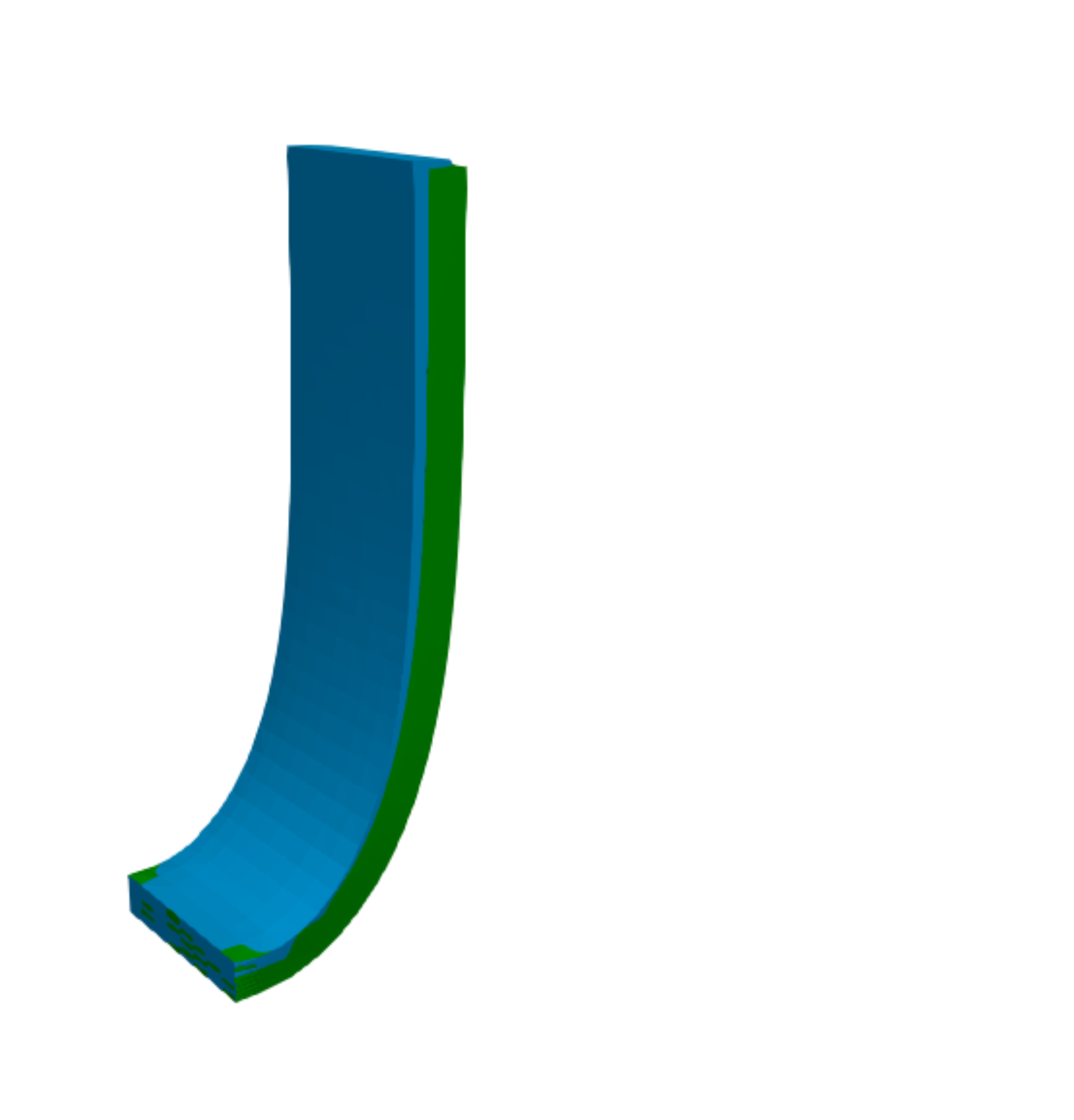}}
 \caption{Cantilever beam: deformed shapes of the beam with $L/C=20.5$ obtained with M3 mesh. Results obtained with BQ2/BQ1 and Q1/Q0 elements are in blue and green, respectively.}
 \label{fig-hm-beam-shapes-lc20p5}
\end{figure}

\begin{figure}[H]
 \centering
 \subfloat[$\vert \mathbbm{B}^{a} \vert = 5 \; \milliTesla$]{\includegraphics[trim=10mm 10mm 10mm 10mm, clip, scale=0.35]{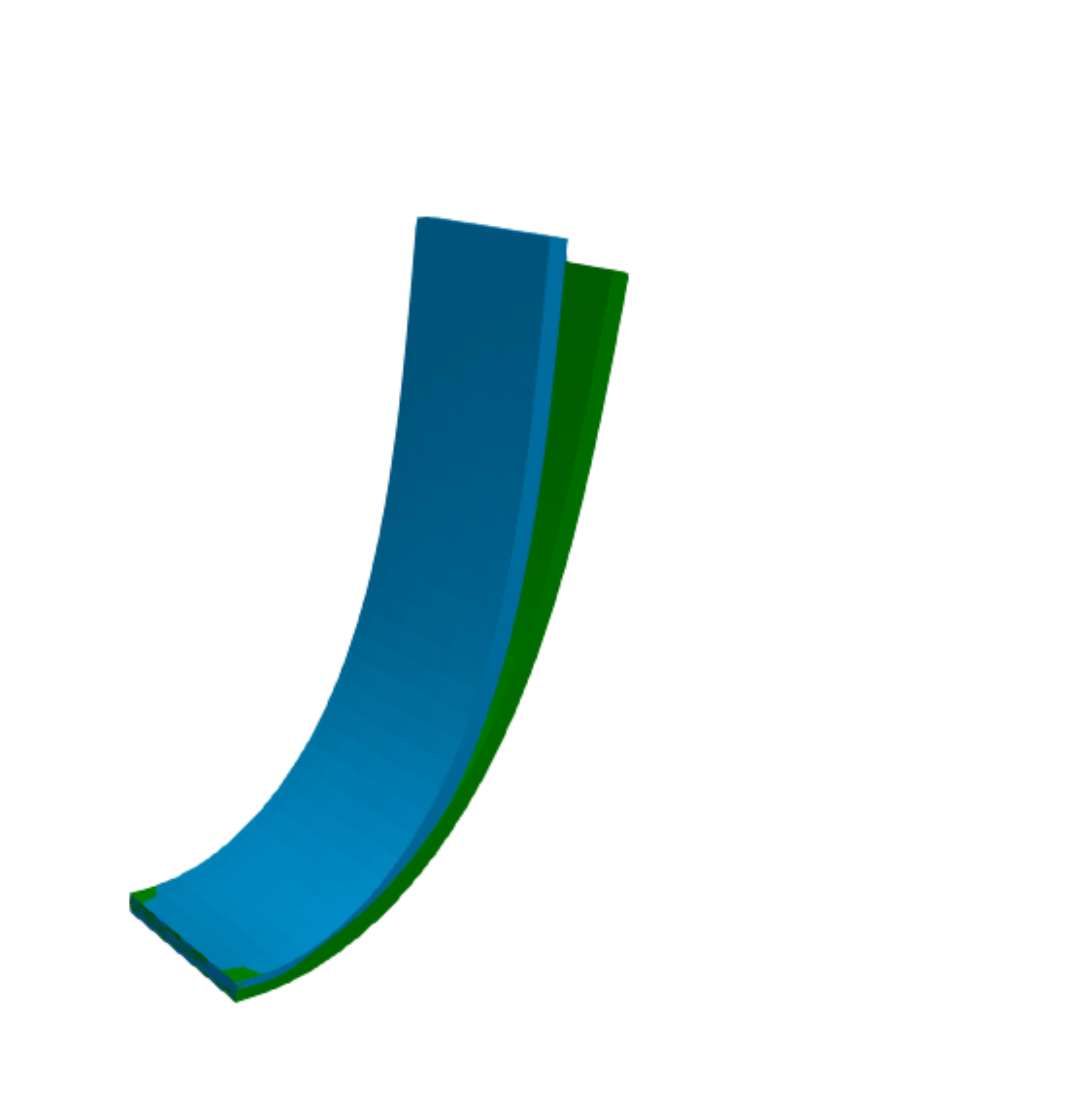}}
 \subfloat[$\vert \mathbbm{B}^{a} \vert = 10 \; \milliTesla$]{\includegraphics[trim=10mm 10mm 10mm 10mm, clip, scale=0.35]{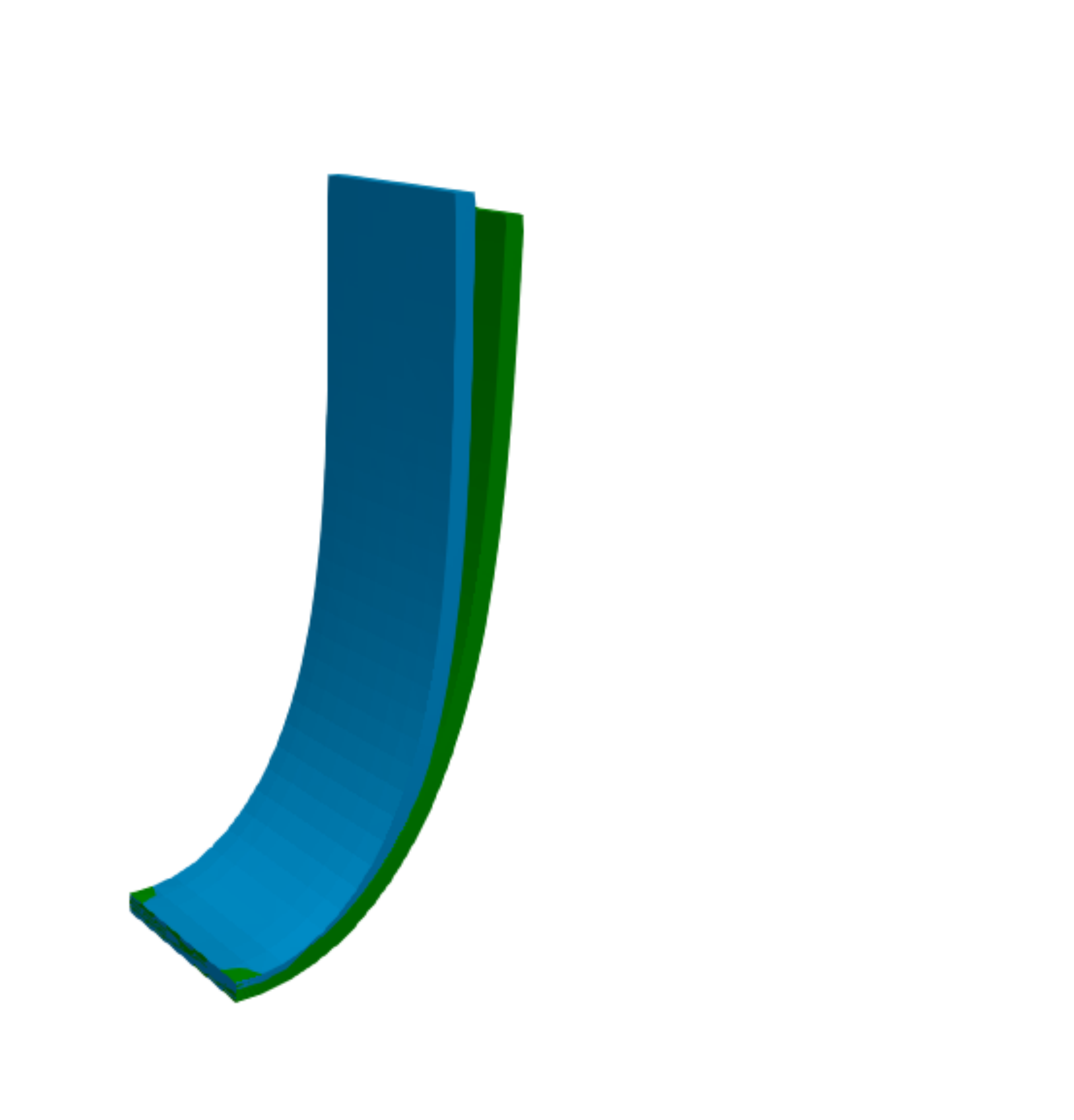}}
 \subfloat[$\vert \mathbbm{B}^{a} \vert = 50 \; \milliTesla$]{\includegraphics[trim=10mm 10mm 40mm 10mm, clip, scale=0.35]{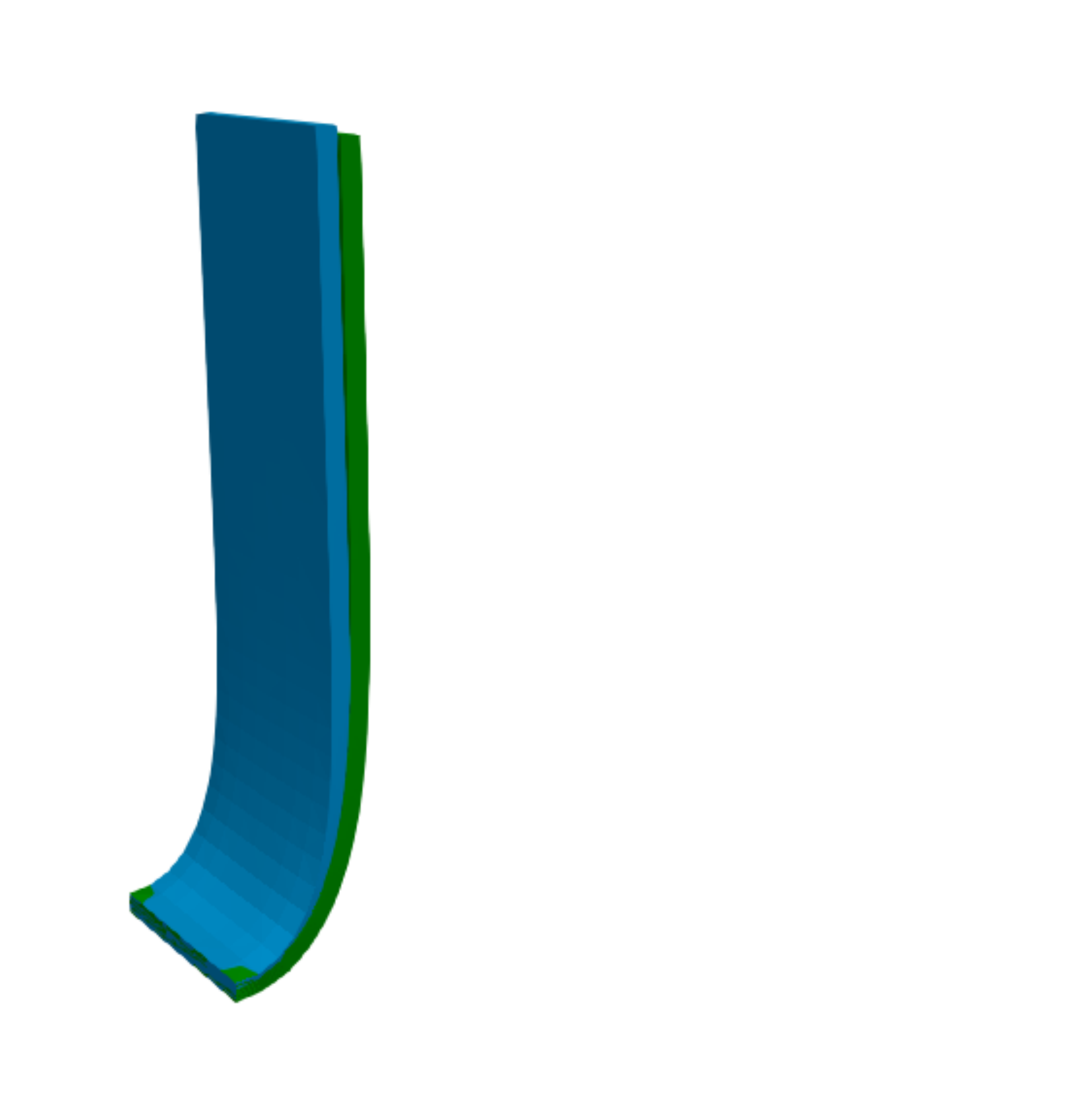}}
 \caption{Cantilever beam: deformed shapes of the beam with $L/C=41$ obtained with M3 mesh. Results obtained with BQ2/BQ1 and Q1/Q0 elements are in blue and green, respectively.}
 \label{fig-hm-beam-shapes-lc41}
\end{figure}

A comparison with the experimental results taken from Zhao et al. \cite{ZhaoJMPS2019} as presented in Fig. \ref{fig-hm-beam-graphs-comp} demonstrates good agreements between the present simulation results and the experimental values. The deformed shapes of the beam at three different values of the applied magnetic field are shown in Figs. \ref{fig-hm-beam-shapes-lc20p5} and \ref{fig-hm-beam-shapes-lc41}, respectively, for $L/C=20.5$ and $L/C=41$. These deformed shapes reveal that, although the final deformed shapes obtained with the two element types are not significantly different, particularly in the case of the thin beam, considerable differences can be observed in the deformed shapes at intermediate load values. This example thus demonstrates the advantages of using higher-order elements, especially for thin beams.

\subsubsection{Pattern formations}
\begin{figure}[H]
 \centering
 \subfloat[]{\includegraphics[clip, scale=0.45]{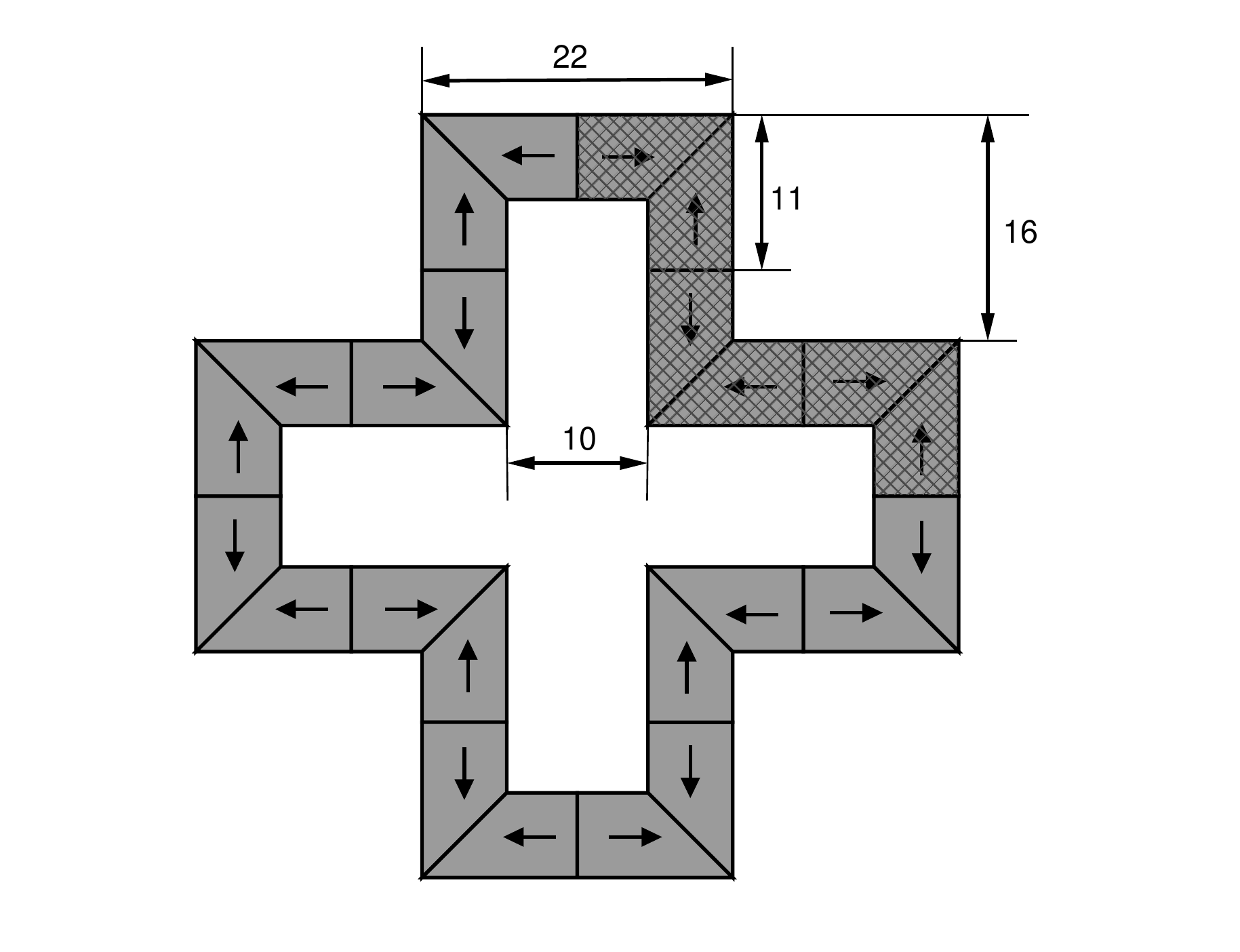} \label{fig-hm-pattern-geom}}
 \subfloat[]{\includegraphics[trim=10mm 0mm 0mm 0mm, clip, scale=0.5]{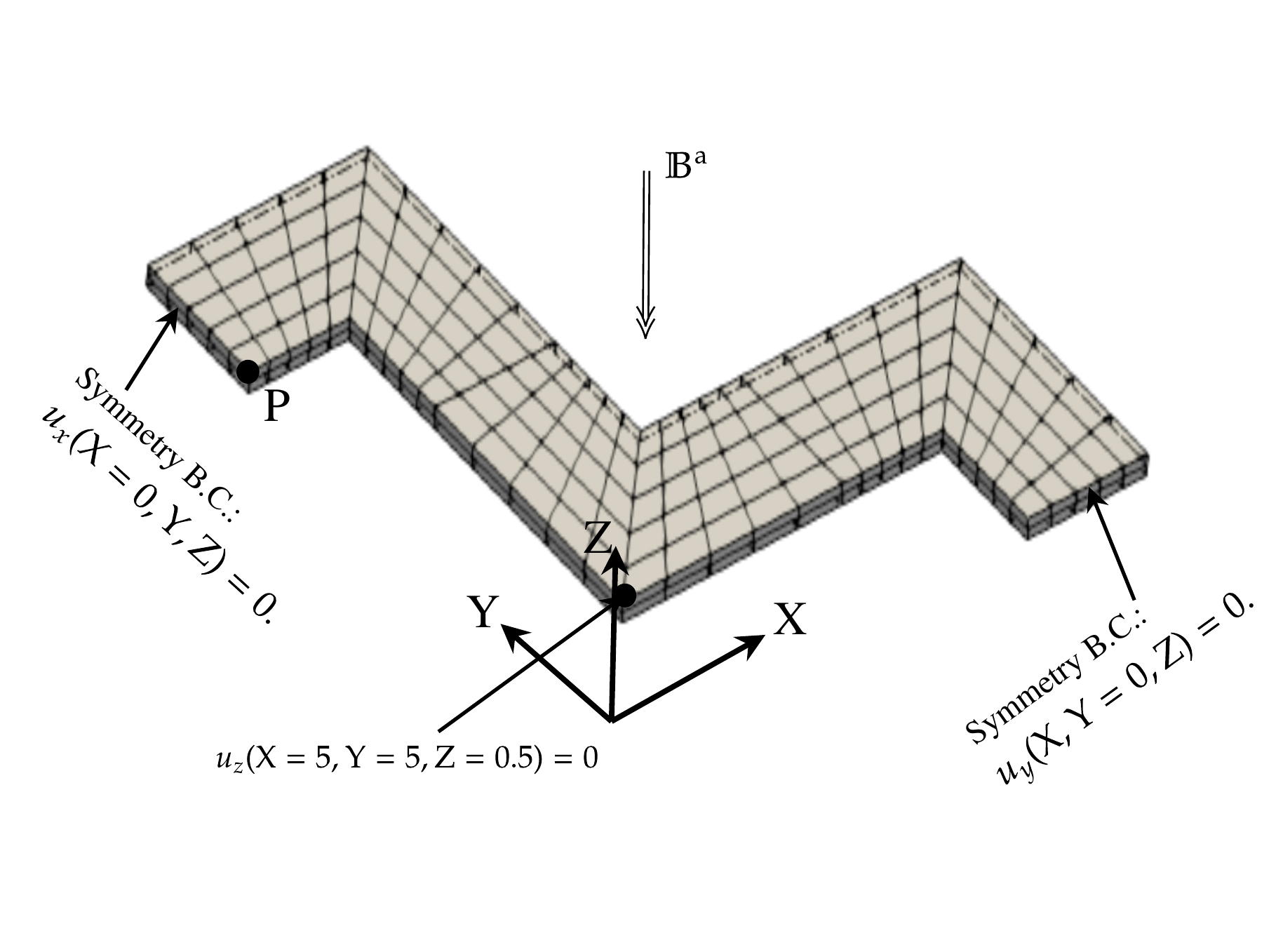} \label{fig-hm-pattern-mesh}}
 \caption{Pattern: (a) problem set up and (b) finite element mesh of a quarter of the model (corresponding to the hatched portion in Fig. \ref{fig-hm-pattern-geom}) along with the boundary conditions and loading.}
\end{figure}

\begin{figure}[H]
 \centering
 \subfloat[$\mathbb{B}^{a}_{z} = - \, 10$ mT]{\includegraphics[clip, scale=0.3]{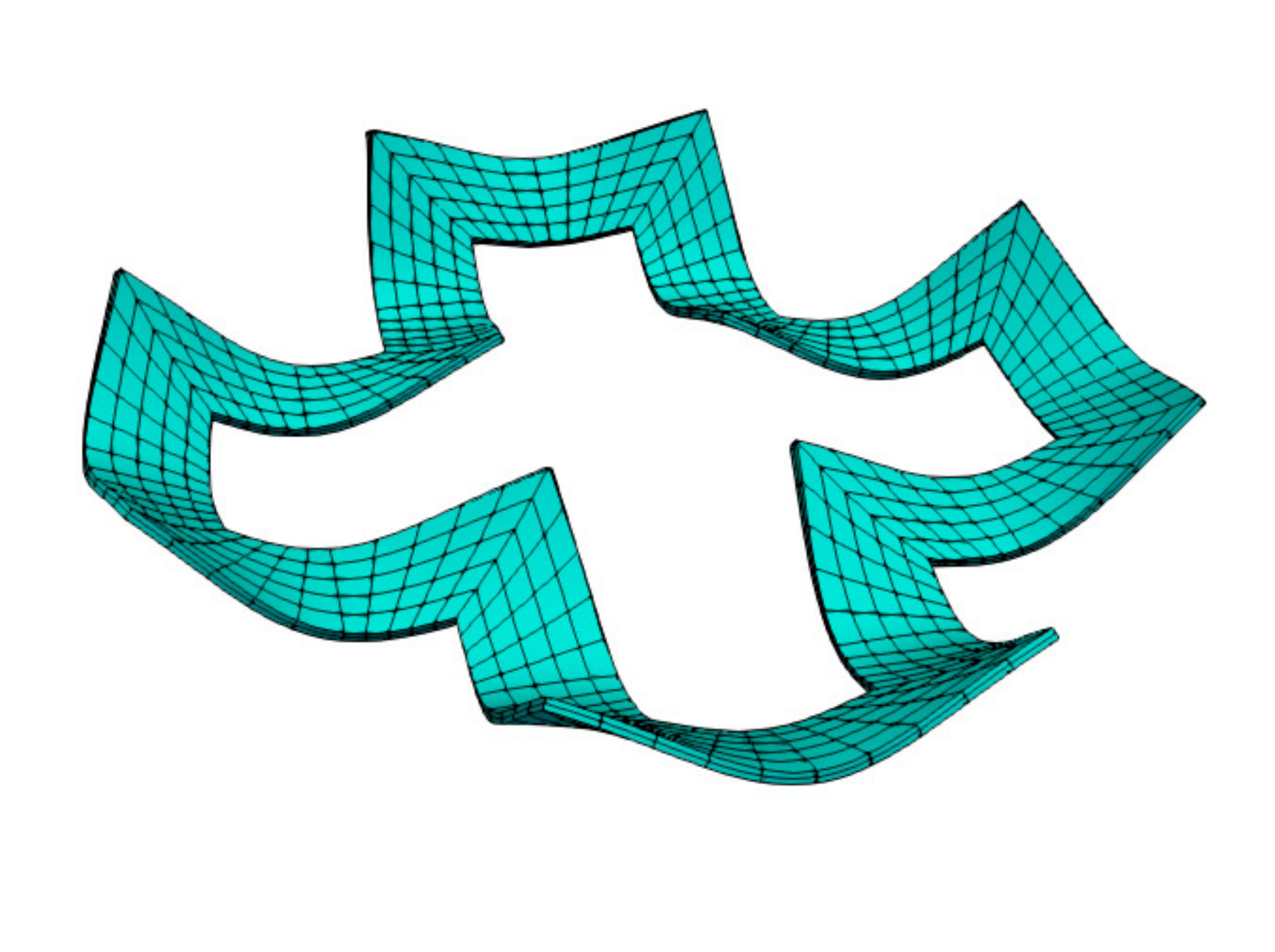}}
 \subfloat[$\mathbb{B}^{a}_{z} = - \, 50$ mT]{\includegraphics[clip, scale=0.3]{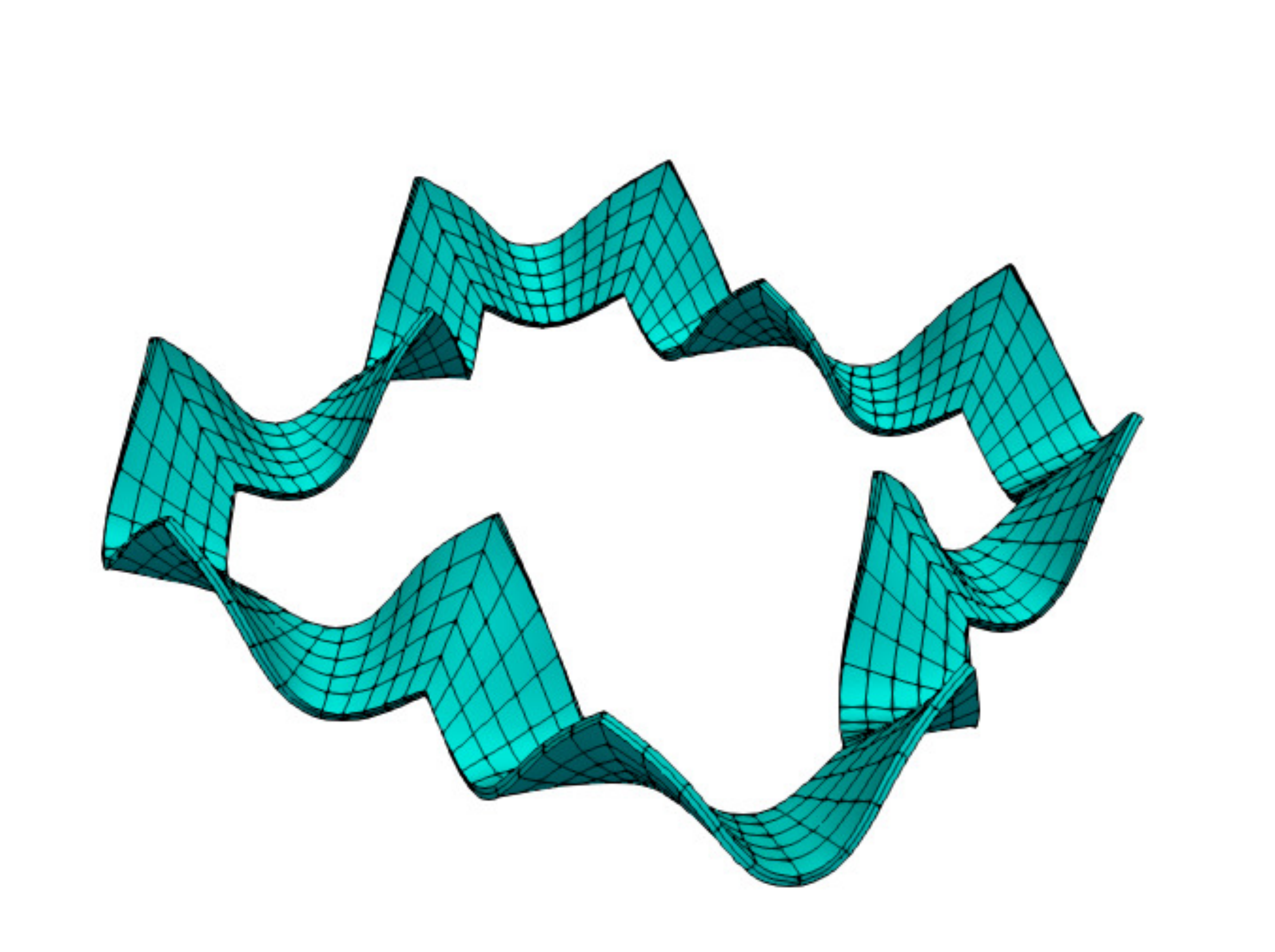}} \\
 \subfloat[$\mathbb{B}^{a}_{z} = - \, 100$ mT]{\includegraphics[clip, scale=0.3]{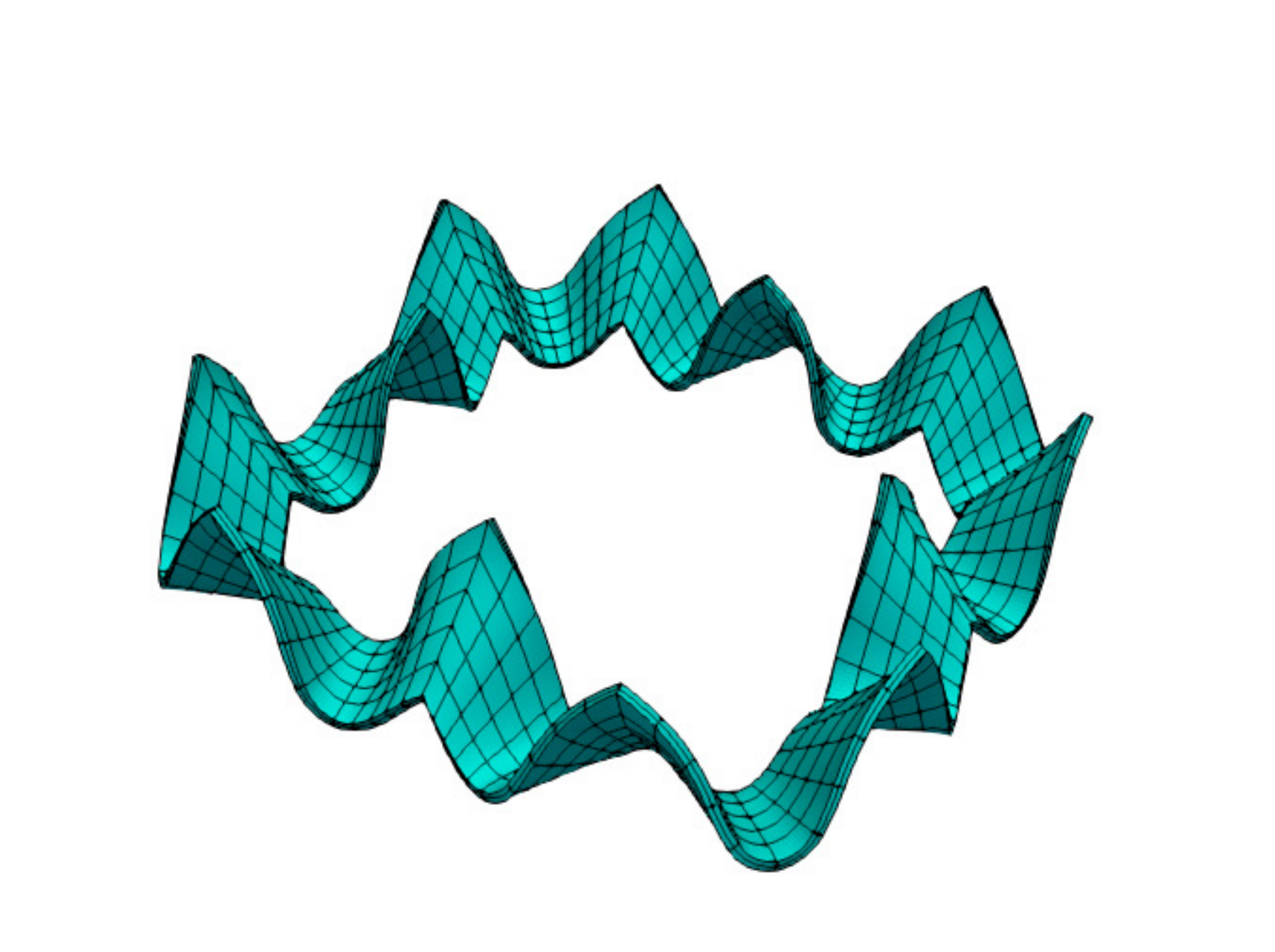}}
 \subfloat[$\mathbb{B}^{a}_{z} = - \, 200$ mT]{\includegraphics[clip, scale=0.3]{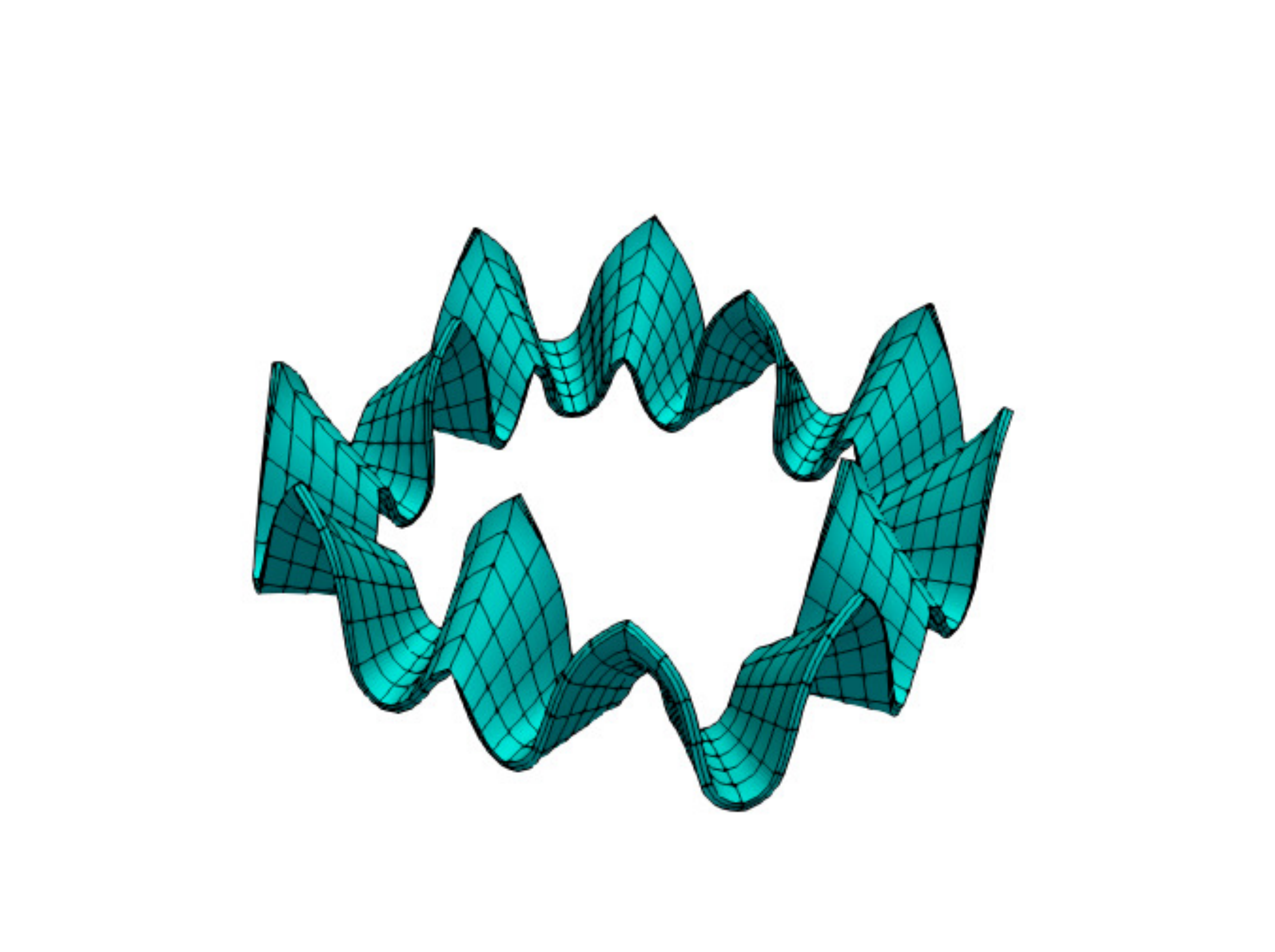}}
 \caption{Pattern: deformed configurations at four different values of the applied magnetic field obtained with the hyperelastic case only.}
 \label{fig-hm-pattern-defshapes}
\end{figure}

In this example, we simulate one of the patterns originating from the response of configurable hard magnetic fillers under an external magnetic field as illustrated in Zhao et al. \cite{ZhaoJMPS2019}. The geometry of the model and the directions of the residual magnetic fields are shown in Fig. \ref{fig-hm-pattern-geom}. The magnitude of the residual magnetic field, $\mathbbm{B}^r$, is $102 \; \milliTesla$ and the maximum value of the applied magnetic field is $200 \; \milliTesla$ in the negative Z-direction, i.e., $\mathbbm{B}^{a}_{z}=-200 \; \milliTesla$. Due to the symmetry of geometry and loading conditions, only a quarter of the model is simulated using a mesh consisting of 300 BQ2/BQ1 elements and 3355 nodes, see Fig. \ref{fig-hm-pattern-mesh}. The hyperelastic material model is the truly incompressible Neo-Hookean one with a shear modulus of $\mu=330$ kPa, an experimentally-driven parameter from Zhao et al. \cite{ZhaoJMPS2019}. The deformed shapes of the configuration for four different values due to the applied magnetic field shown in Fig. \ref{fig-hm-pattern-defshapes} demonstrate the ability of the proposed numerical framework to capture the complex deformations using even coarse meshes.

\begin{figure}[H]
 \centering
 \subfloat[$\mu_{v}=165$ kPa]{\includegraphics[clip, scale=0.55]{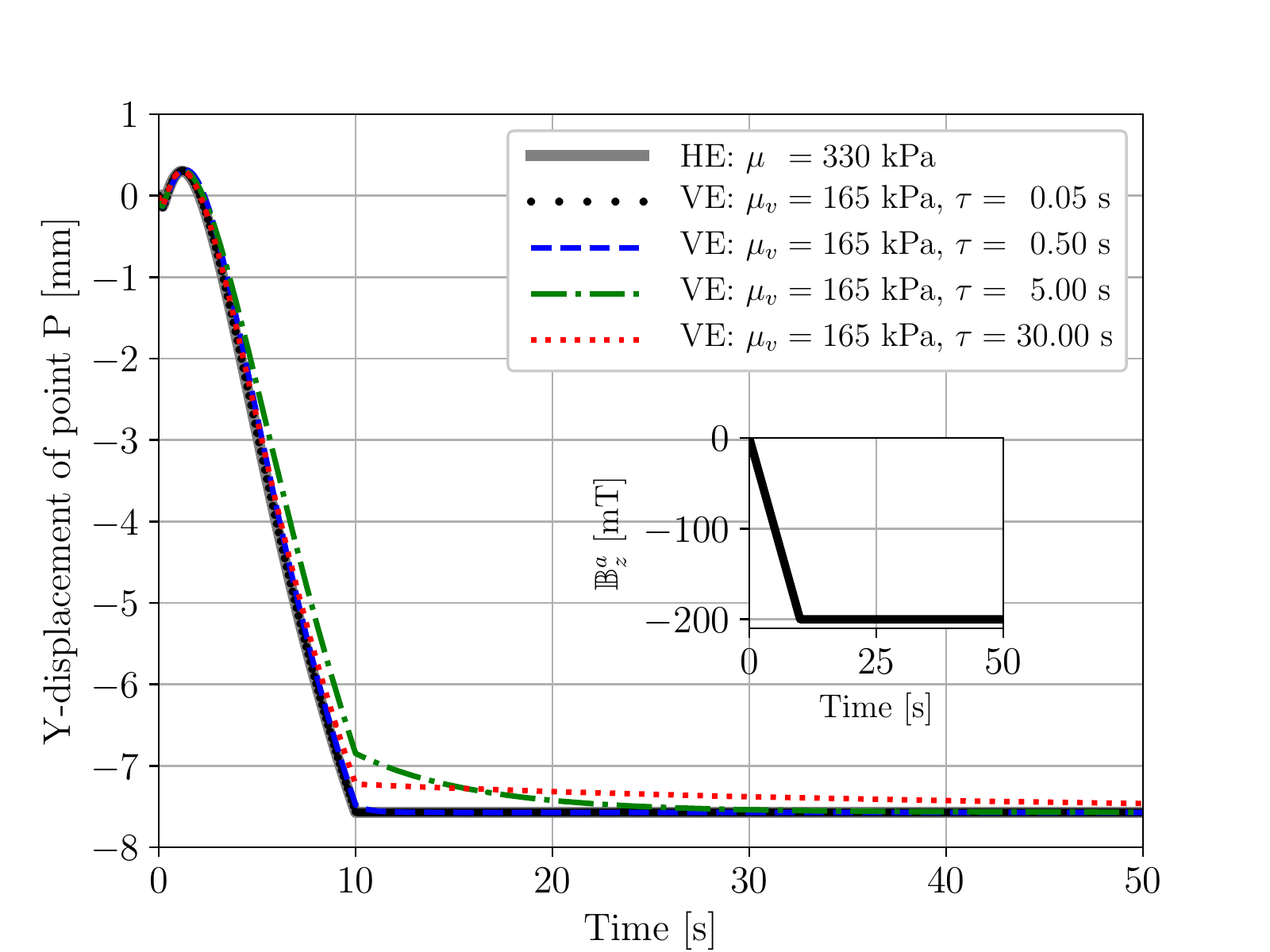}} \\
 \subfloat[$\mu_{v}=330$ kPa]{\includegraphics[clip, scale=0.55]{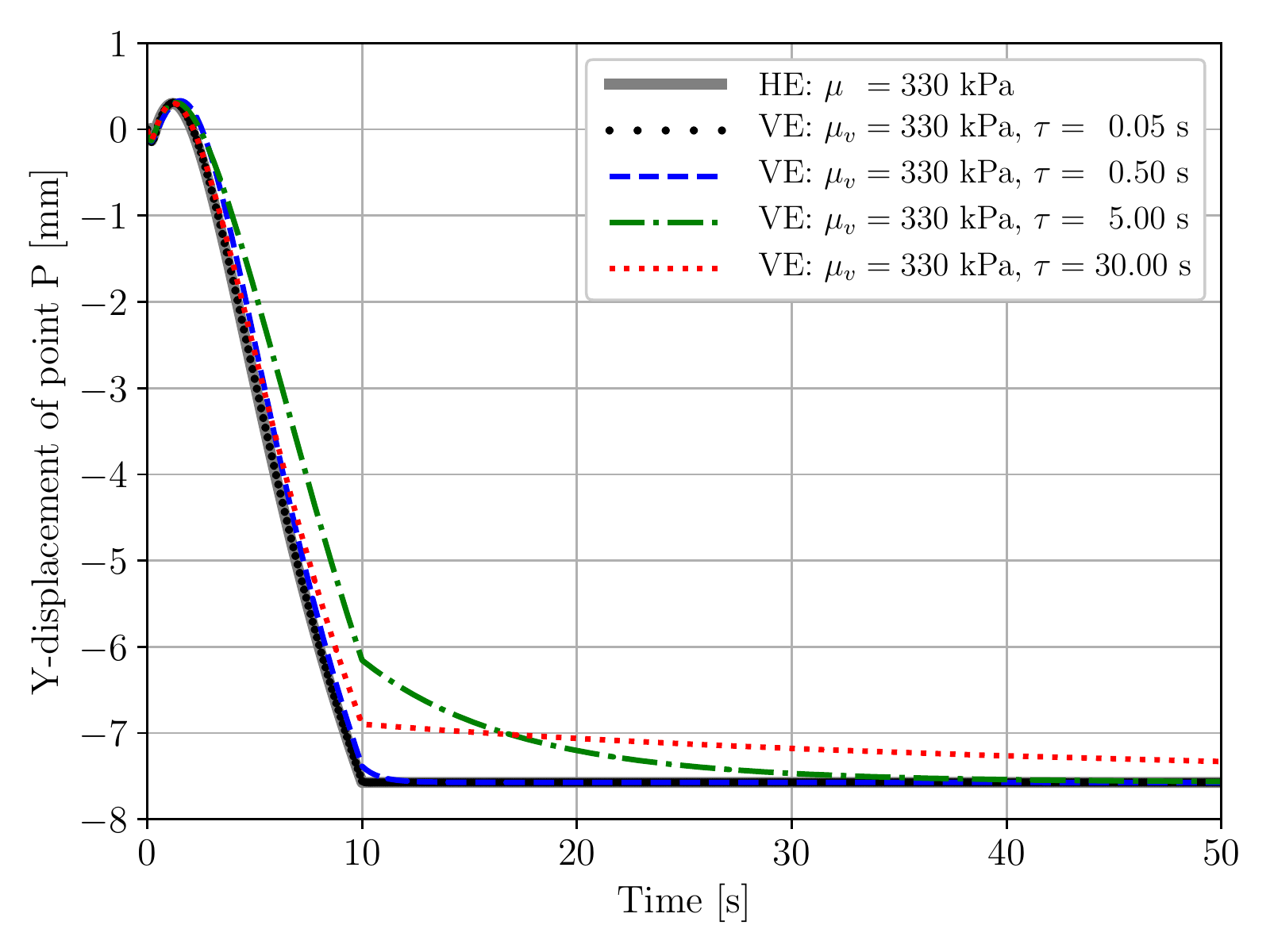}} \\
 \subfloat[$\mu_{v}=660$ kPa]{\includegraphics[clip, scale=0.55]{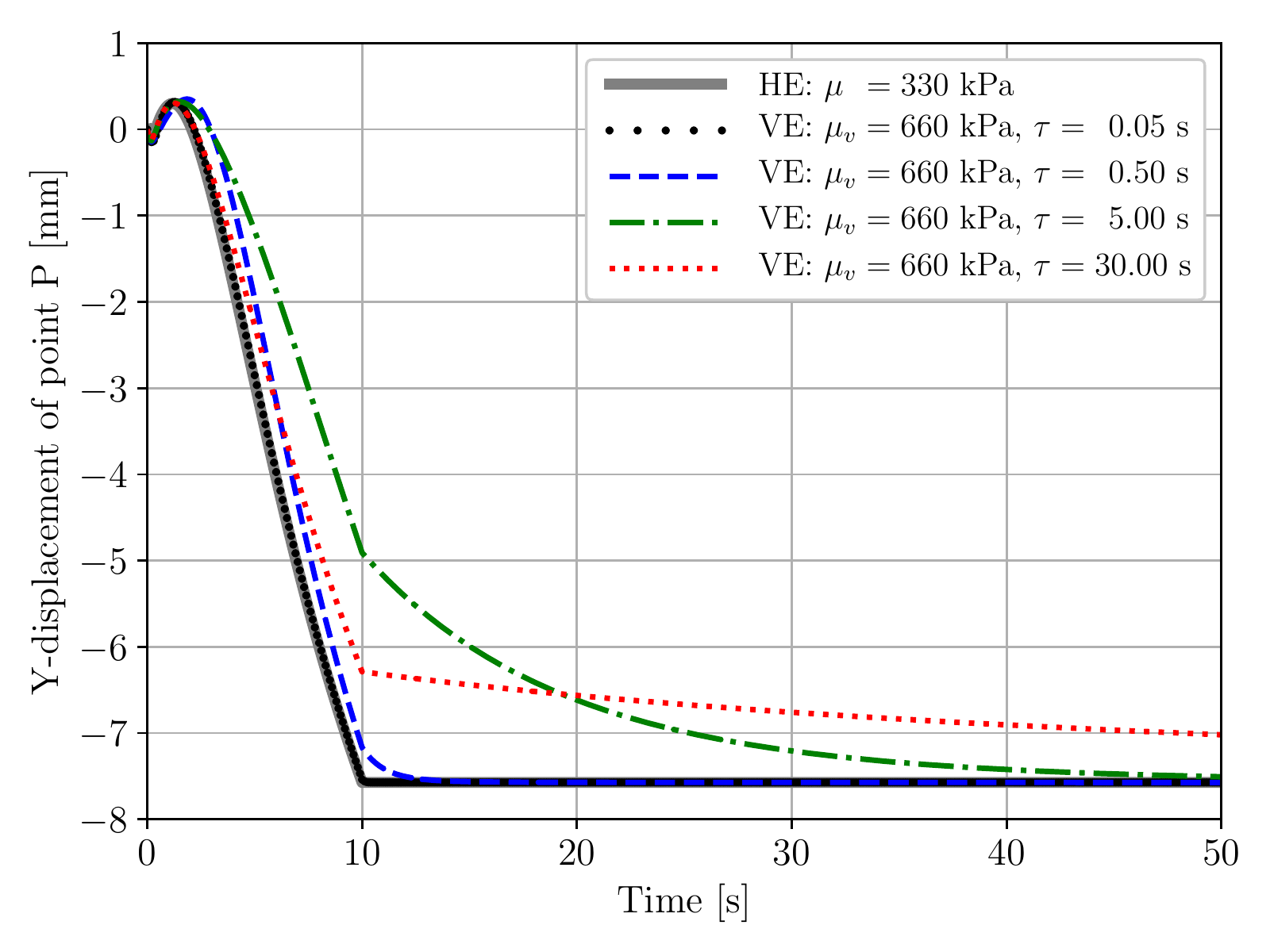}}
 \caption{Pattern: time evolution of Y-displacement of the point P obtained with the hyperelastic (HE) model and viscoelastic (VE) model for different material parameters.}
 \label{fig-hm-pattern-graphs}
\end{figure}

\begin{figure}[H]
 \centering
 \subfloat[At $t=10$s with $\mu_{v} = 165$ kPa, $\, \tau=5$ s]{\includegraphics[clip, scale=0.4]{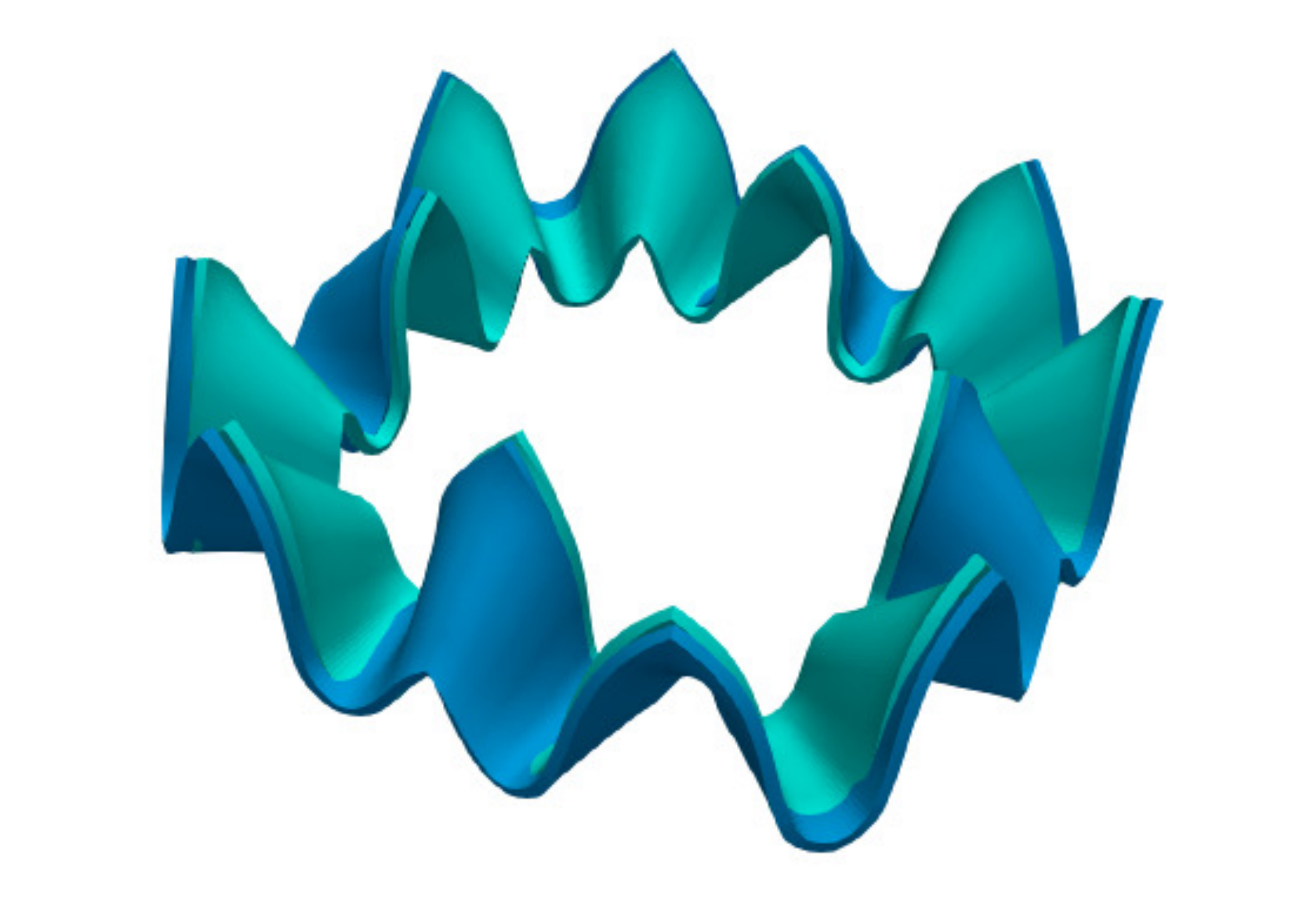}}
 \subfloat[At $t=20$s with $\mu_{v} = 165$ kPa, $\, \tau=5$ s]{\includegraphics[clip, scale=0.4]{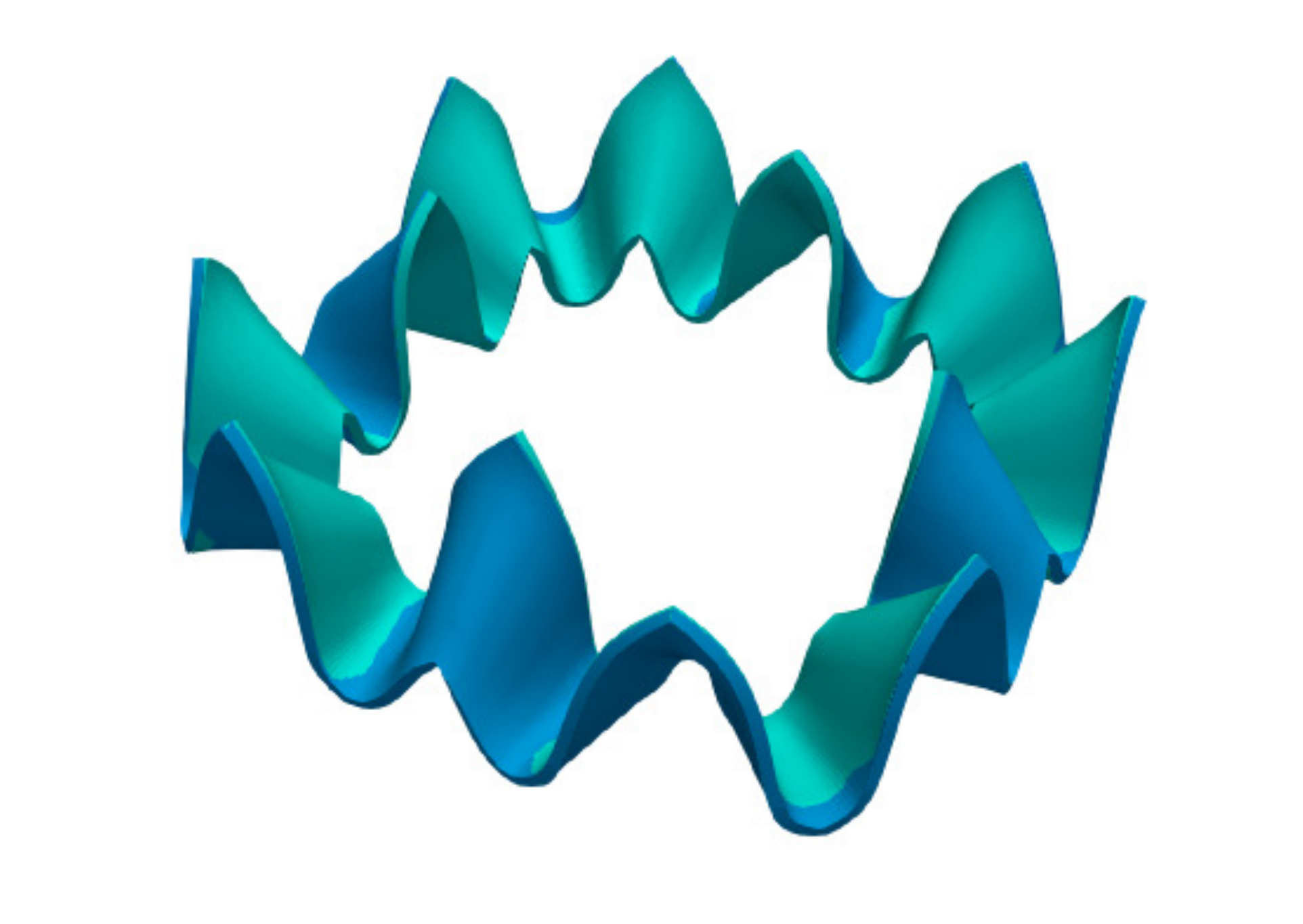}} \\
 \subfloat[At $t=10$s with $\mu_{v} = 660$ kPa, $\, \tau=5$ s]{\includegraphics[clip, scale=0.4]{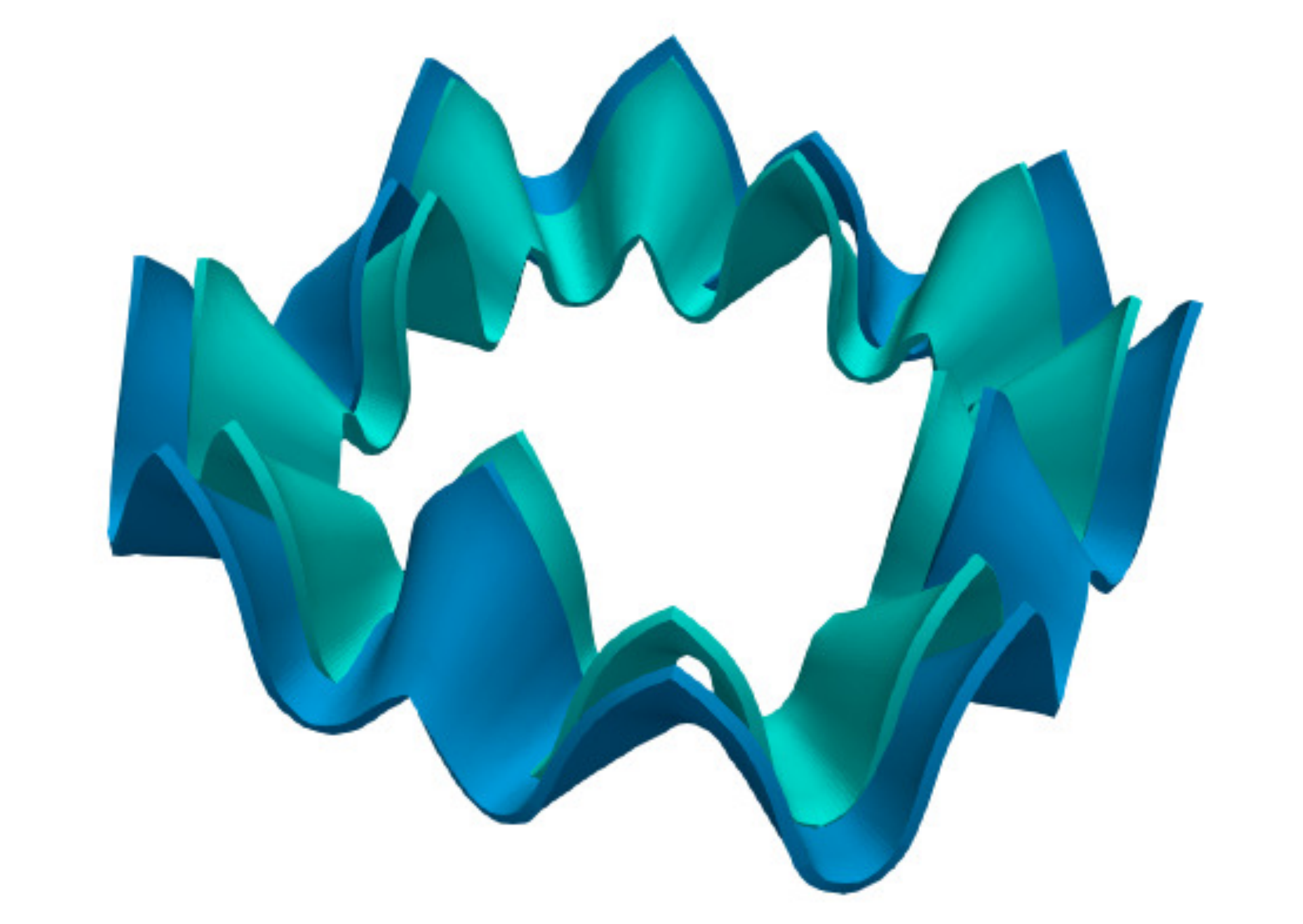}}
 \subfloat[At $t=20$s with $\mu_{v} = 660$ kPa, $\, \tau=5$ s]{\includegraphics[clip, scale=0.4]{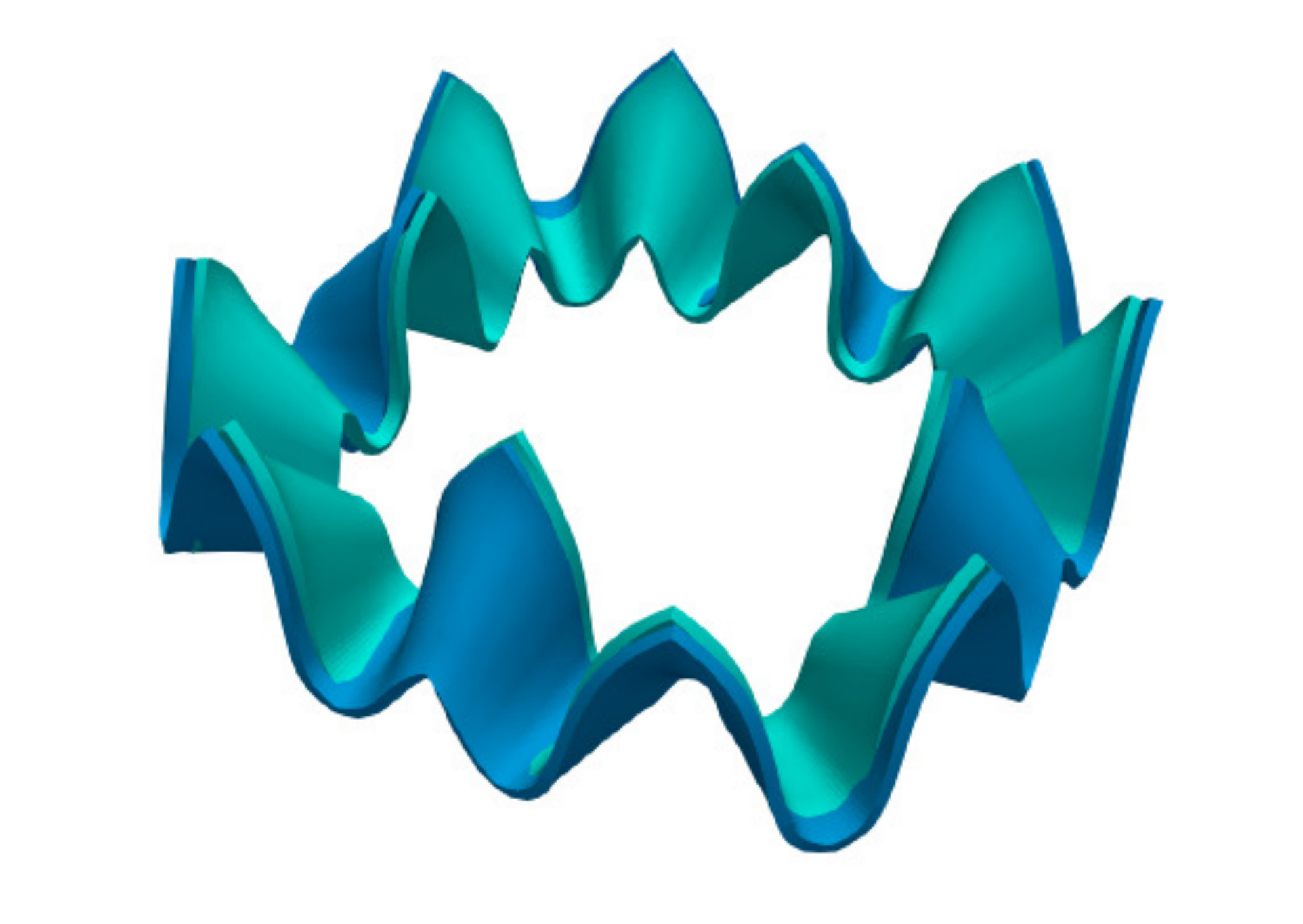}}
 \caption{Pattern: comparison of deformed shapes at $t=10$ s and $t=20$ s obtained with pure hyperelastic model (cyan) and viscoelastic model (blue) with $\mu_{v} = 165$ kPa and $\mu_{v} = 660$ kPa and $\, \tau=5$ s.}
 \label{fig-hm-pattern-defshapes-2}
\end{figure}

Now, we assess the effect of viscoelasticity on the deformation response on the patterns. For this purpose,  a viscoelastic model with a single Maxwell element is chosen, and the effect of the viscous shear modulus ($\mu_{v}$) and the relaxation time ($\tau$) is studied on the response. The applied magnetic field is ramped up linearly during the first ten seconds to a maximum value of $- \, 200 \; \milliTesla$ and then is held constant at the maximum value. With a constant time step of $\Delta t = 0.1$ s, the simulations are performed for various combinations of $\mu_{v} = \{165, 330, 660\}$ kPa and $\tau=\{0.05,0.5,5.0,30.0\}$ s and the results are presented in the form of temporal evolution of Y-displacement of the point P in Fig. \ref{fig-hm-pattern-graphs}. The graphs also include the displacement response obtained with the pure hyperelastic case for comparison. As shown, the creep response of the model due to the viscoelastic effects is captured. These graphs indicate that the time lag between the response obtained with the pure hyperelastic case and the viscoelastic case increases with either the viscous shear modulus or the relaxation time, or both. We can also observe that the effect of the relaxation time is more pronounced when compared with the effect of the viscous shear modulus. For both $\tau=0.05$ s and $\tau=0.5$ s, the viscous shear modulus has a negligible effect on the response of the model when compared with the pure hyperelastic case. Differences in the response appear only for $\tau>=5$ s. That means, for a higher relaxation time, the structure very slowly attains a relaxed position compared to a shorter relaxation time in which it will go to the relaxed state very quickly, cf. Fig.  \ref{fig-hm-pattern-graphs}. 
Figure \ref{fig-hm-pattern-defshapes-2} shows differences in the deformed shapes at $t=10$ s and $t=20$ s obtained with $\tau=5$ s and $\mu_{v}=165$ kPa and $\mu_{v}=660$ kPa. 
If we compare purely elastic (cyan) and viscoelastic (blue) shapes in Fig. \ref{fig-hm-pattern-defshapes-2}, it is clear that for a given applied magnetic field and residual magnetic flux, the viscoelastic material deforms less as both elastic and viscous parts of the stress eventually greater than the purely elastic part resulting in a stiffer behaviour of the material.


\subsubsection{A hard magneto-active gripper}
\begin{figure}[H]
 \centering
 \subfloat[]{\includegraphics[clip, scale=0.5]{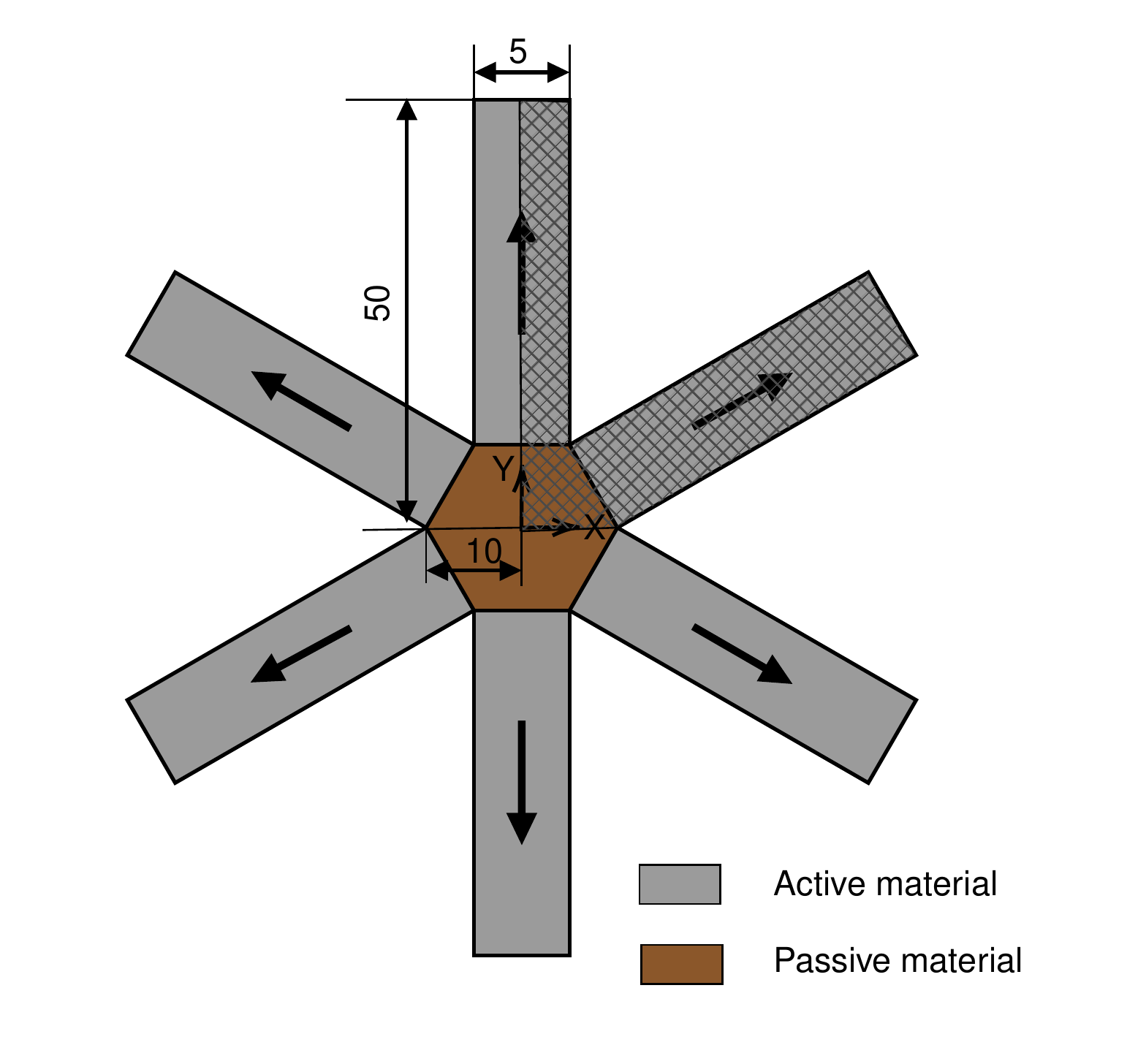} \label{fig-hm-gripper-geom} }
 \subfloat[]{\includegraphics[clip, scale=0.5]{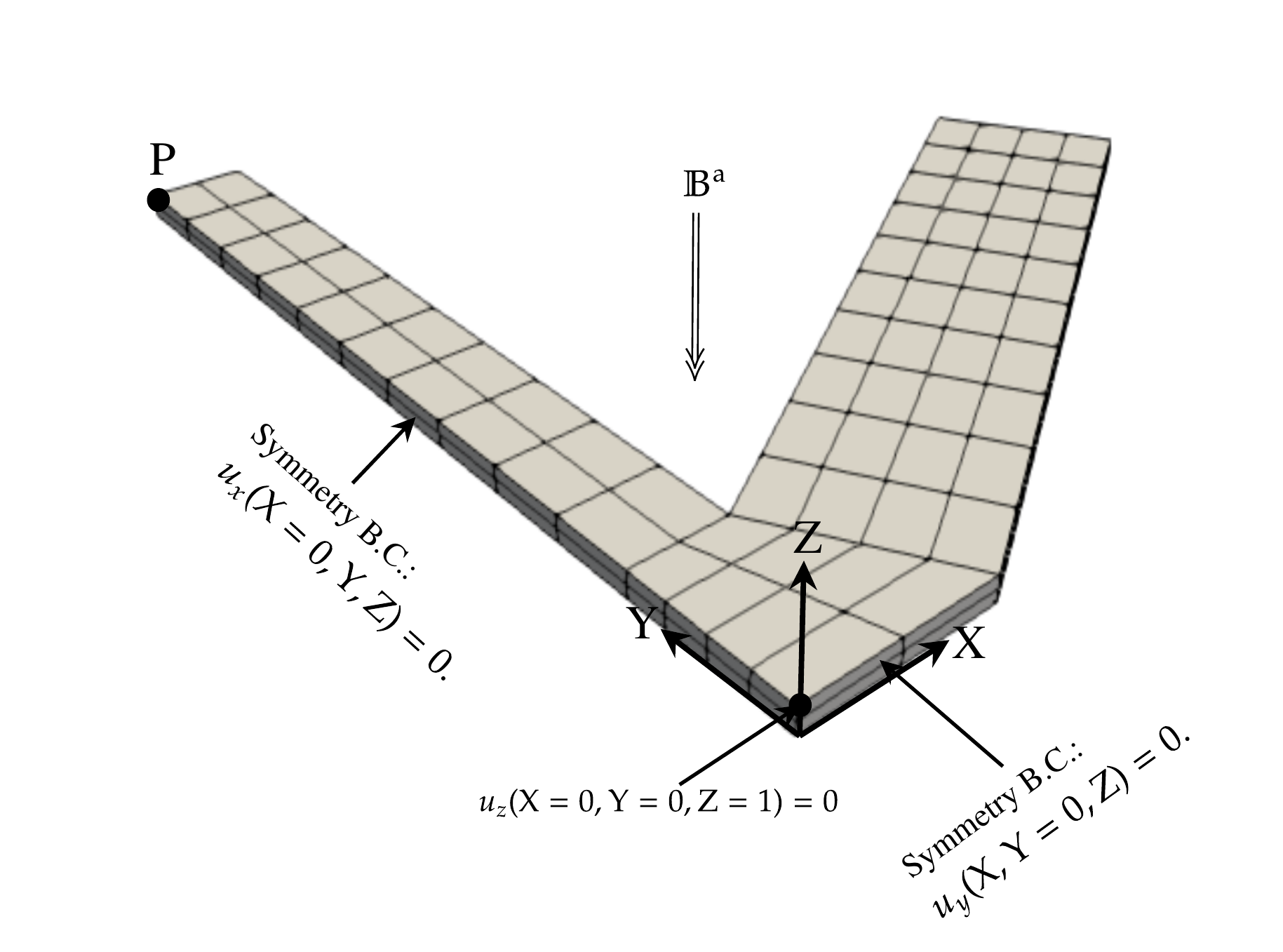} \label{fig-hm-gripper-mesh} }
 \caption{Hard magnetic gripper: (a) configuration of the gripper with arrows indicate the direction of residual magnetic field, and (b) finite element mesh of the quarter model (corresponding to the hatched portion in Fig. \ref{fig-hm-gripper-geom}), along with the boundary conditions and loading.}
\end{figure}

\begin{figure}[H]
 \centering
 \subfloat[loading case 1]{\includegraphics[clip, scale=0.5]{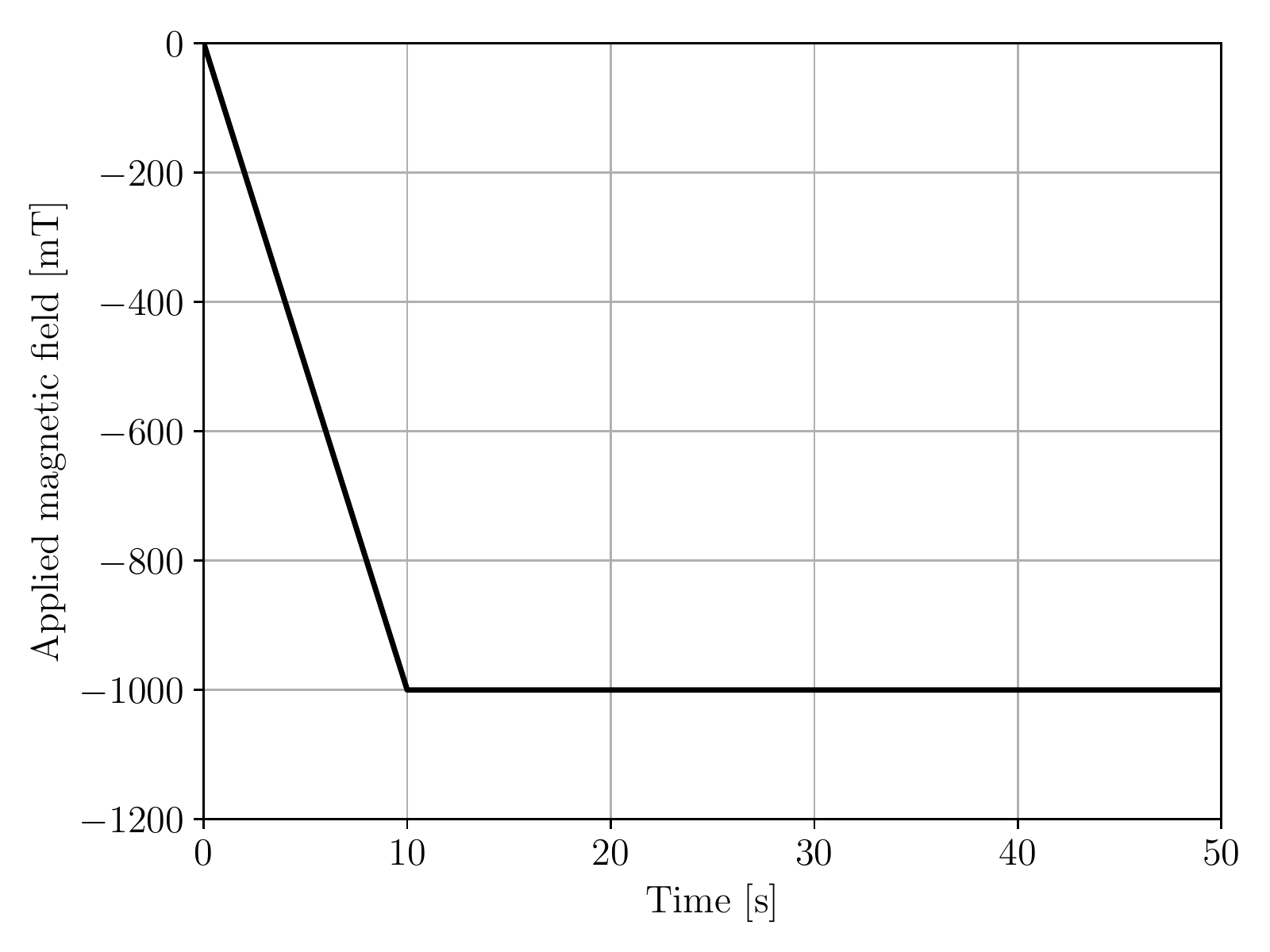} \label{fig-hm-gripper-loading1} }
 \subfloat[loading case 2]{\includegraphics[clip, scale=0.5]{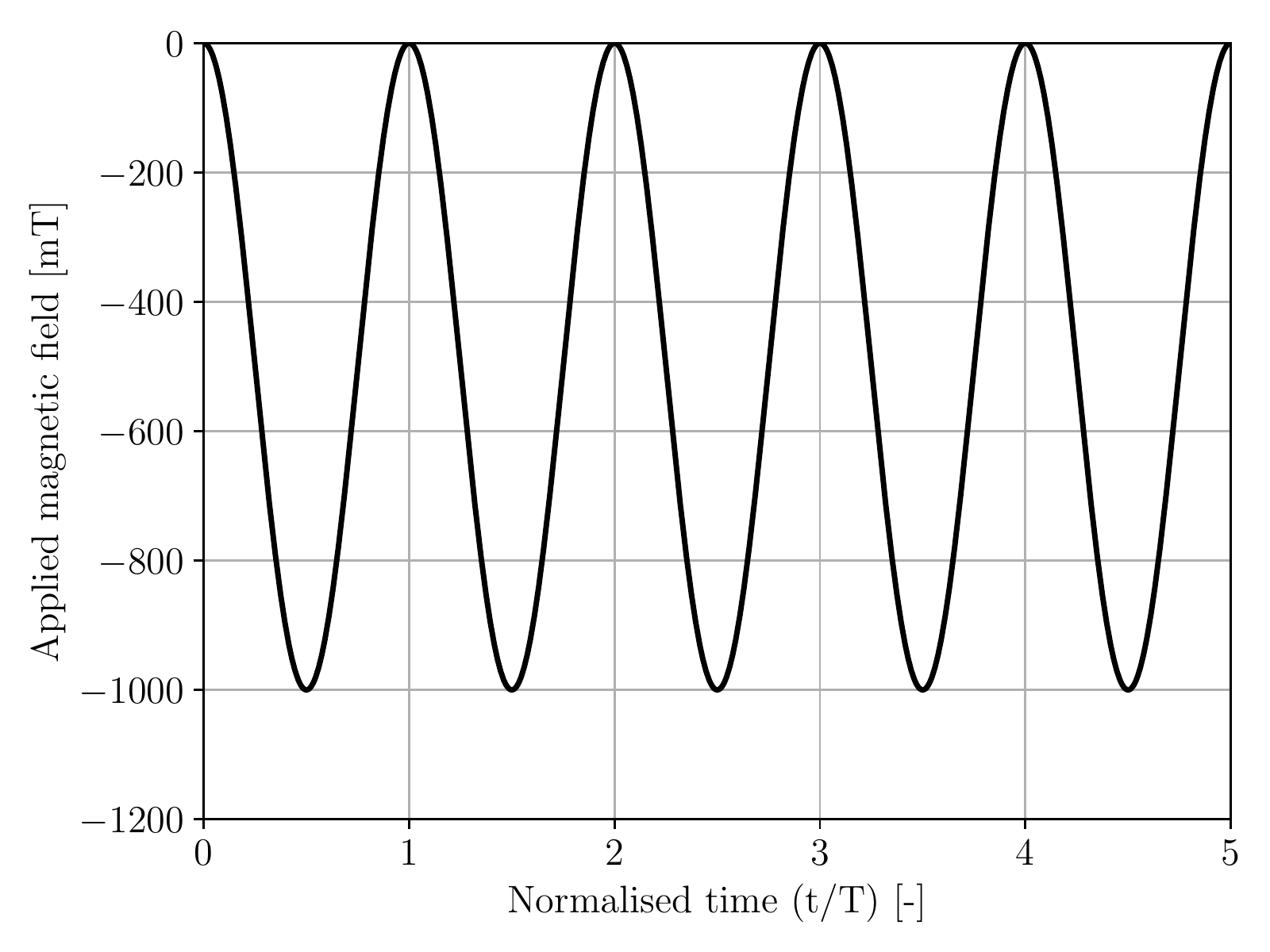} \label{fig-hm-gripper-loading2} }
 \caption{Hard magnetic gripper: loading cases.}
\label{fig-hm-gripper-loading}
\end{figure}

Soft robotic grippers based on MAPs are becoming popular due to the ease of their actuation \cite{JuNE2021,CarpenterAIS2021,LiSMS2021}. In this example, we consider a hard magnetic gripper consisting of six fingers, as shown in Fig.  \ref{fig-hm-gripper-geom}. The thickness is assumed to be 1 mm. The material model is the truly incompressible Neo-Hookean one with a shear modulus of $\mu=330$ kPa. The residual magnetic field is assumed to be of unit magnitude, and the maximum value of the applied magnetic field is $1000 \; \milliTesla$ in the negative Z-direction. The finite element mesh (see Fig. \ref{fig-hm-gripper-mesh}) consists of 544 BQ2/BQ1 elements and 6165 nodes. With the Neo-Hookean model for the base hyperelastic material,  the gripper configuration is studied without and with viscoelastic effects to assess the influence of the material parameters appearing in the viscoelastic model on the response characteristics of gripper under the influence of the applied magnetic field. Two types of loading cases, as shown in Fig. \ref{fig-hm-gripper-loading} are considered. In the first loading case, the applied magnetic field is ramped up during the first ten seconds and then held constant while the applied magnetic field is varied sinusoidally in the second loading case.

\begin{figure}[H]
 \centering
 \subfloat[$\mu_{v} = 165$ kPa]{\includegraphics[clip, scale=0.5]{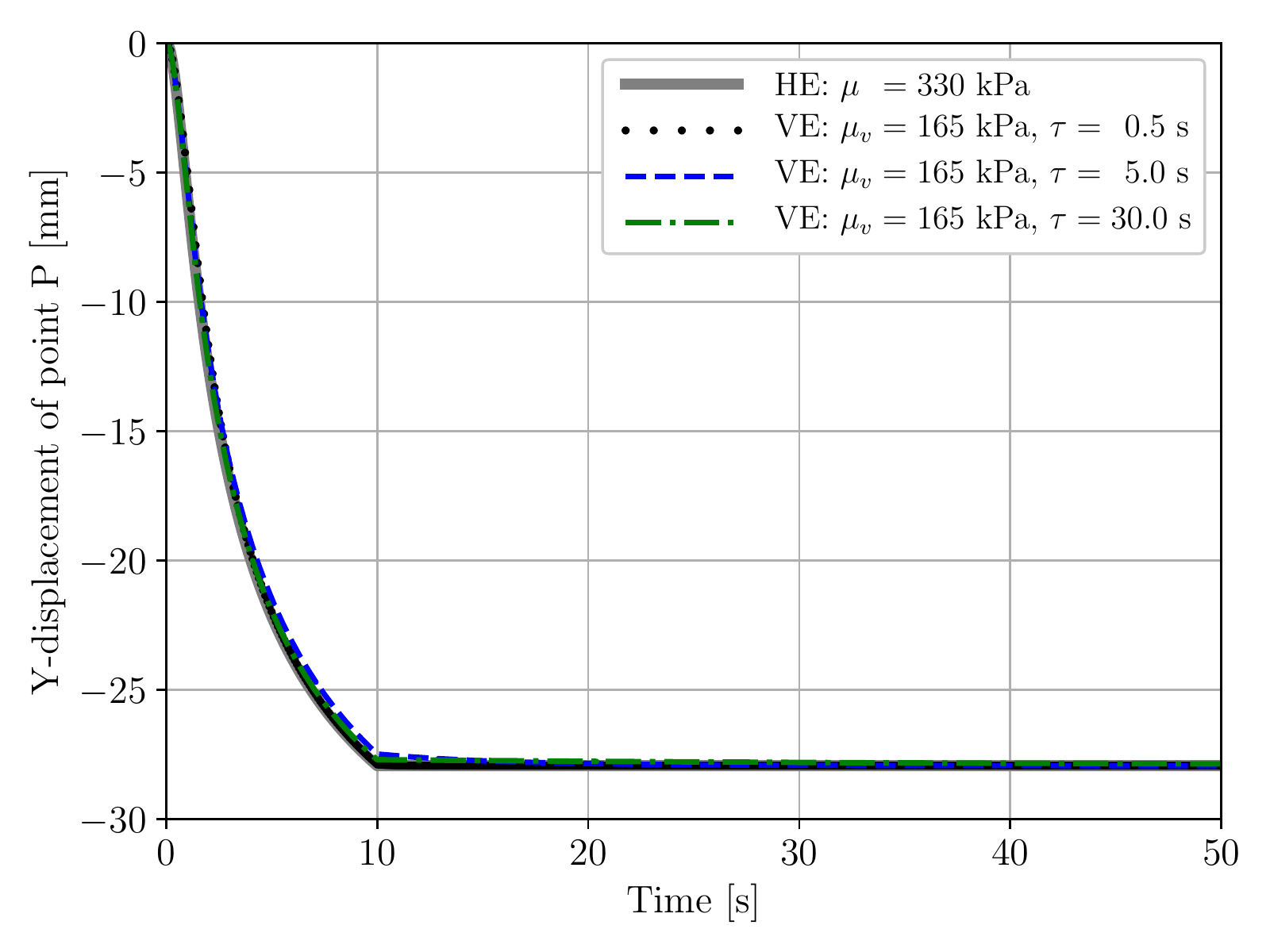}}
 \subfloat[$\mu_{v} = 330$ kPa]{\includegraphics[clip, scale=0.5]{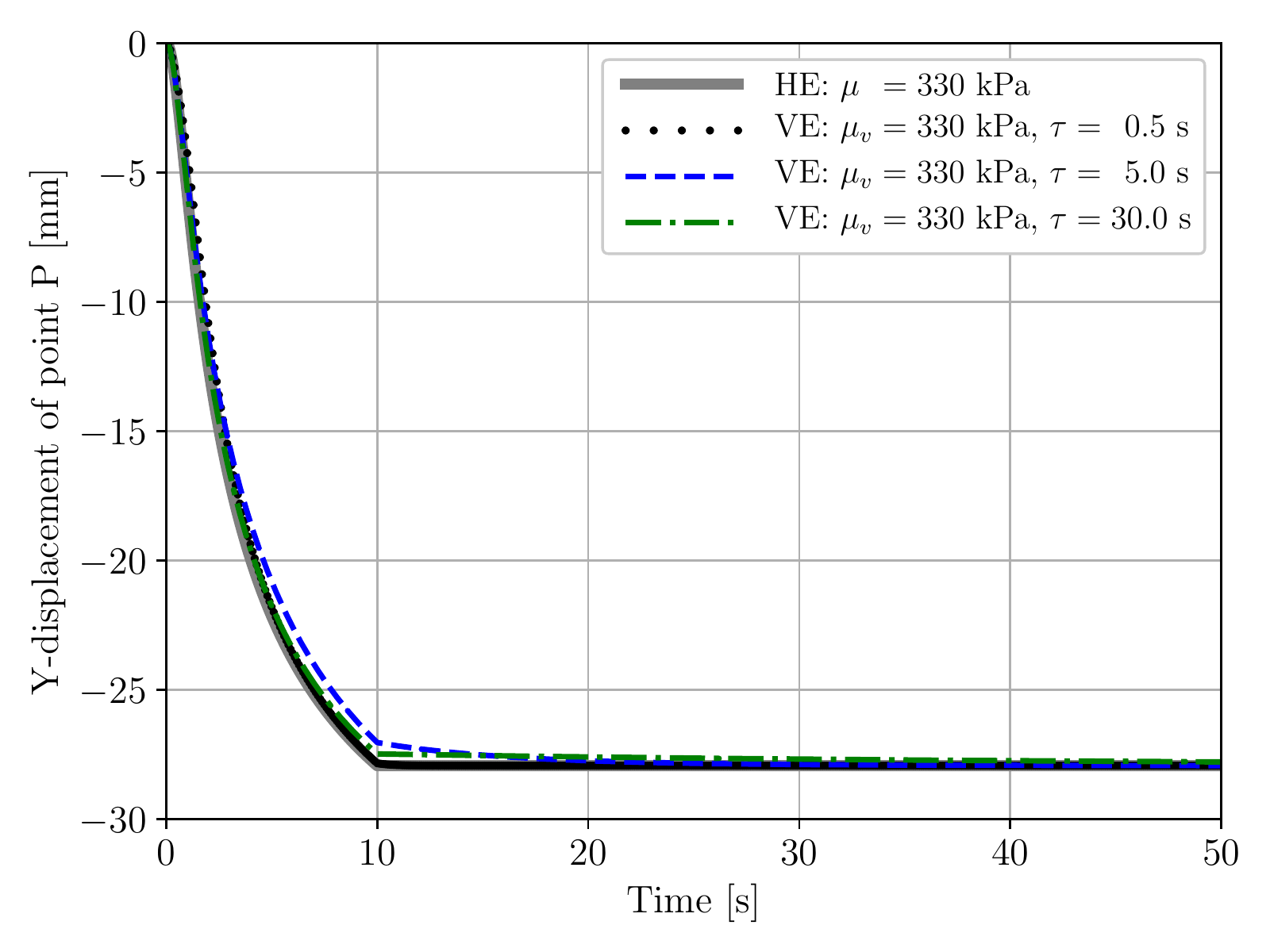}} \\
 \subfloat[$\mu_{v} = 660$ kPa]{\includegraphics[clip, scale=0.5]{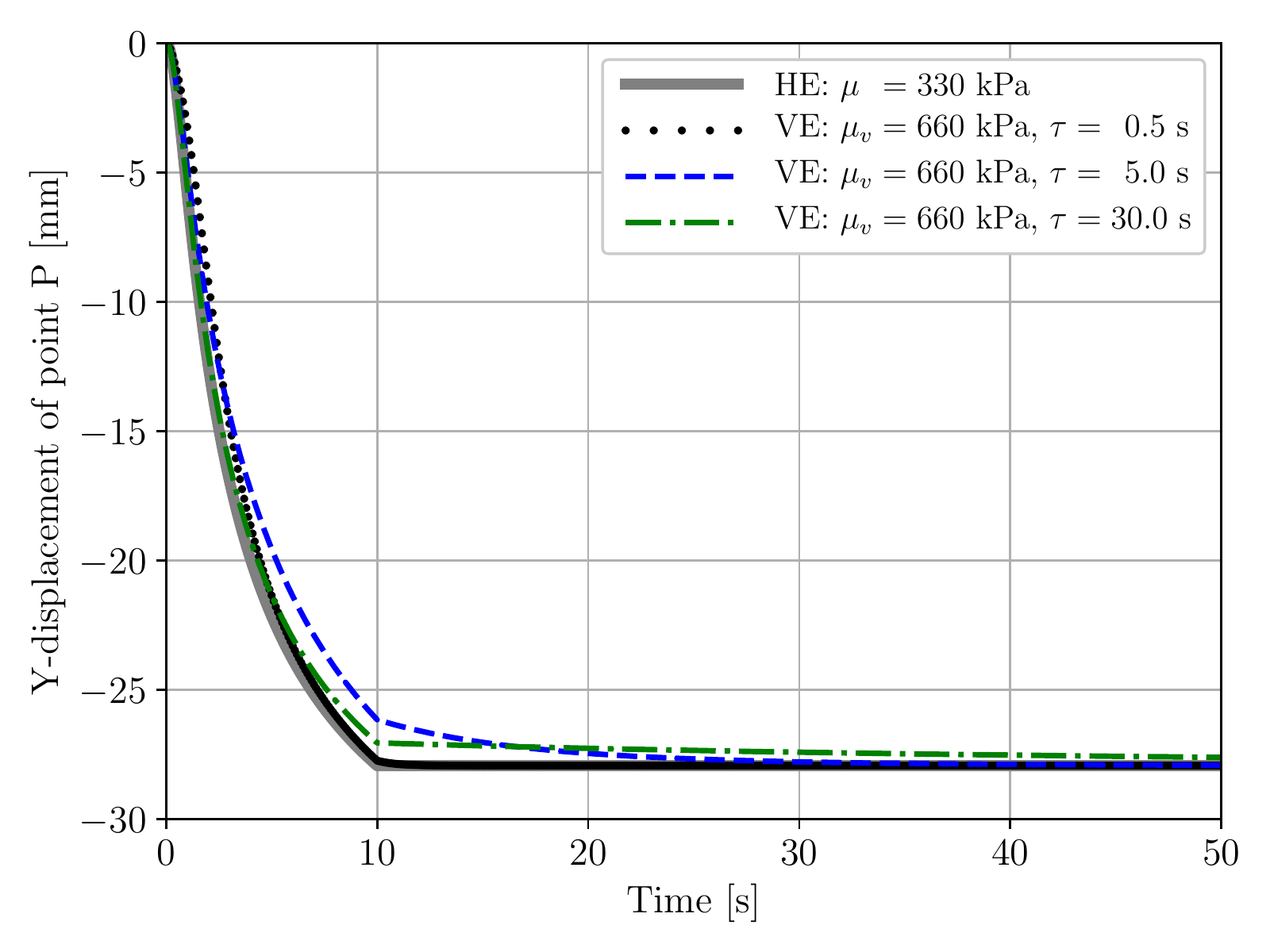}}
 \subfloat[$\mu_{v} = 3300$ kPa]{\includegraphics[clip, scale=0.5]{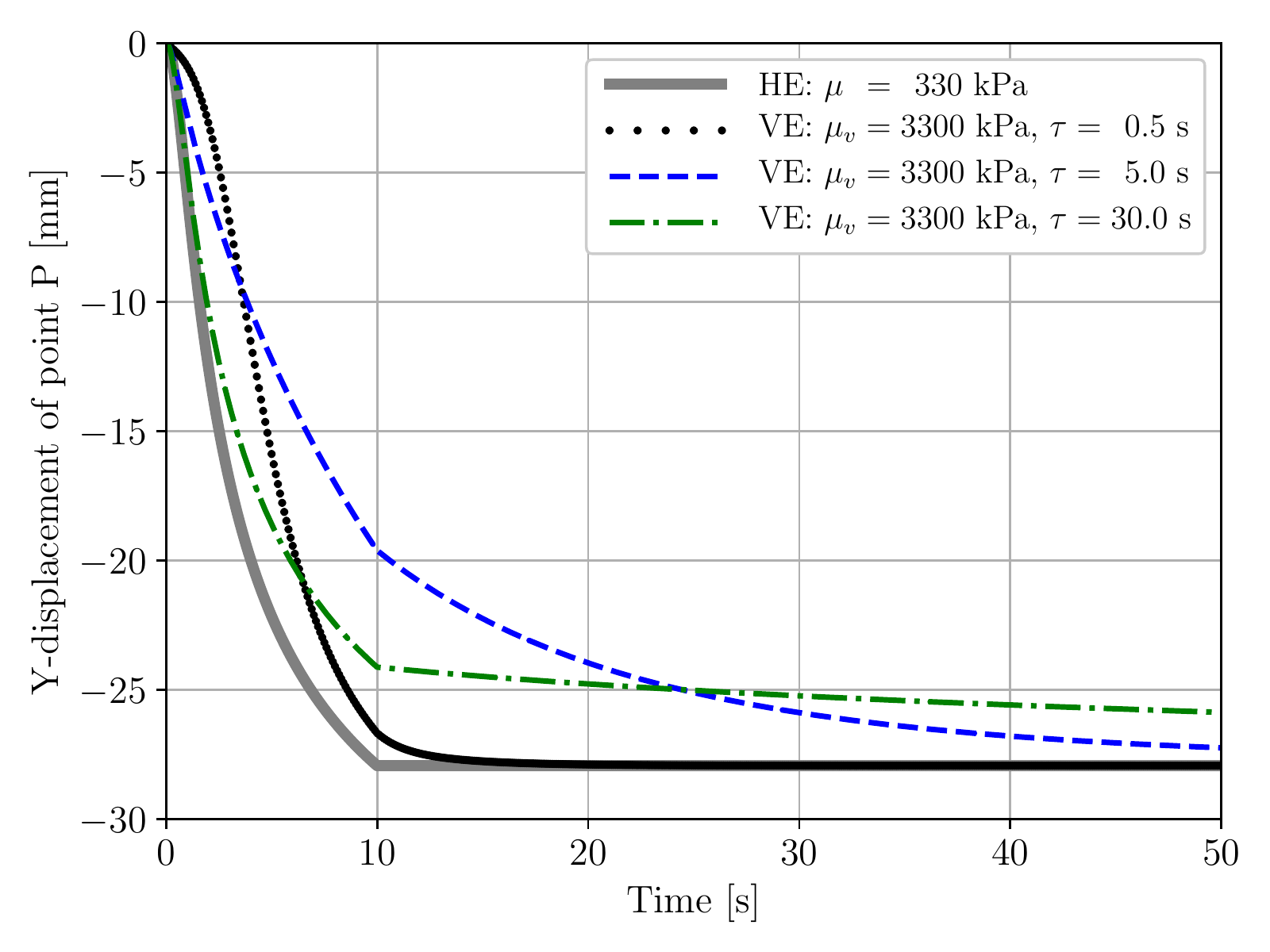}}
 \caption{Hard magnetic gripper: evolution of Y-displacement of point P for different values of viscous shear modulus ($\mu_{v}$) and relaxation time ($\tau$).}
 \label{fig-hm-gripper-graphs-1}
\end{figure}

\begin{figure}[H]
 \centering
 \subfloat[At $t=10$s with $\mu_{v} = 330$ kPa, $\, \tau=5.0$ s]{\includegraphics[clip, scale=0.5]{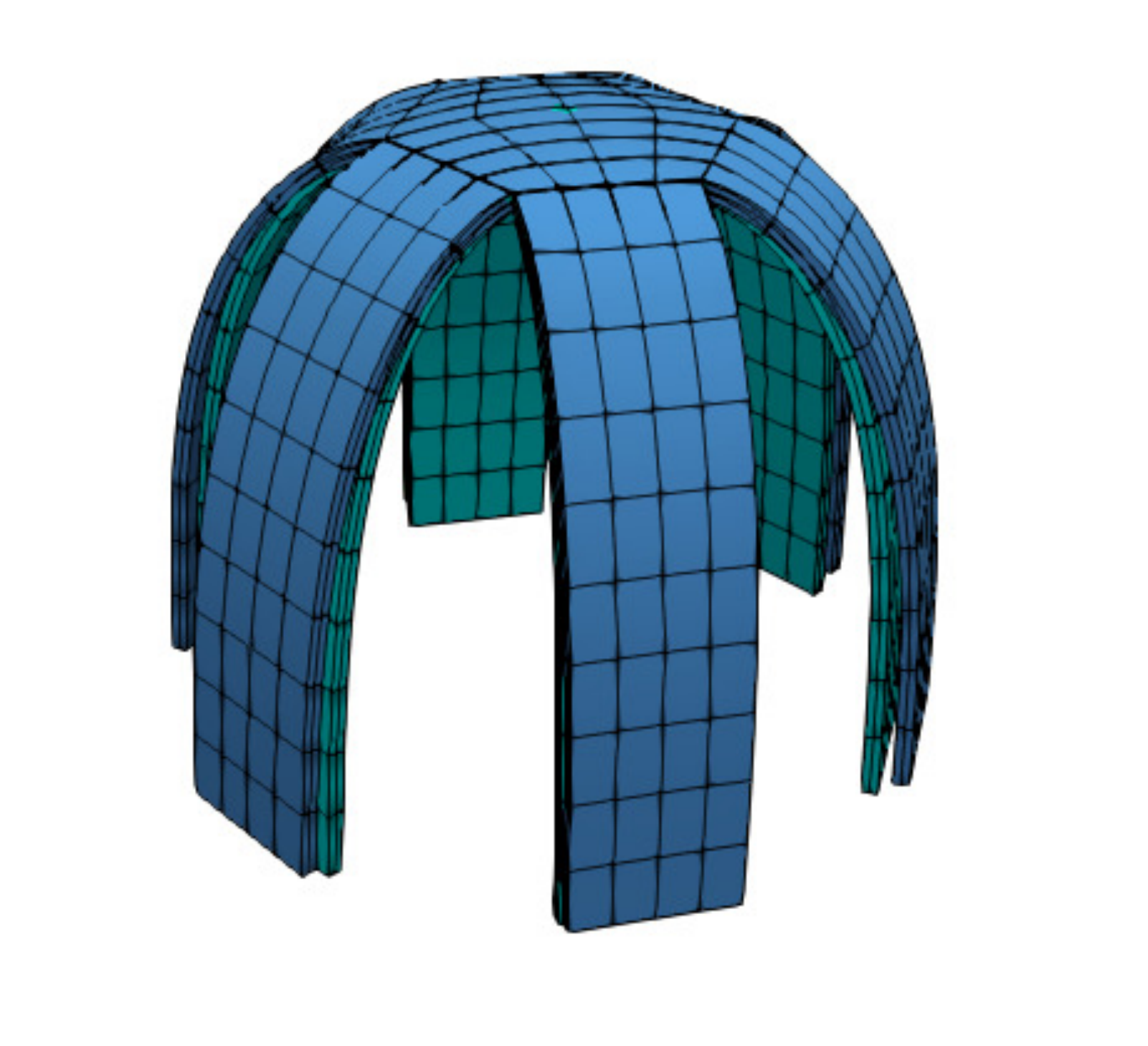}}
 \subfloat[At $t=30$s with $\mu_{v} = 330$ kPa, $\, \tau=5.0$ s]{\includegraphics[clip, scale=0.5]{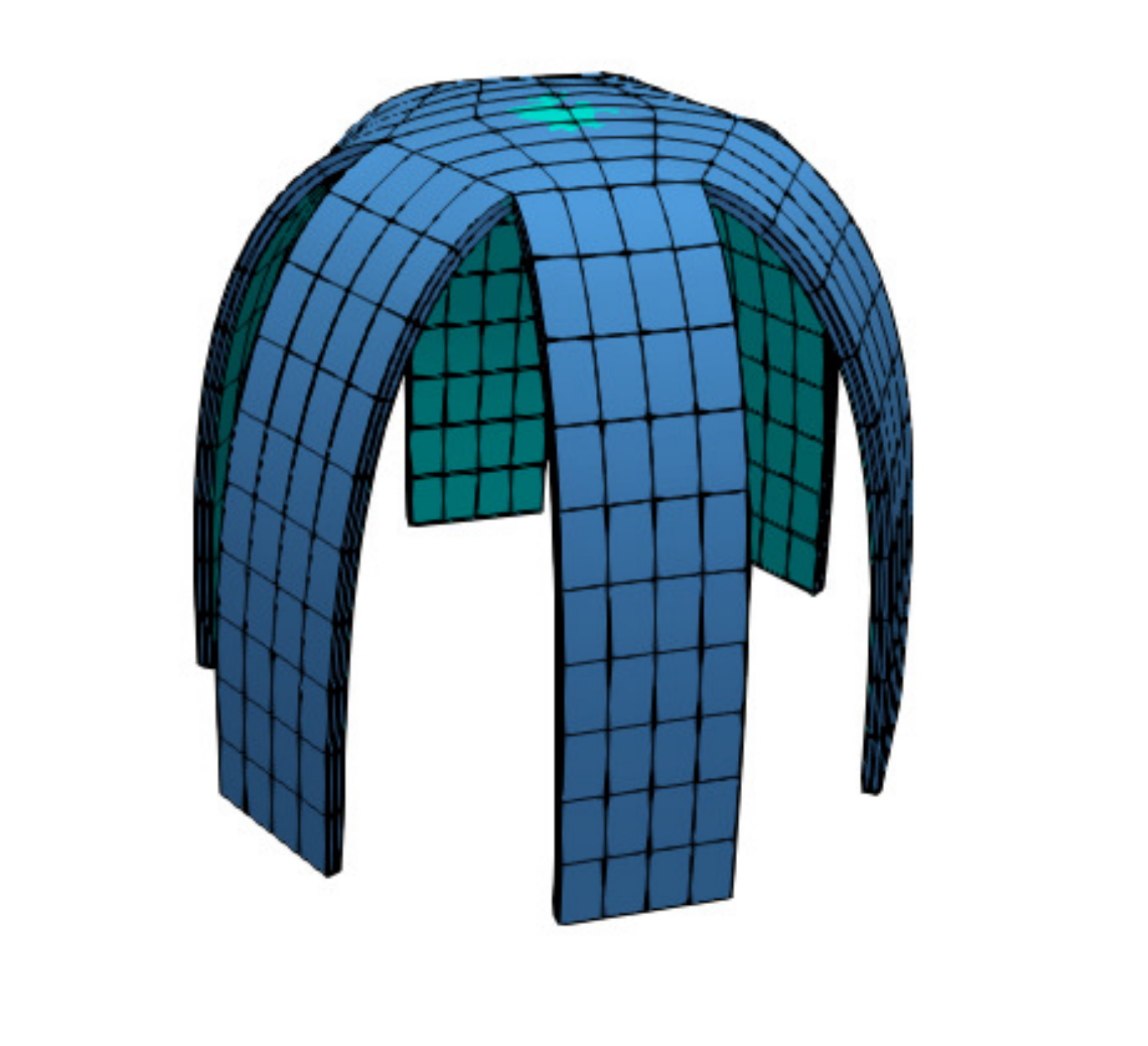}} \\
 \subfloat[At $t=10$s with $\mu_{v} = 3300$ kPa, $\, \tau=5.0$ s]{\includegraphics[clip, scale=0.5]{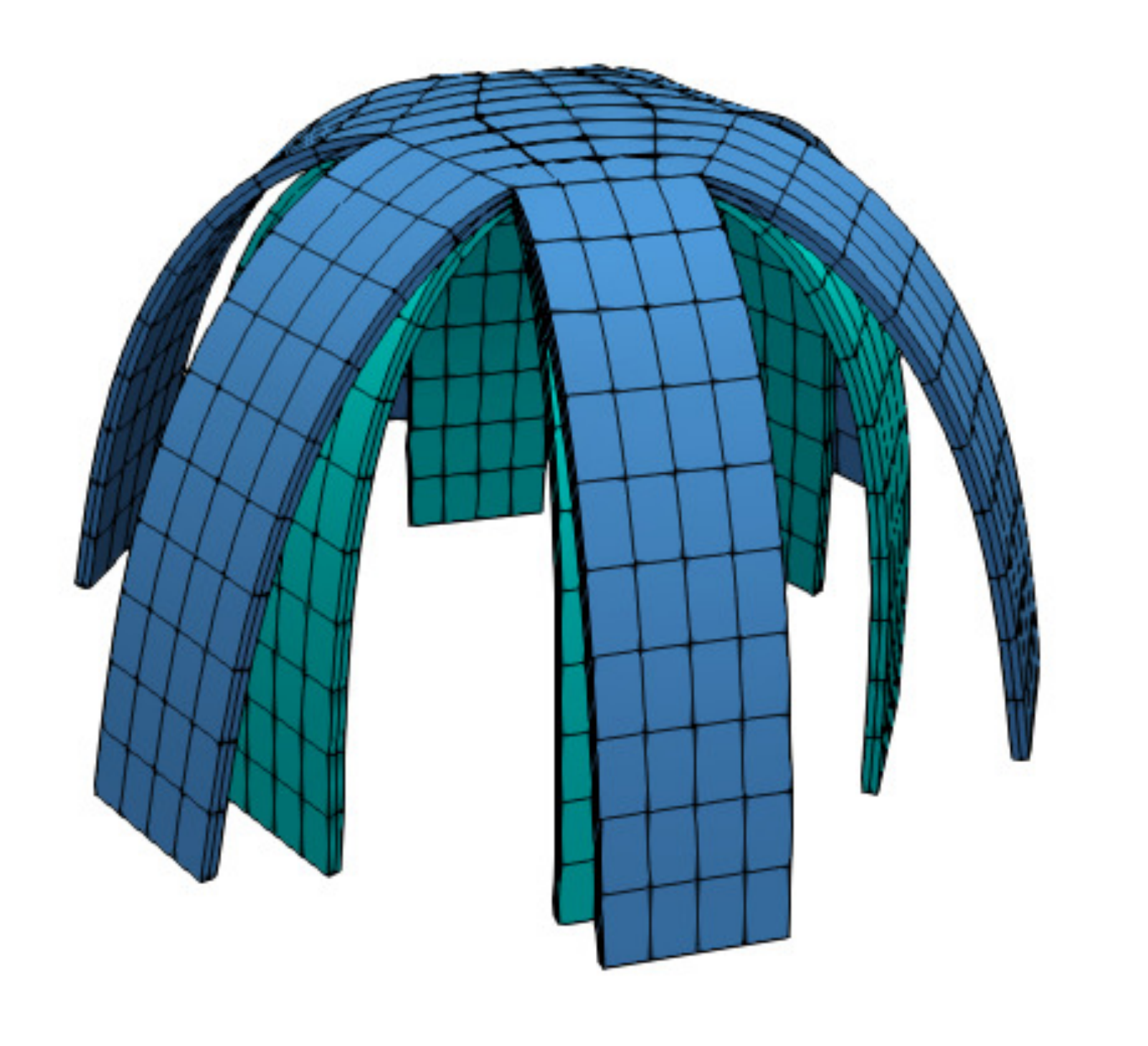}}
 \subfloat[At $t=30$s with $\mu_{v} = 3300$ kPa, $\, \tau=5.0$ s]{\includegraphics[clip, scale=0.5]{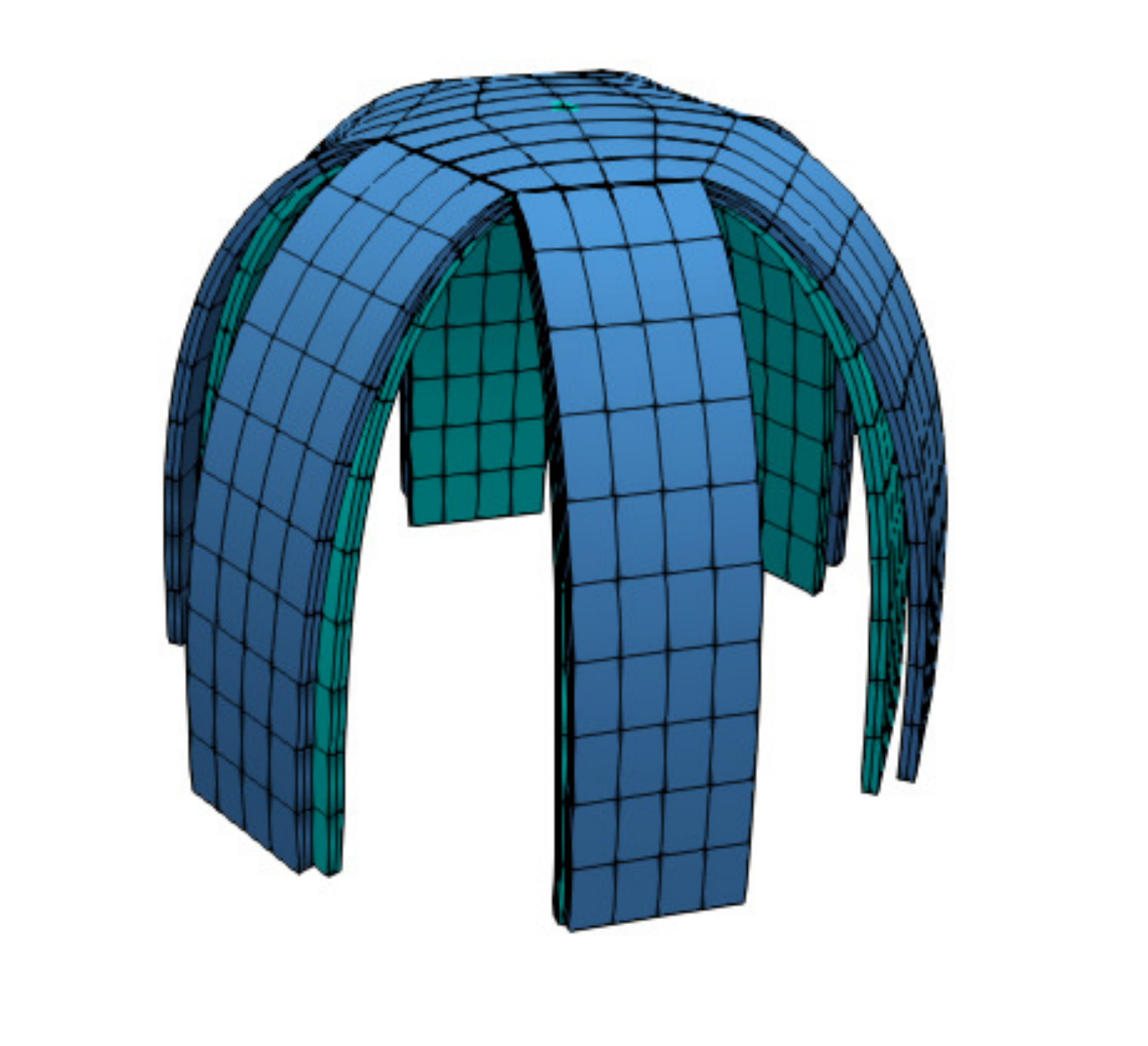}}
 \caption{Hard magnetic gripper: comparison of deformed shaped at $t=10$ s and $t=30$ s obtained with pure hyperelastic model (cyan) and viscoelastic model (blue).}
 \label{fig-hm-gripper-defshapes-1}
\end{figure}

Evolution of Y-displacement of the point P (see Fig. \ref{fig-hm-gripper-mesh}) under the first loading case for the parameters $\mu_{v} = \{165, 330, 660,3300\}$ kPa and $\tau=\{0.5,5.0,30.0\}$ s is plotted in Fig. \ref{fig-hm-gripper-graphs-1}. These graphs indicate that the time taken by the gripper to reach the desired gripping position increases with the increasing the values of $\mu_{v}$ or $\tau$ or both. Significant differences in the deformed shapes of the gripper at time instants $t=10$ s and $t=30$ s are presented in Fig. \ref{fig-hm-gripper-defshapes-1}.

\begin{figure}[H]
 \centering
 \subfloat[$\mu_{v} = 330$ kPa, $\, \tau=0.5$ s]{\includegraphics[clip, scale=0.4]{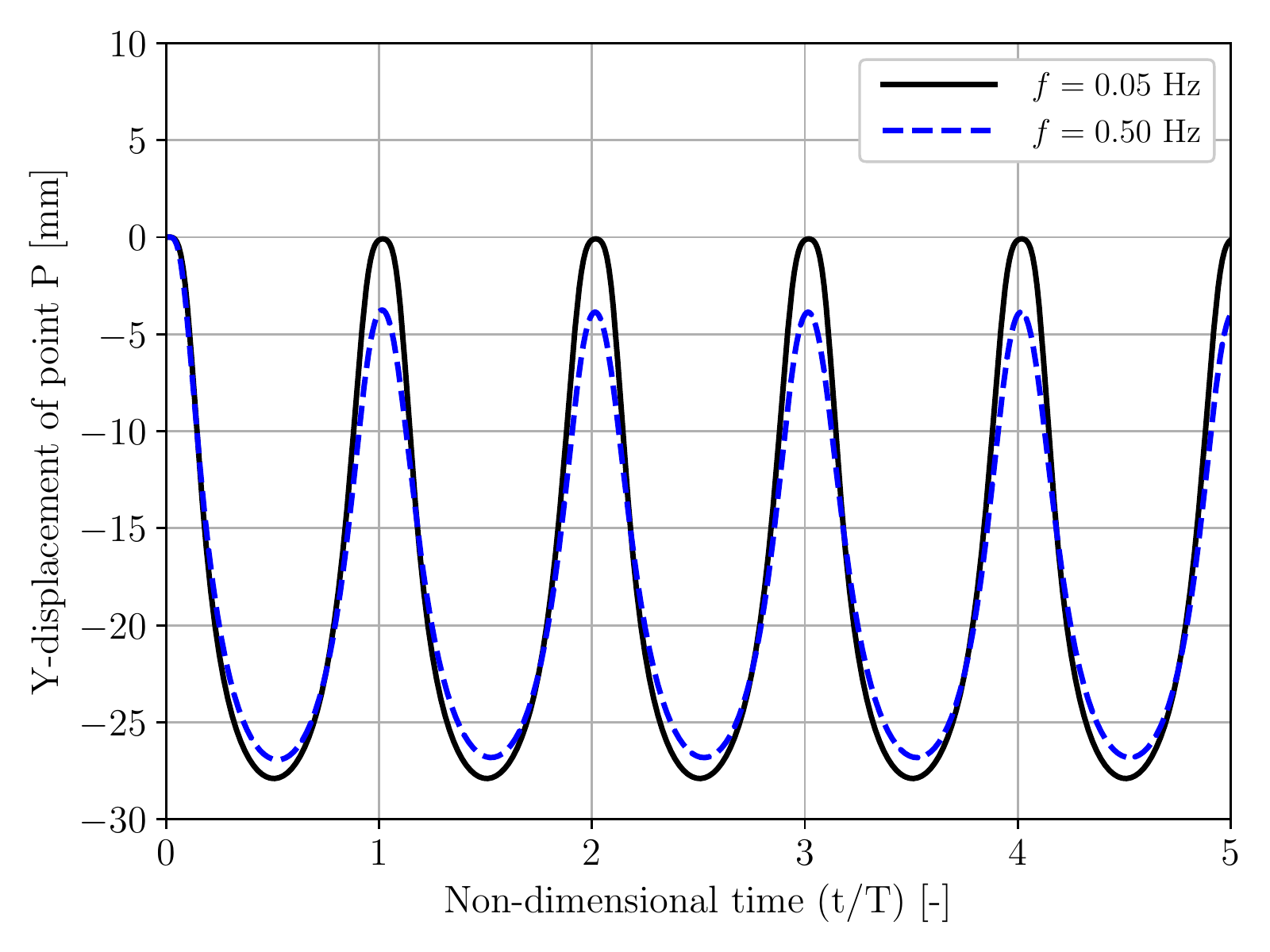}}
 \subfloat[at $t/T=3.0$]{\includegraphics[clip, scale=0.35]{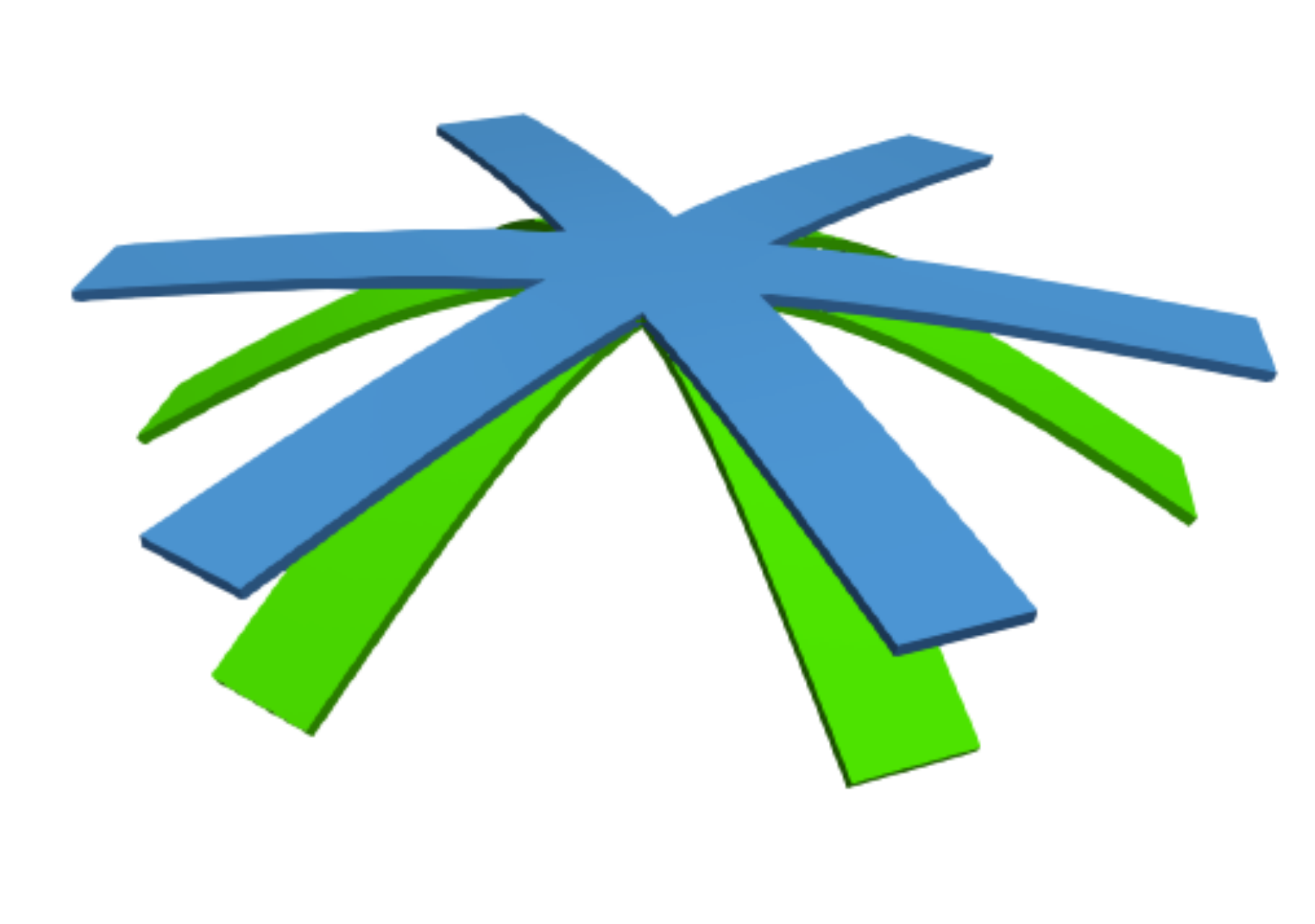}}
 \subfloat[at $t/T=4.5$]{\includegraphics[trim=15mm 0mm 15mm 0mm. clip, scale=0.35]{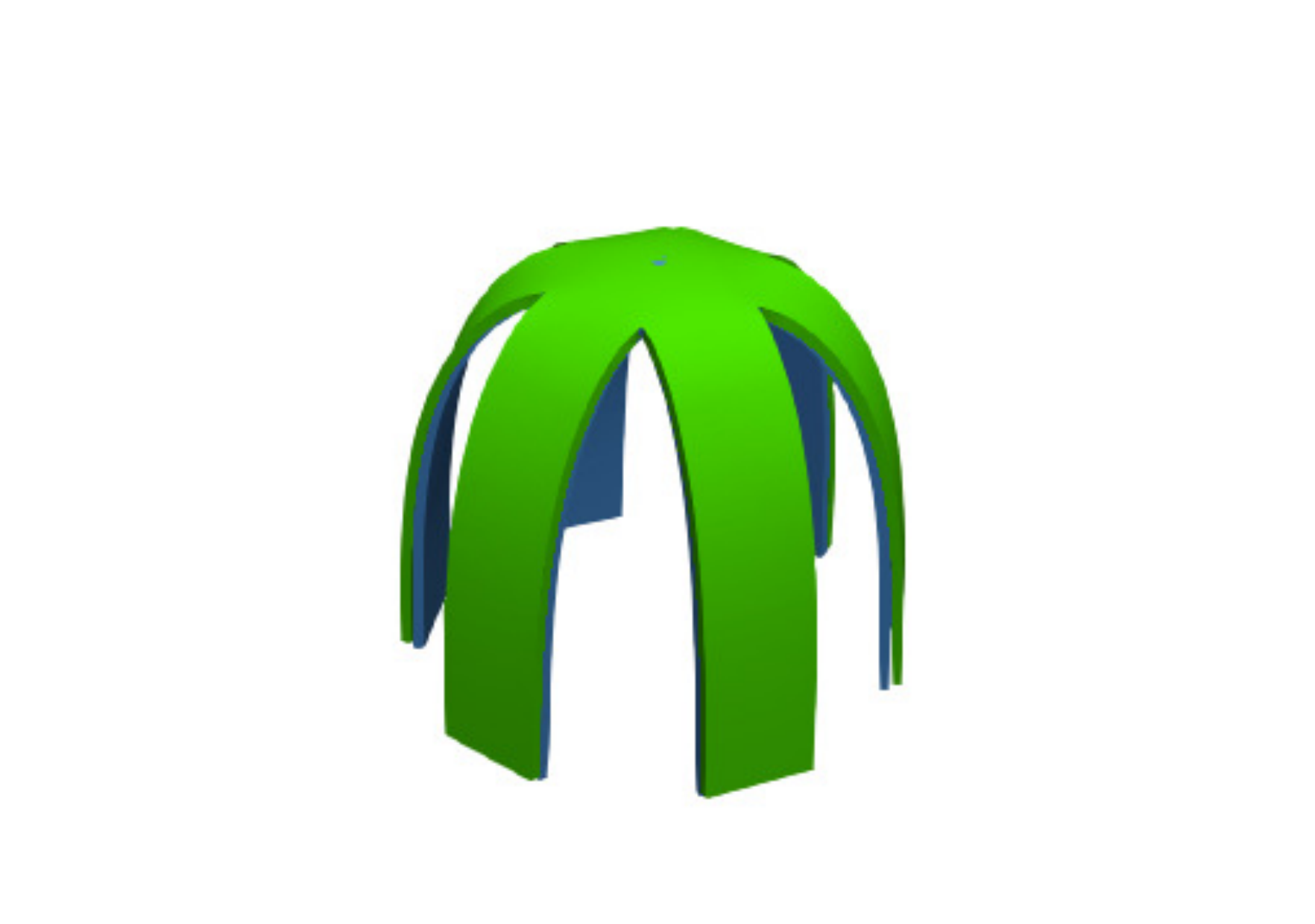}} \\
 \subfloat[$\mu_{v} = 330$ kPa, $\, \tau=5.0$ s]{\includegraphics[clip, scale=0.4]{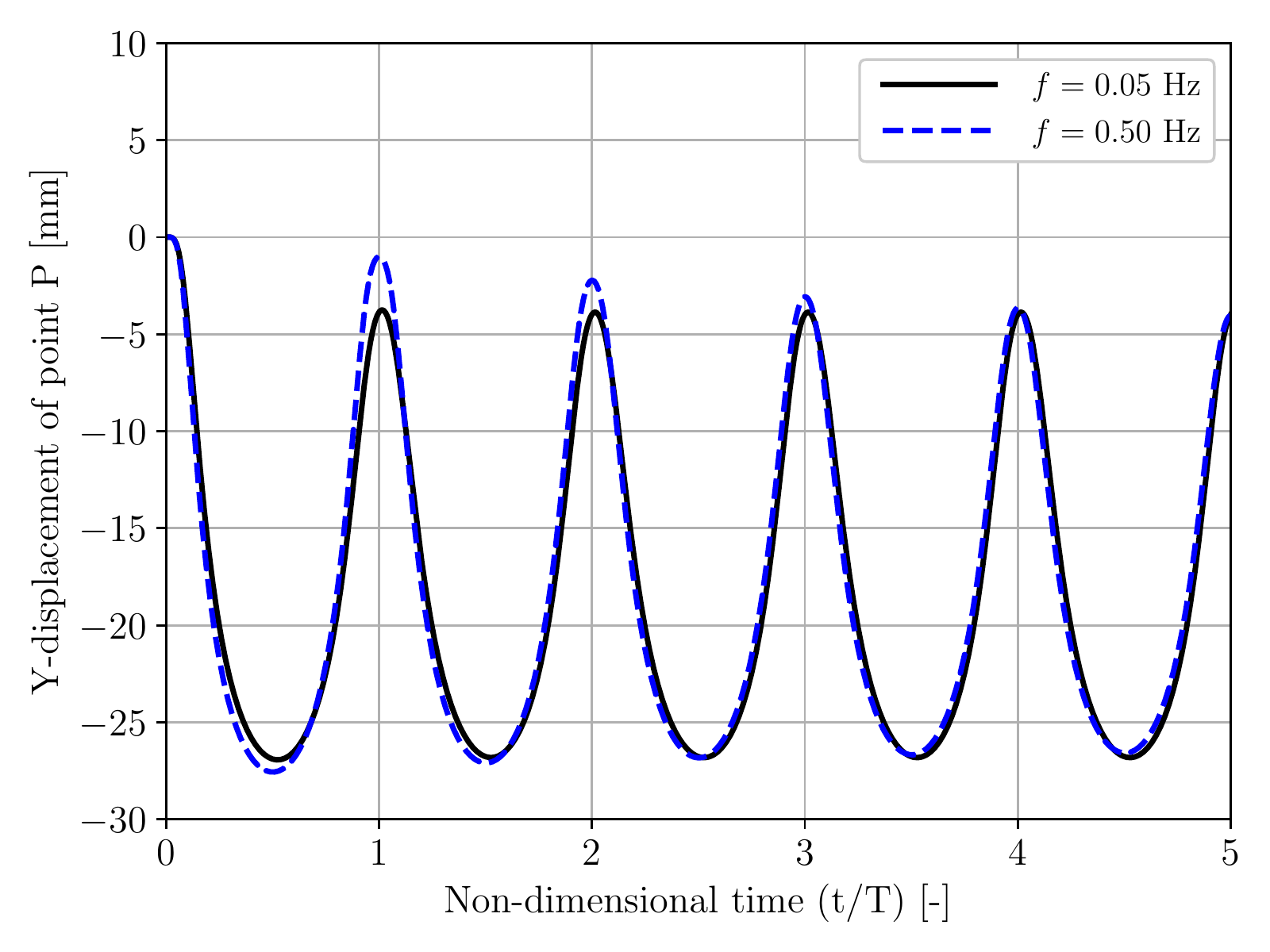}}
 \subfloat[at $t/T=3.0$]{\includegraphics[clip, scale=0.35]{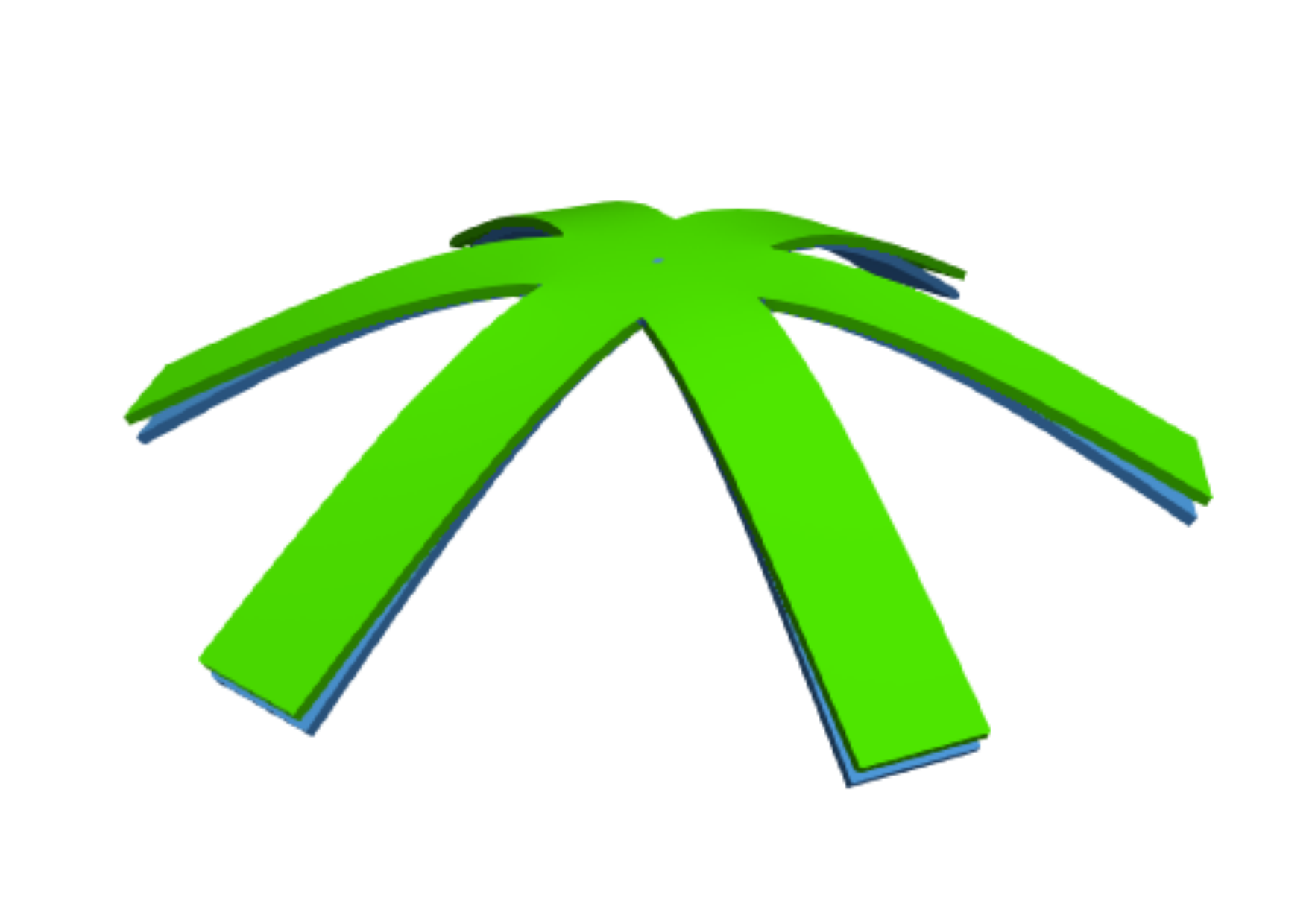}}
 \subfloat[at $t/T=4.5$]{\includegraphics[trim=15mm 0mm 15mm 0mm. clip, scale=0.35]{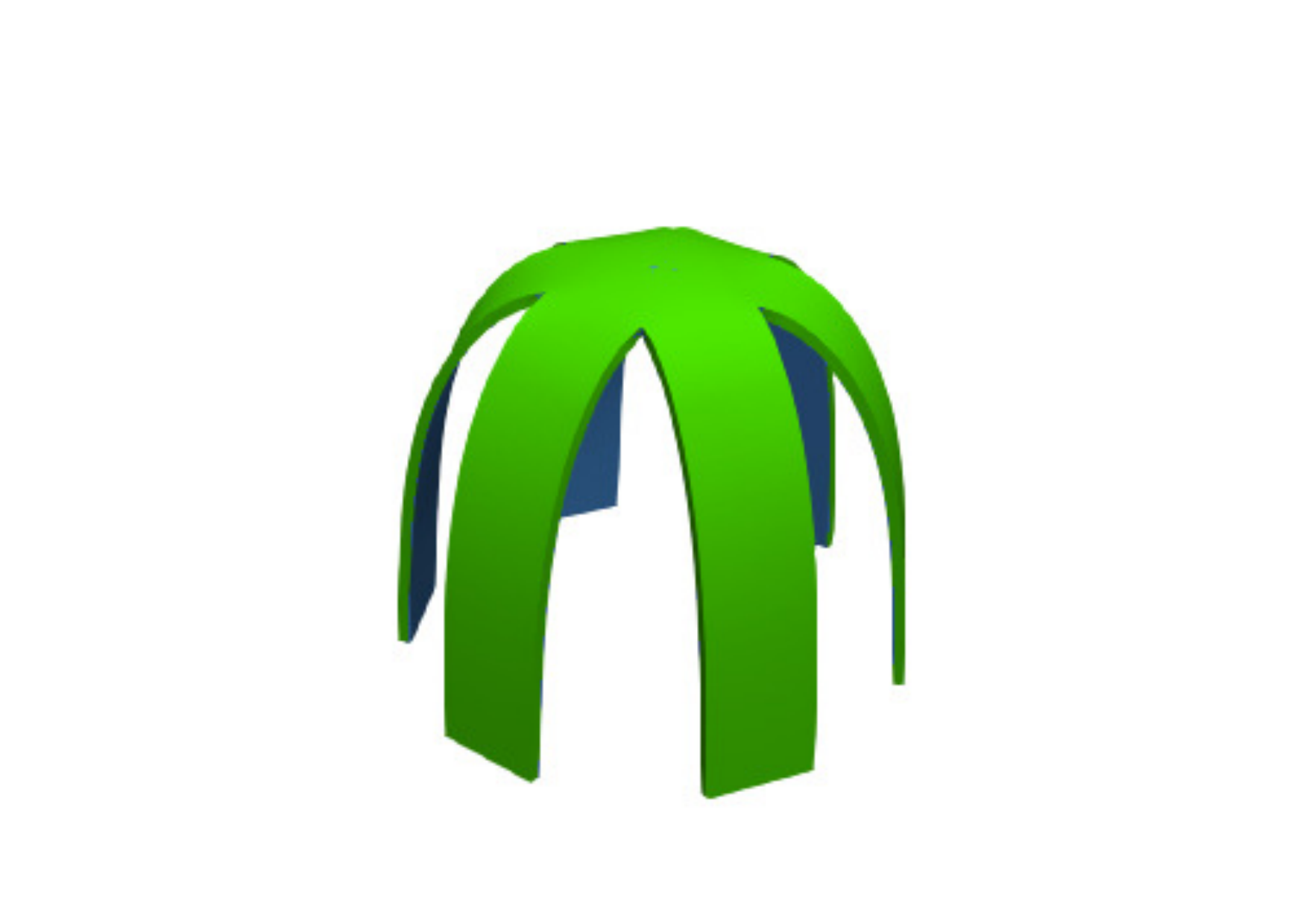}} \\
 \subfloat[$\mu_{v} = 3300$ kPa, $\, \tau=0.5$ s]{\includegraphics[clip, scale=0.4]{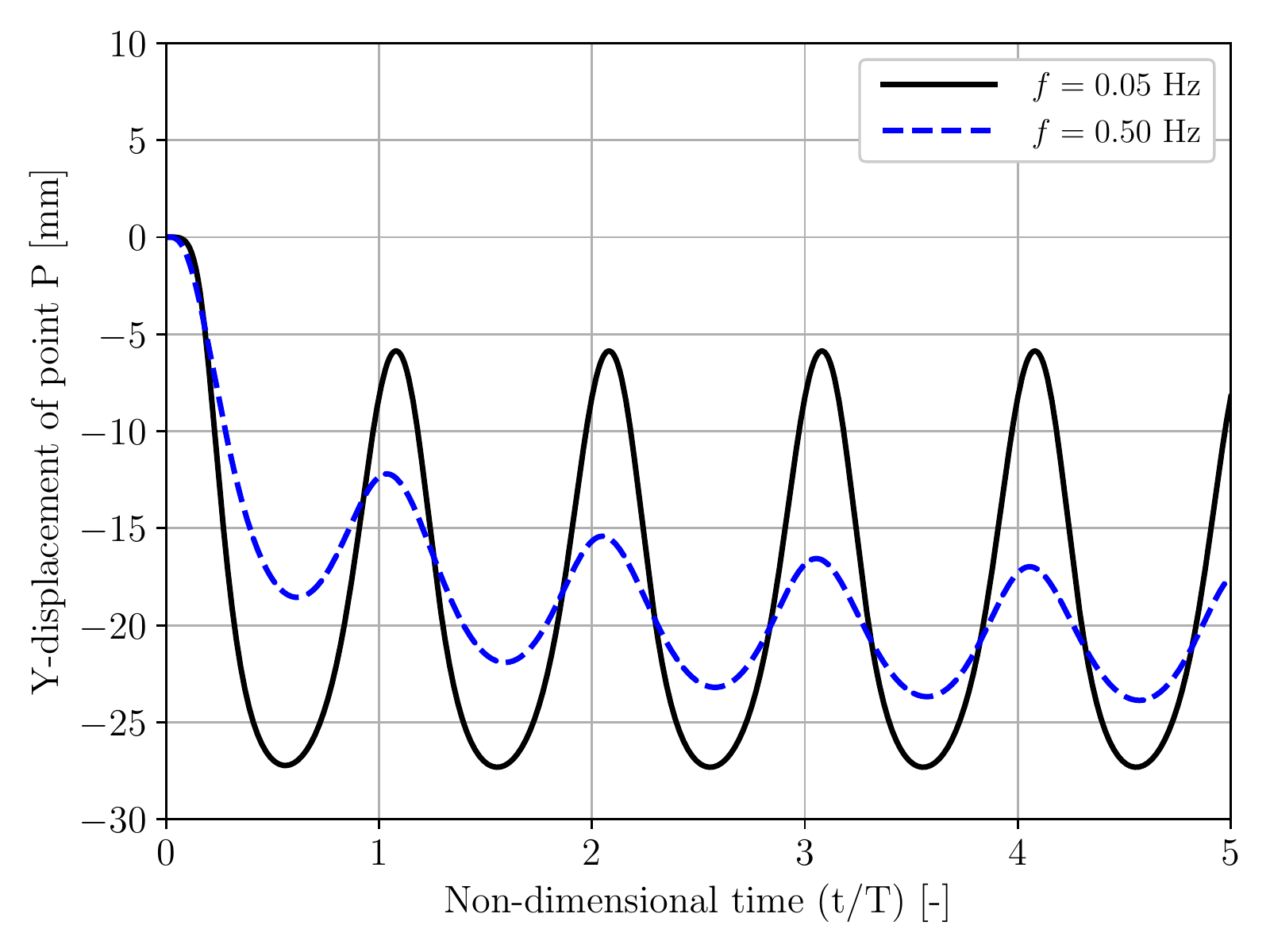}}
 \subfloat[at $t/T=3.0$]{\includegraphics[clip, scale=0.35]{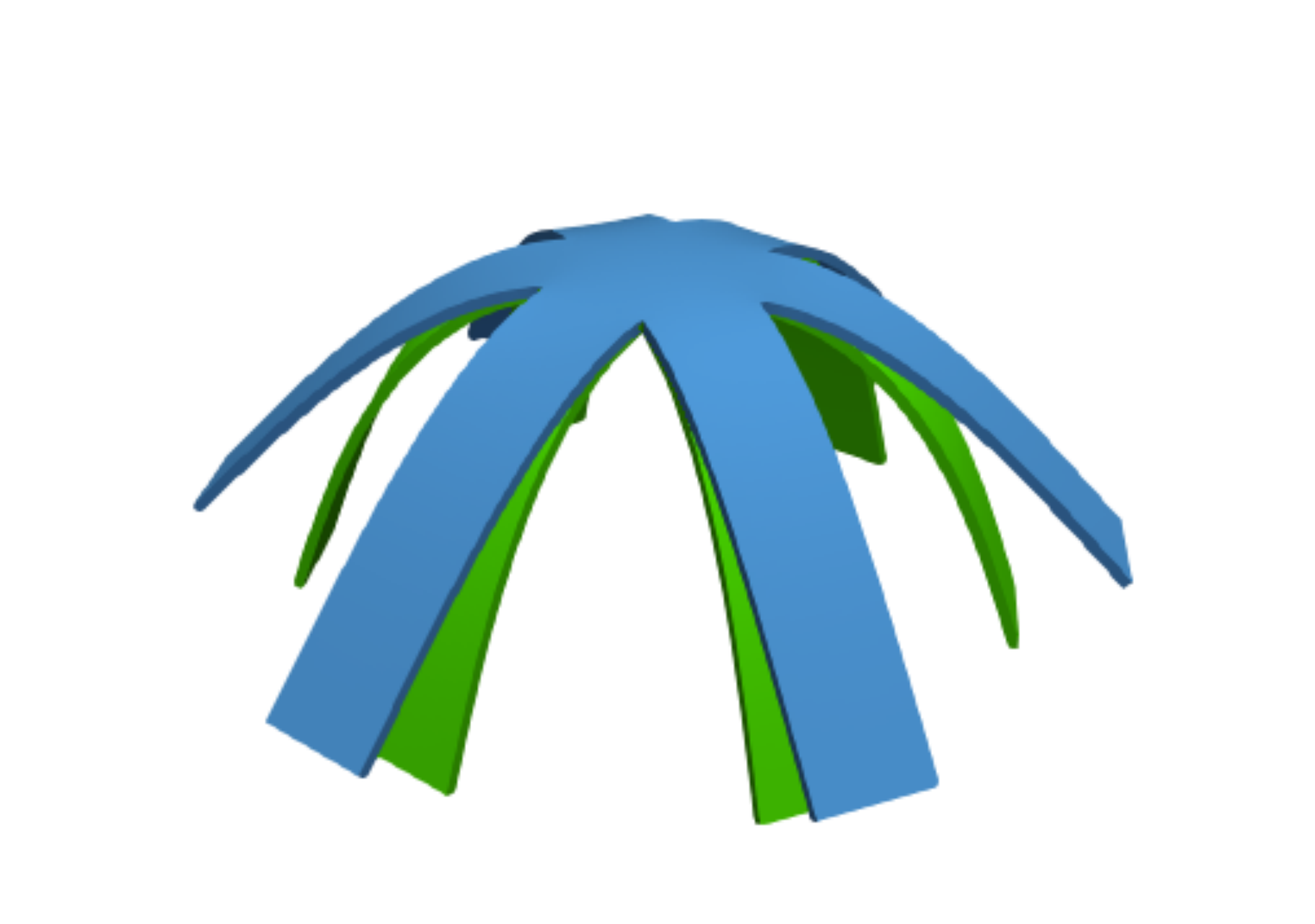}}
 \subfloat[at $t/T=4.5$]{\includegraphics[trim=15mm 0mm 15mm 0mm. clip, scale=0.35]{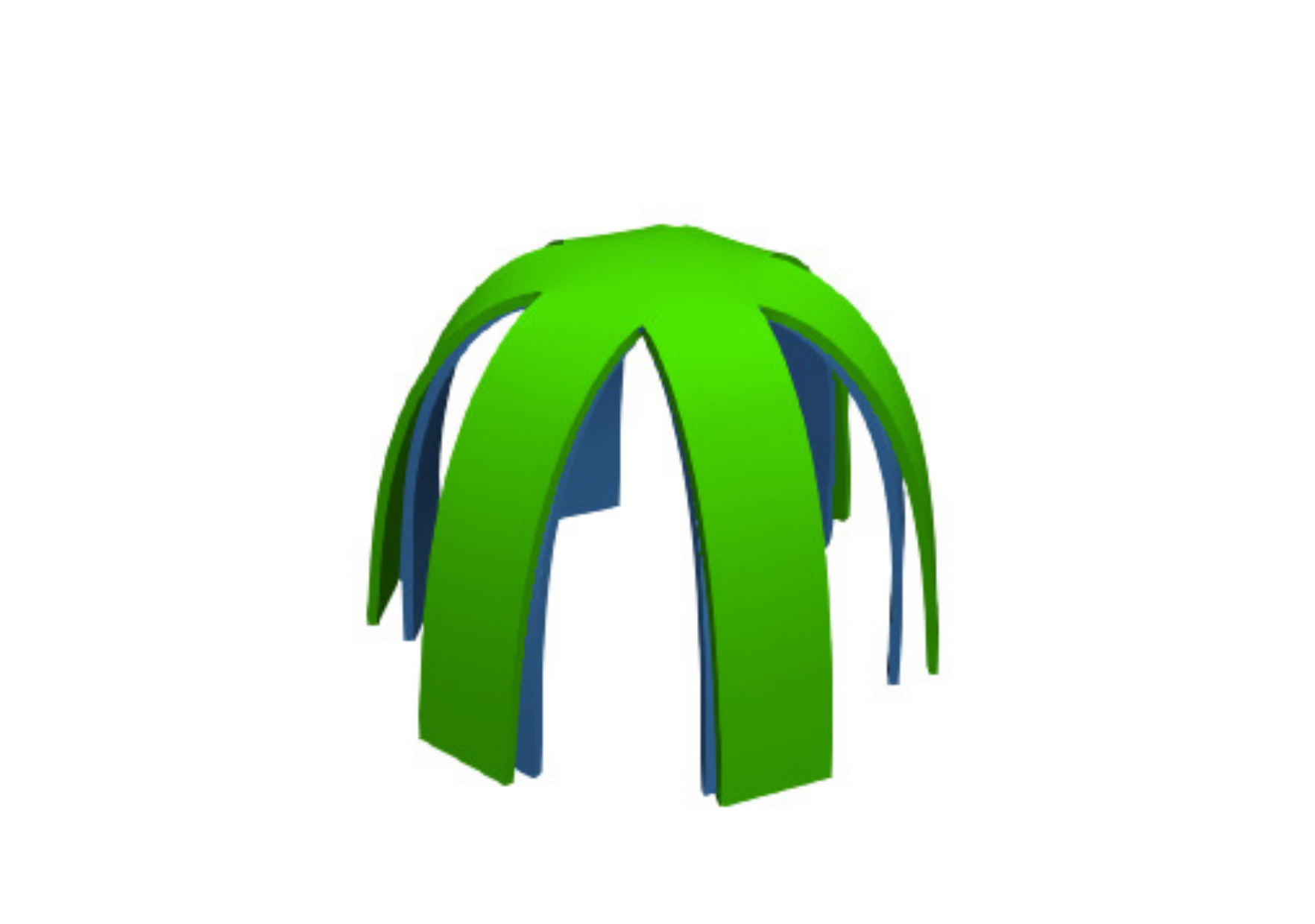}} \\
 \subfloat[$\mu_{v} = 3300$ kPa, $\, \tau=5.0$ s]{\includegraphics[clip, scale=0.4]{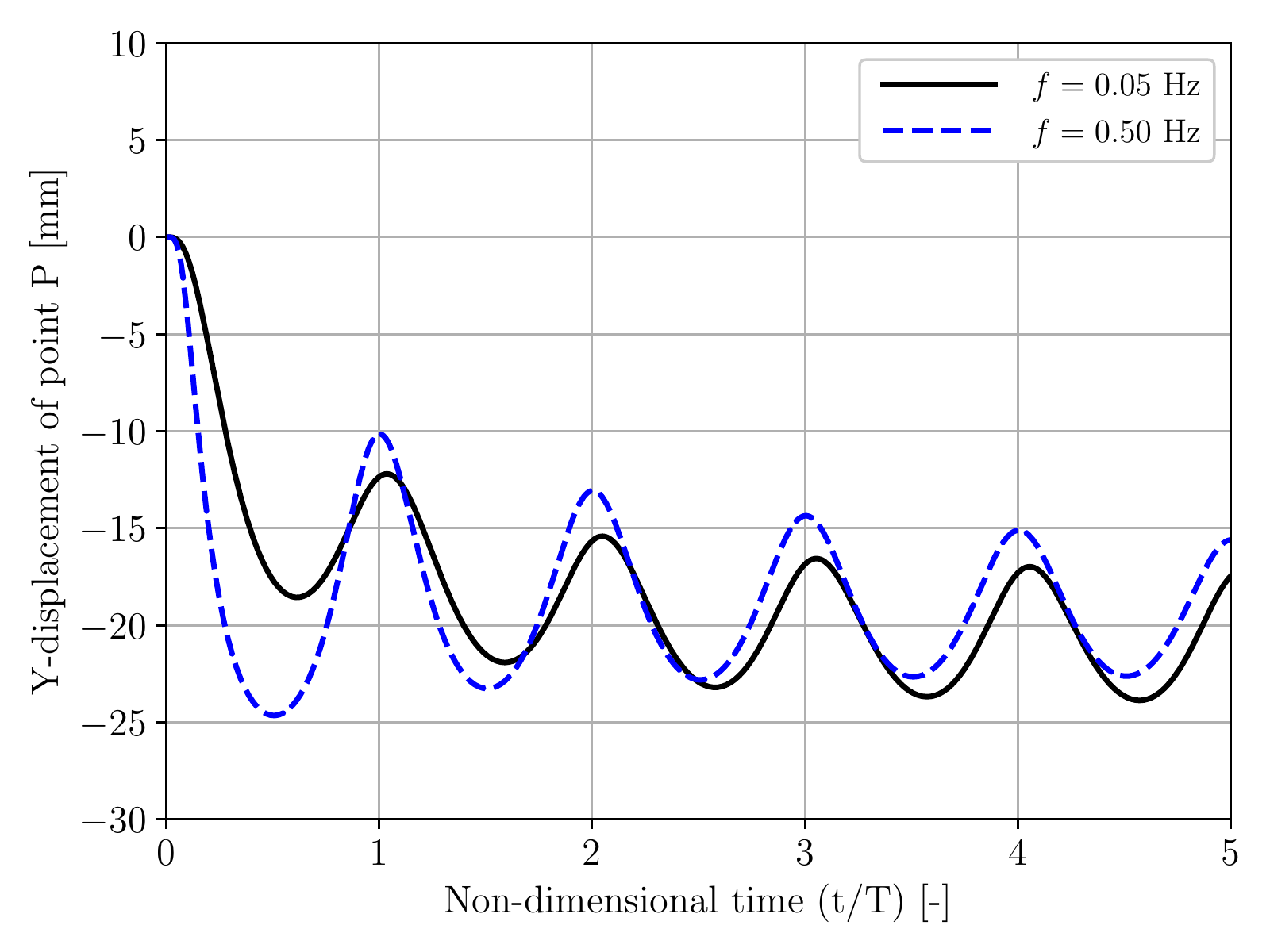}}
 \subfloat[at $t/T=3.0$]{\includegraphics[clip, scale=0.35]{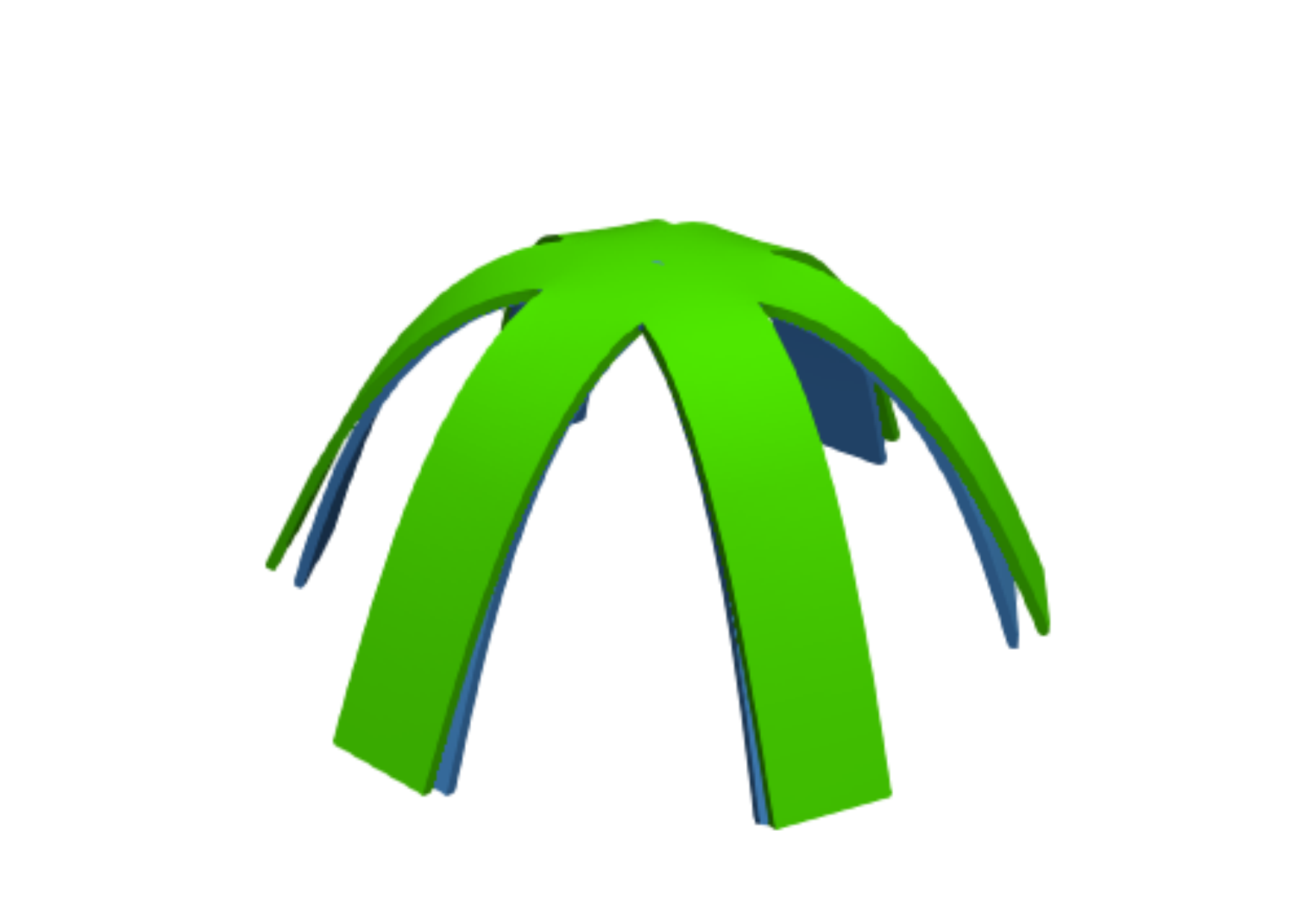}}
 \subfloat[at $t/T=4.5$]{\includegraphics[trim=15mm 0mm 15mm 0mm. clip, scale=0.35]{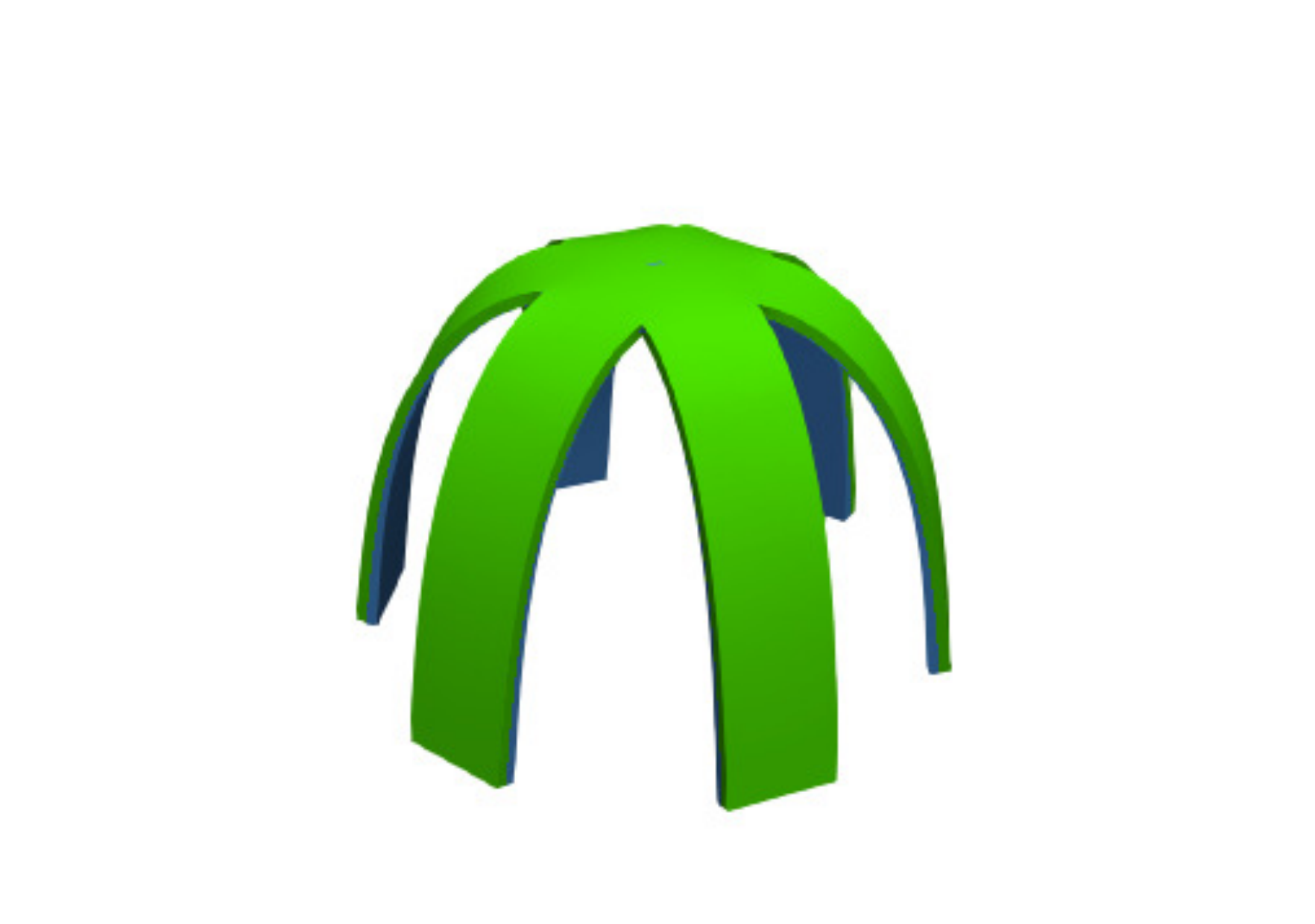}} \\
 \caption{Hard magnetic gripper: displacement response under sinusoidal loading along with the deformed shapes at time instants $t/T=3$ and $t/T=4.5$ obtained with $f=0.05$ Hz (Blue) and $f=0.5$ Hz (Green).}
 \label{fig-hm-gripper-graphs-2}
\end{figure}

Next, we assess the response of the gripper under a sinusoidal loading. The applied magnetic field is varied such that $\mathbb{B}^{a}_{z} = - \, 500 \, [1-\cos(2 \, \pi \, f \, t)]$. Two different frequencies,$f=0.5$ Hz and $f=5$ Hz, are considered. Simulations are performed for different combinations of the viscoelastic model parameters and the Y-displacement of point P is plotted in Fig. \ref{fig-hm-gripper-graphs-2}, for the combinations of $\mu_{v}=\{330, 3000\}$ kPa and $\tau=\{0.5,5.0\}$ s. For $\mu_{v}=330$ kPa, the gripper still attains the desired final shape for different values of $\tau$ and $f$ considered. However, for $\mu_{v}=3300$ kPa, the gripper loses its ability to attain the desired gripping shape. 
This is because when the gripper is simulated taking the viscoelastic material model, it has contributions both from elastic and viscous parts of the stress. That means the system becomes stiffer, resulting in a reduced deformation for the same amount of applied magnetic field. As also shown in Fig. \ref{fig-hm-gripper-graphs-2}, there is a significant difference in the deformed shapes of the gripper obtained with loading at different frequencies.

\subsection{Numerical examples - soft magnetic materials}

\subsubsection{Verification against analytical solution}
\begin{figure}[H]
 \centering
  \subfloat[]{\includegraphics[clip, scale=0.65]{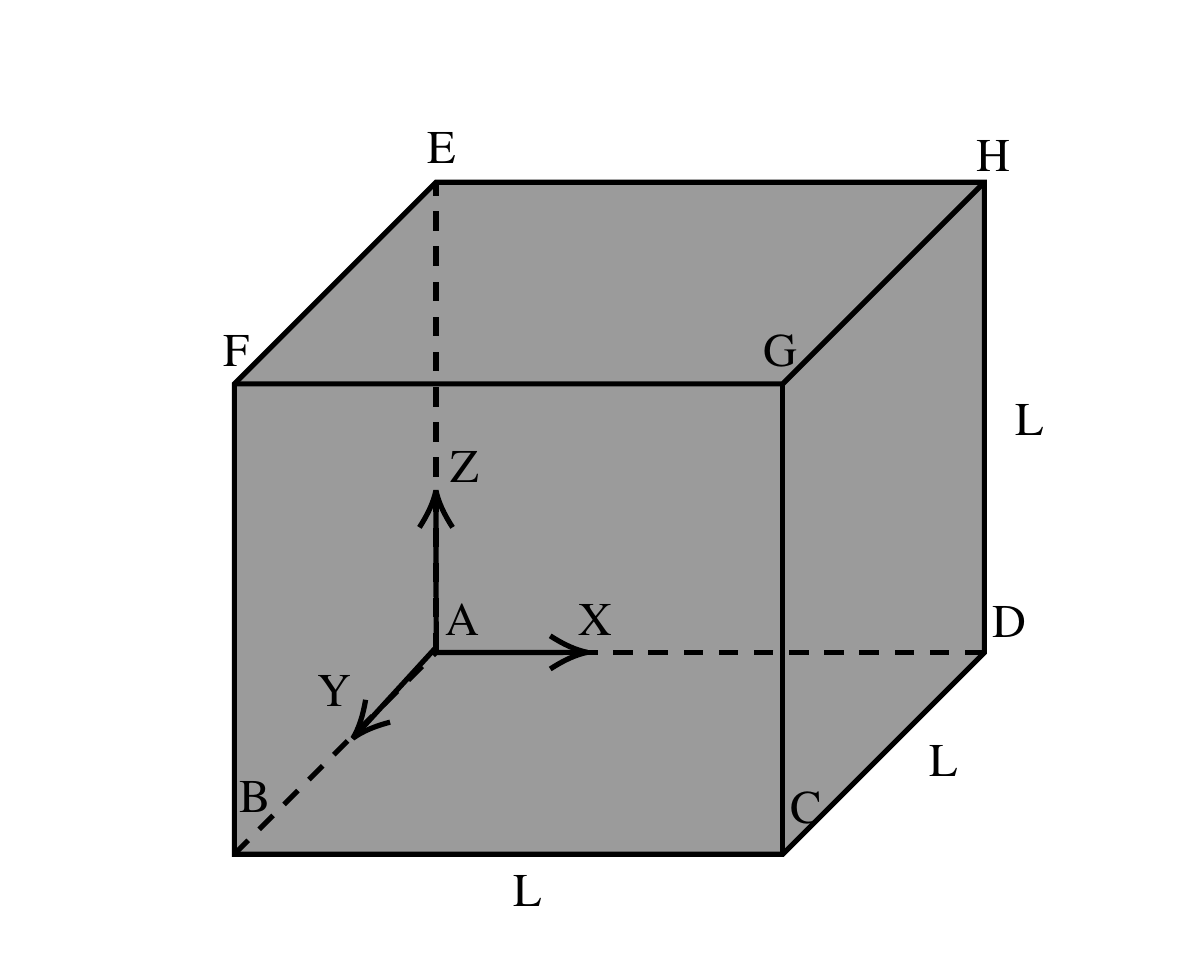} \label{fig-sm-cube-geom} }
  \subfloat[]{\includegraphics[clip, scale=0.5]{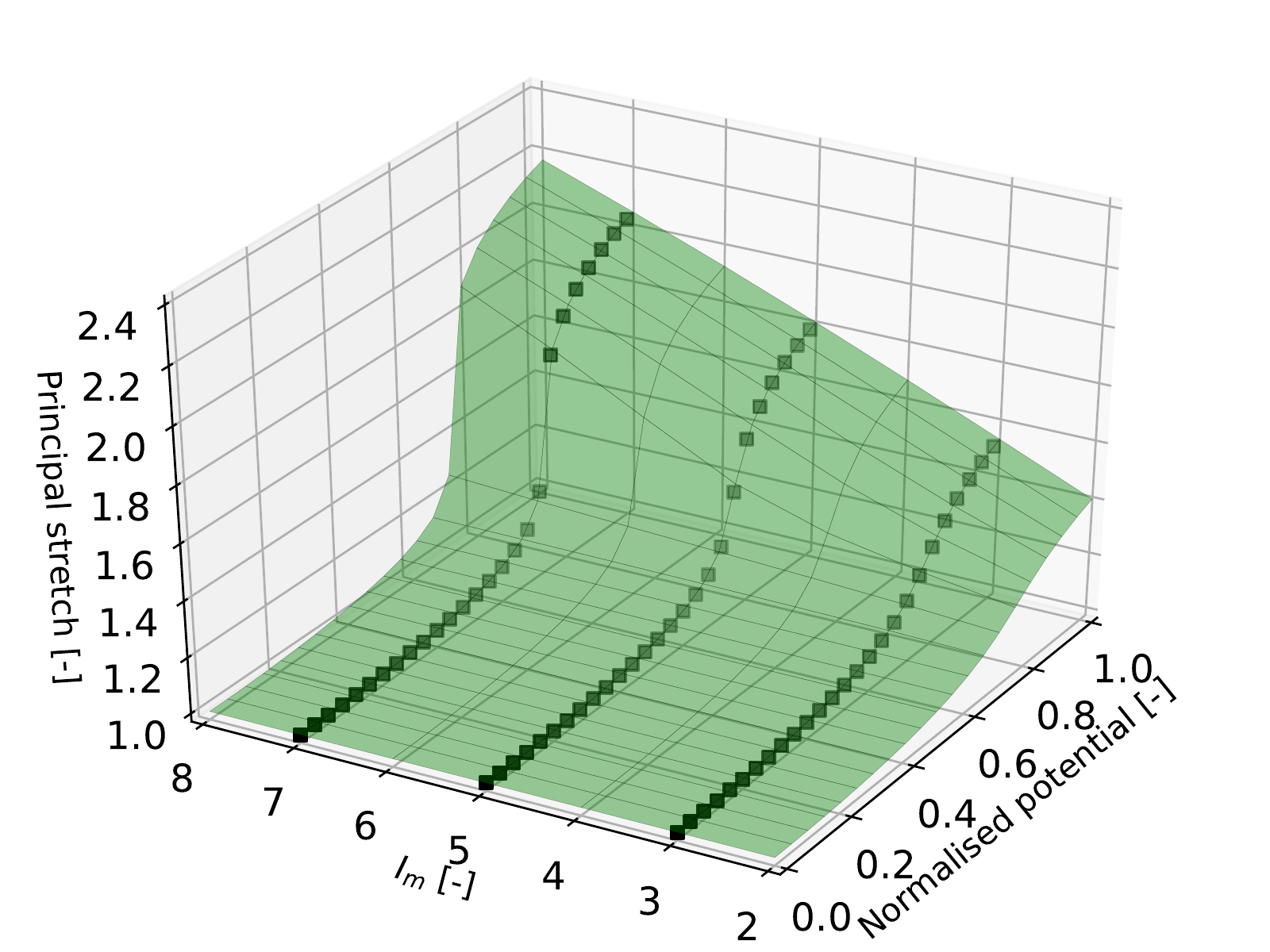} \label{fig-sm-cube-graph} }
 \caption{Unit cube: (a) problem setup and (b) comparison of analytical and numerical solutions. The surface represents the analytical solution and the square markers denoted the numerical solution.}
\end{figure}

To verify the computer implementation of the proposed framework for coupled magneto-mechanical problems, we consider the example of a unit cube ($L=1$ mm) subjected to a magnetic potential difference across one of its axes, as shown in Fig. \ref{fig-sm-cube-geom}. This problem is similar to the one previously studied in the context of computational electromechanics in Henann et al.  \cite{HenannJMPS2013}, Kadapa and Hossain \cite{KadapaCMAME2020}. 

The material constitutive form is the truly incompressible Gent model with a shear modulus of $\mu = 1$ Pa and three different values for the parameter $I_m$ are chosen. The free-space permeability is taken as $\mu_0 = 1.0 \; \frac{\gram \millimeter}{\second^2 \, \Ampere^2}$ and the parameters $\alpha$, $\beta$ and $\eta$ in the magnetic part of the energy function (\ref{eqn-Psi-magn-sm}) are zero. The faces ADHE, ABFE, and ADCB are constrained from moving in the normal directions. The magnetic potential on the face ABCD is set to zero and a positive magnetic potential is prescribed on the face EFGH in uniform increments. The plot of the principal stretch ($\lambda$) versus the normalised potential ($\bar{\phi}=\frac{\phi}{L} \, \sqrt{\frac{\mu_{0}}{\mu}}$) for three different values of $I_m$ as shown in Fig. \ref{fig-sm-cube-graph} illustrate that the results obtained with the present numerical approach (square markers) are in good agreement with the analytical solution (surface) given by the relation in Hanann et al. \cite{HenannJMPS2013}
\begin{align}
\frac{\phi}{L} \, \sqrt{\frac{\mu_{0}}{\mu}} = \sqrt{\left[ 1 - \frac{1}{I_m} \, \left[ 2\lambda^2 + \frac{1}{\lambda^4} - 3 \right] \right]^{-1} \, \frac{1}{\lambda^2} \, \left[ 1-\frac{1}{\lambda^6} \right]}.
\end{align}


\subsubsection{A bilayer cantilever beam}
\begin{figure}[H]
 \centering
  \subfloat[]{\includegraphics[clip, scale=0.5]{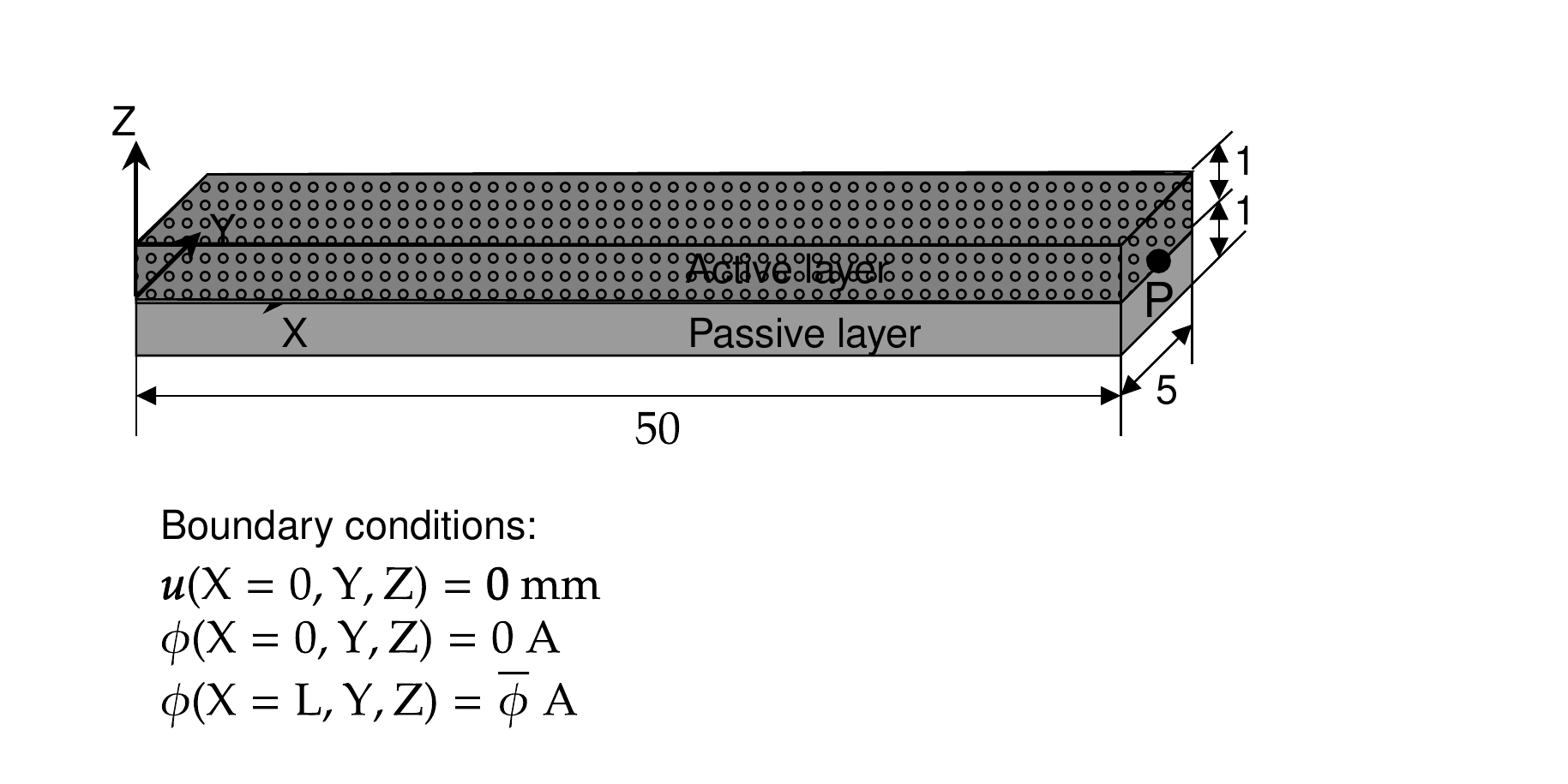}  \label{fig-sm-beam-geom}}
  \subfloat[]{\includegraphics[clip, scale=0.4]{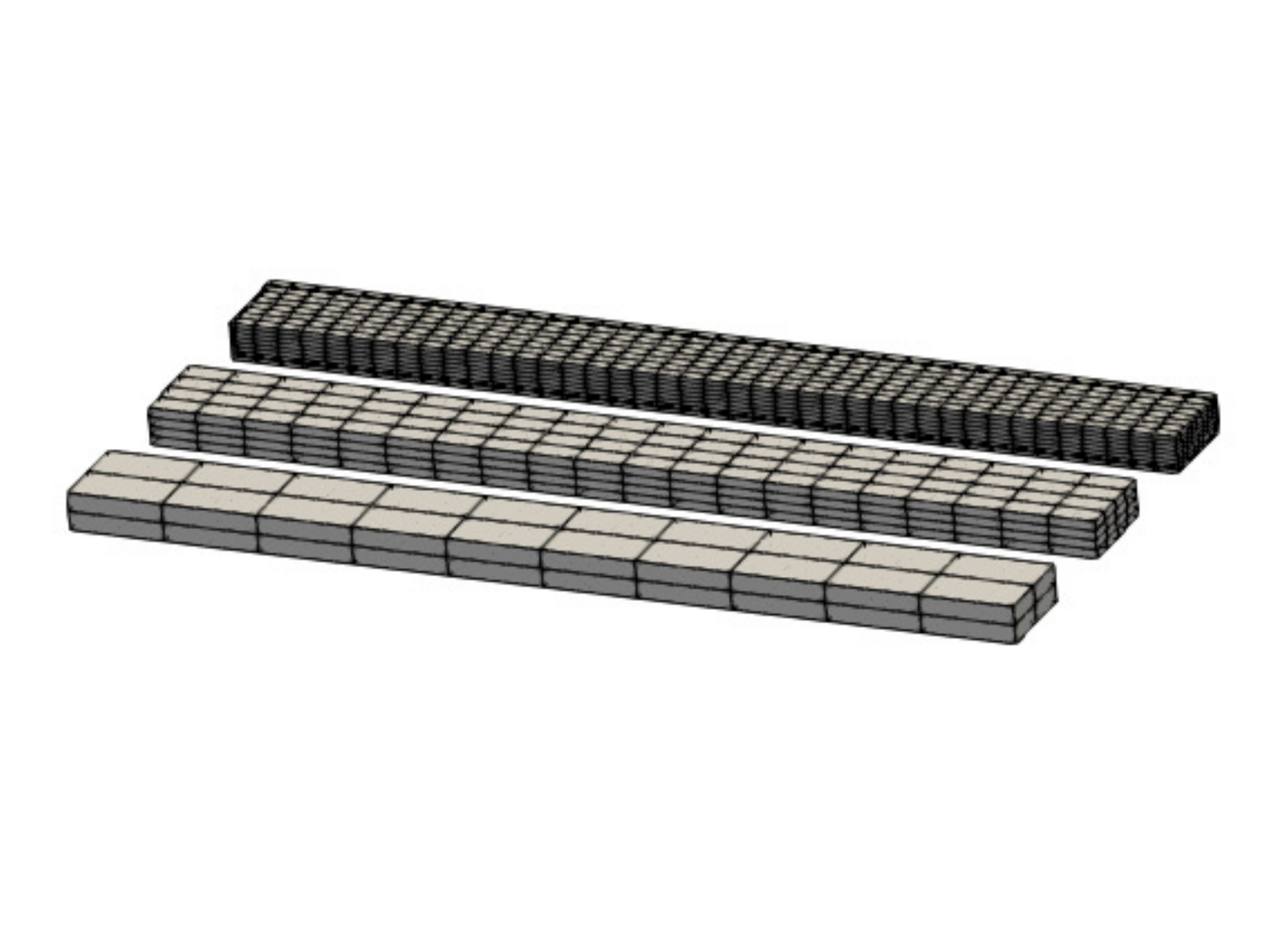} \label{fig-sm-beam-mesh}}
 \caption{A soft magnetic beam: (a) problem setup and (b) finite element meshes.}
\end{figure}

Inspired by Chattock \cite{Chattock1887} for practical application of MAPs in sensing and measurement, this example serves to demonstrate (i) the capability of the proposed displacement-pressure-potential formulation for modelling extremely large deformation behaviour in soft MAPs and (ii) mesh convergence of the proposed finite element formulation for the coupled magneto-mechanics problem. In this example, we consider a bilayer cantilever beam made up of an active smart magnetic polymeric layer (responsive to a magnetic field) and a passive polymer (irresponsive to any externally applied magnetic field), as shown in Fig. \ref{fig-sm-beam-geom}. The two layers are assumed to be perfectly bonded together. As shown, the specified magnetic potential is zero on the fixed faces and a positive on the faces at the free end of the beam. The setup is such that, due to the applied magnetic potential difference along the length of the beam, the active layer expands, thereby making the entire beam curve in a clockwise direction. The hyperelastic model is assumed to be a Neo-Hookean one with $\mu = 30$ kPa.

\begin{figure}[H]
 \centering
  \subfloat[]{\includegraphics[clip, scale=0.5]{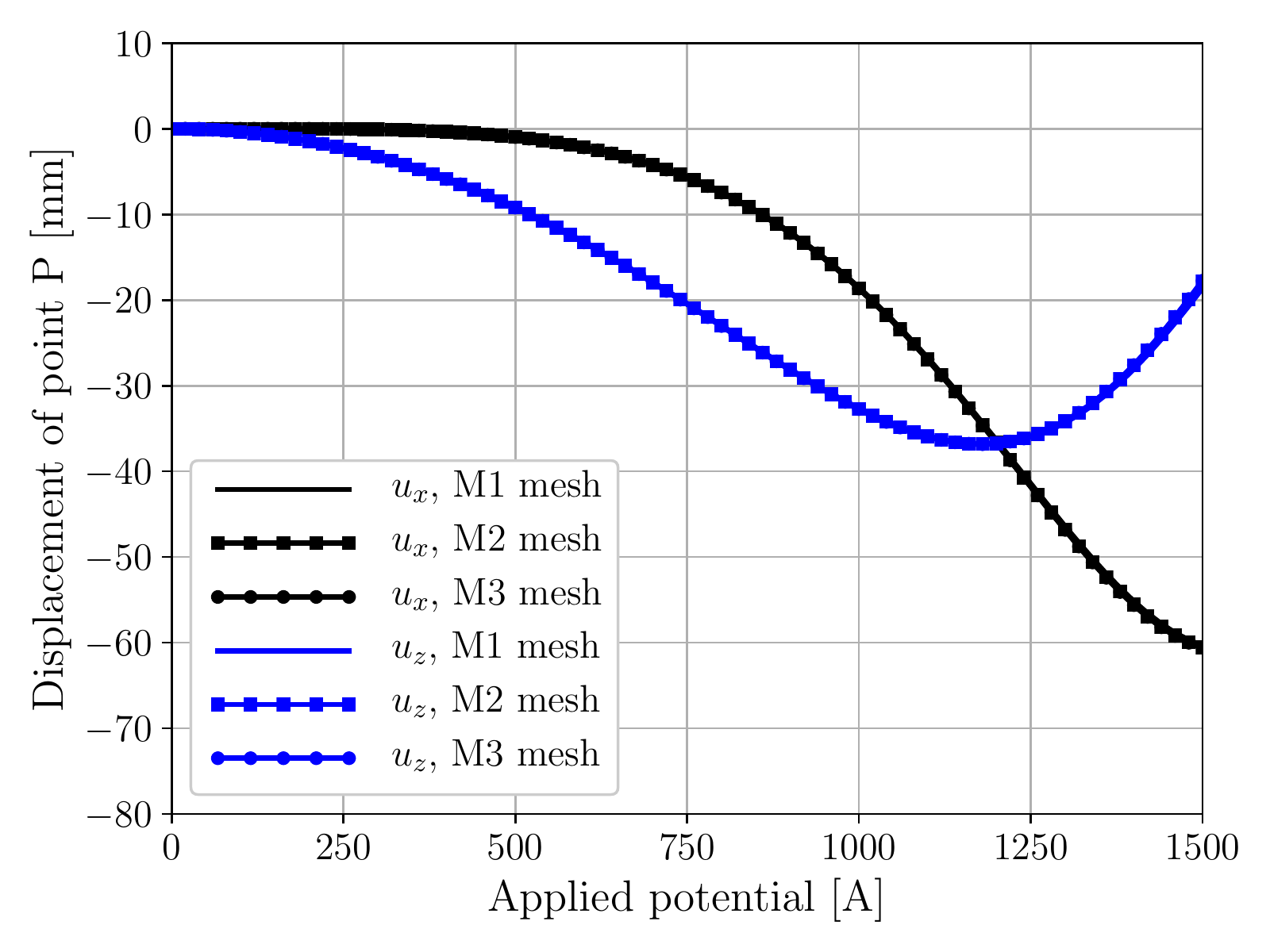}}
  \subfloat[]{\includegraphics[clip, scale=0.5]{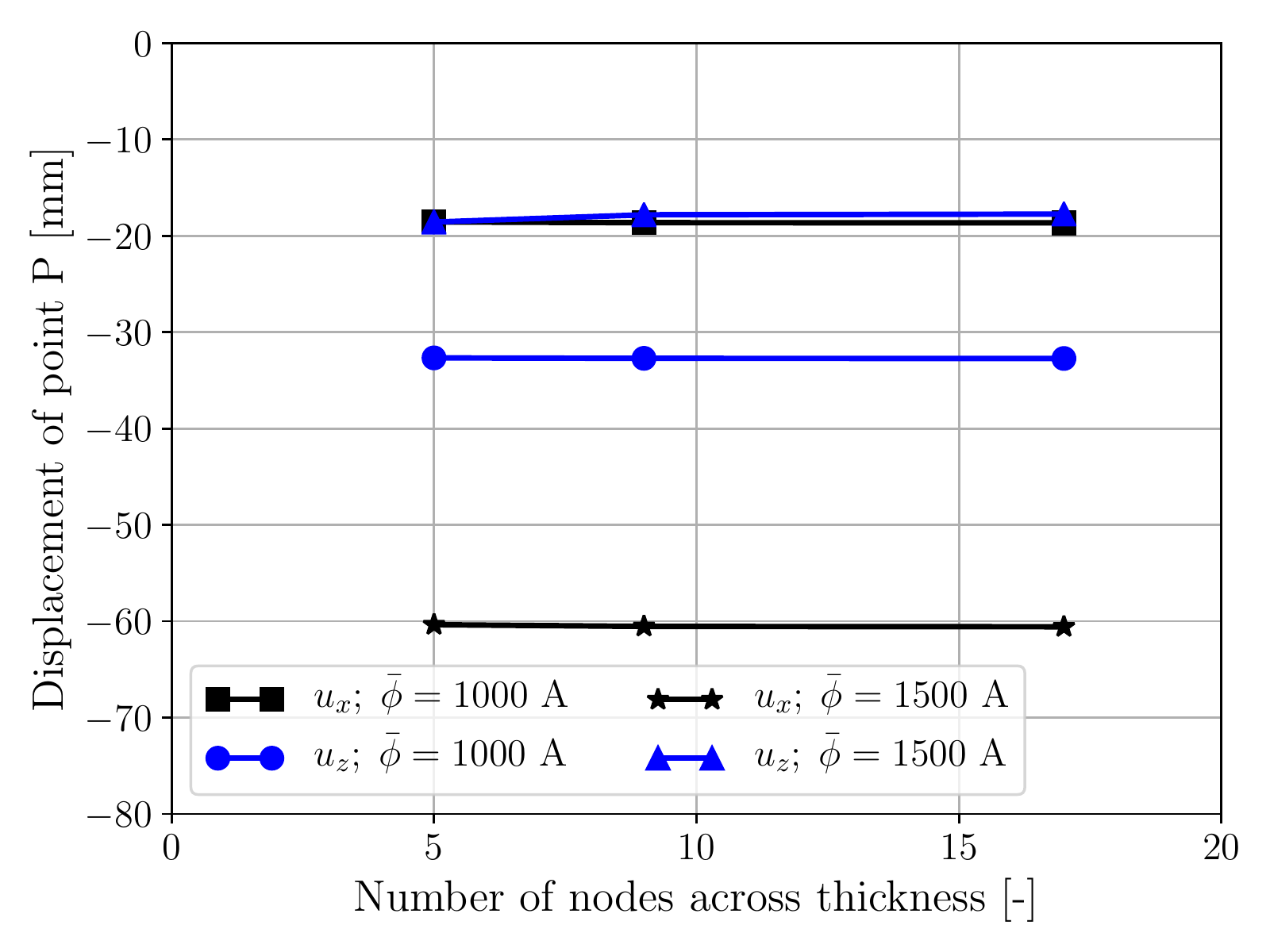}}
 \caption{A soft magnetic beam: (a) evolution of displacement of point P with respect to the applied potential and (b) graphs showing the convergence with respect to mesh refinement.}
 \label{fig-sm-beam-graph}
\end{figure}

\begin{figure}[H]
 \centering
  \includegraphics[clip, scale=0.6]{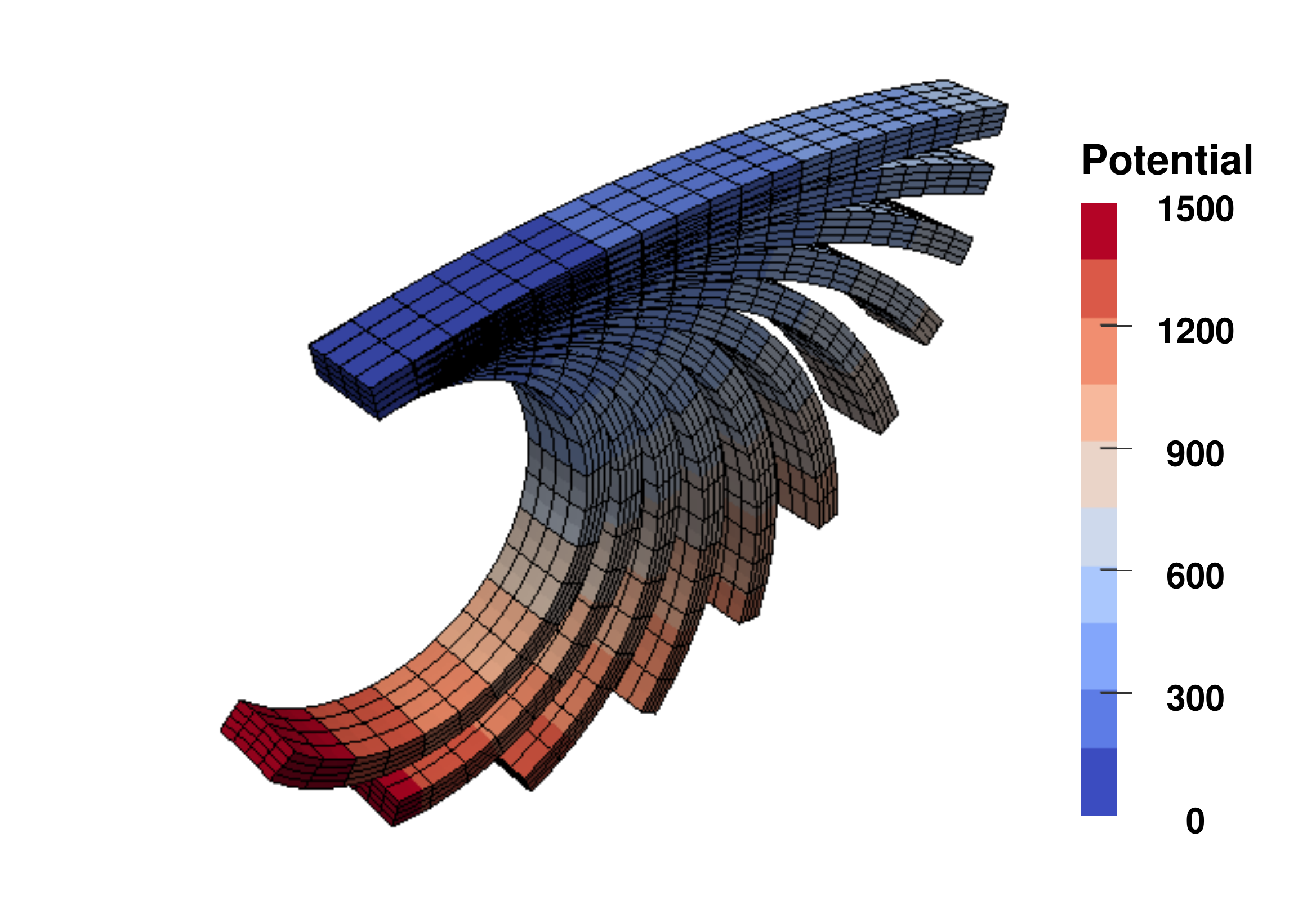}
 \caption{A soft magnetic beam: deformed shapes of the beam along with the contour plot of magnetic potential at different values of applied potential starting with $\bar{\phi}=500 \Ampere$ and increments of $100 \Ampere$.}
 \label{fig-sm-beam-contours}
\end{figure}

A mesh convergence study is performed using three different meshes shown in Fig. \ref{fig-sm-beam-mesh}. The meshes are denoted as M1, M2 and M3, with M1 being the coarsest mesh and M3 has the finest mesh. The coarsest (M1) mesh consists of $10 \times 2 \times 2$ elements, and the subsequent meshes are obtained by successive subdivision. The M1, M2 and M3 consist of 5, 9 and 17 nodes across the total thickness of the beam. Values of X- and Y-displacement of the point P obtained with the three meshes for different values of applied magnetic potential ($\bar{\phi}$) shown in Fig.  \ref{fig-sm-beam-graph} demonstrate an apparent convergence as the mesh is refined. The deformed shapes of the beam at different values of the applied magnetic potential are shown in Fig. \ref{fig-sm-beam-contours}. The results obtained for this example are consistent with the observations made in the example of a hard magnetic beam: accurate numerical results can be obtained with coarse meshes using the proposed numerical framework.

\subsubsection{A magneto-active gripper}
\begin{figure}[H]
 \centering
  \subfloat[]{\includegraphics[clip, scale=0.45]{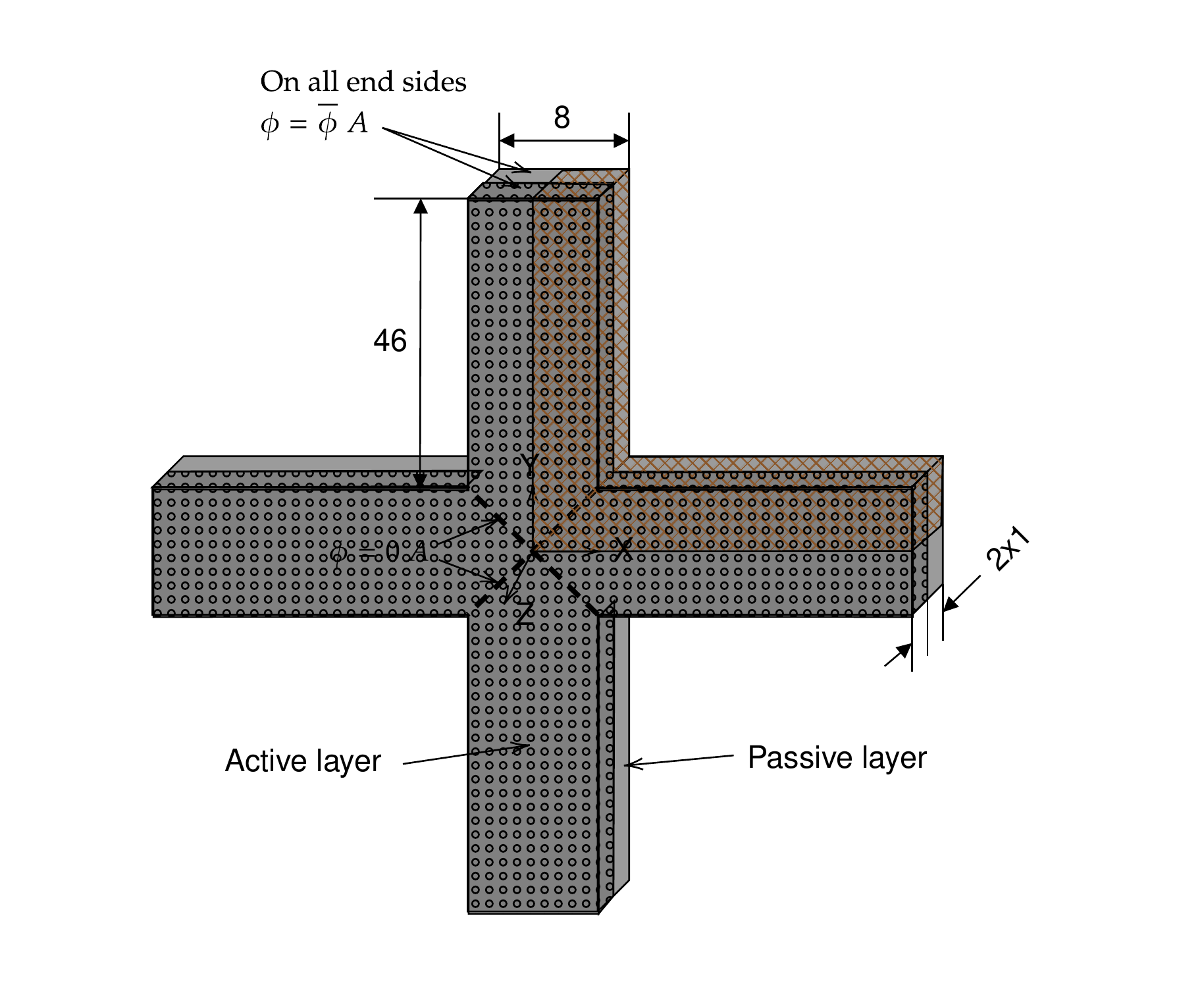}  \label{fig-sm-gripper-geom} }
  \subfloat[]{\includegraphics[trim=10mm 0mm 10mm 0mm, clip, scale=0.45]{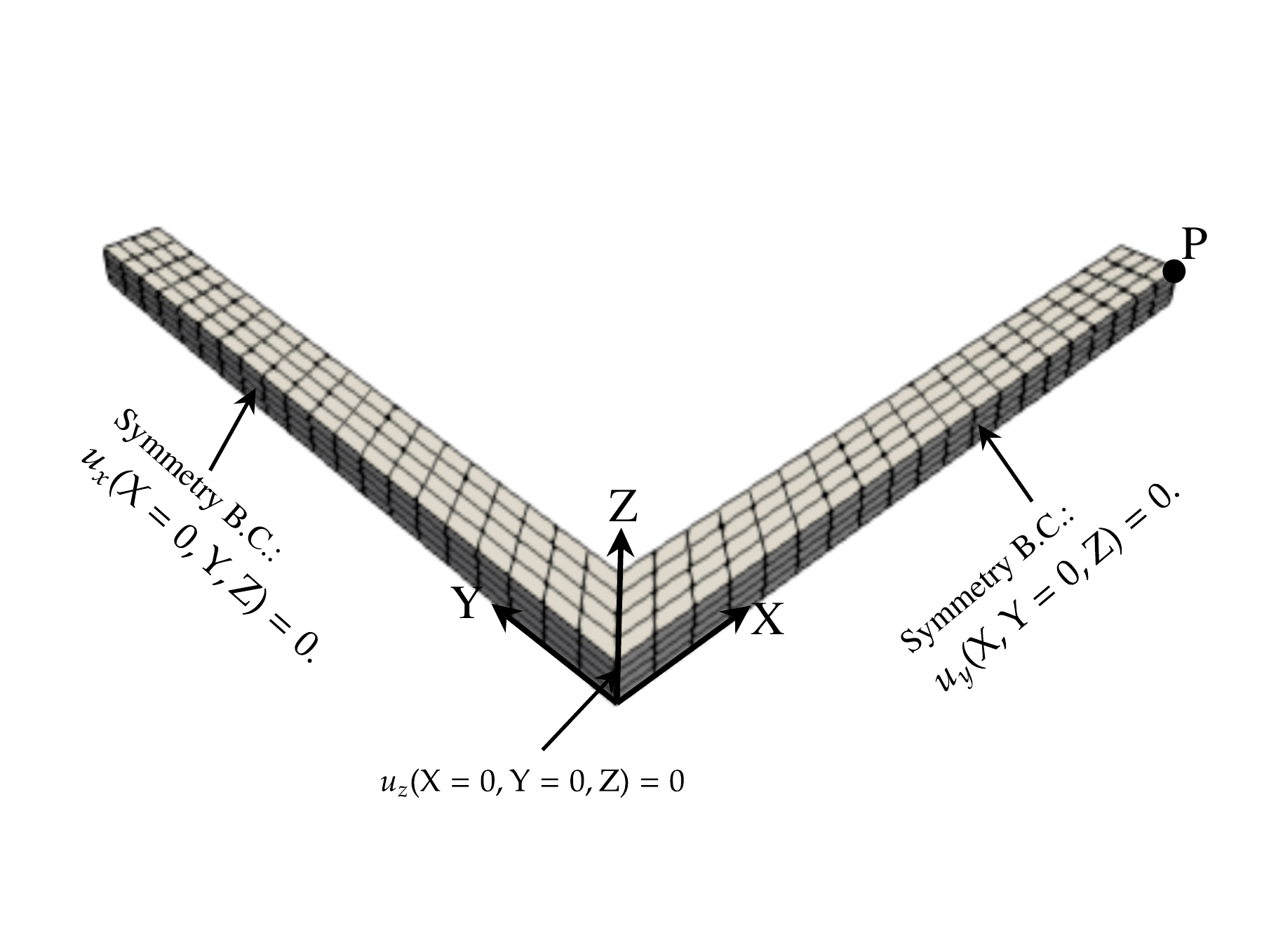} \label{fig-sm-gripper-mesh} }
 \caption{Soft-magnetic gripper: (a) problem setup and (b) finite element mesh of the quarter model (corresponding to the hatched portion in Fig. \ref{fig-sm-gripper-geom}), along with the displacement boundary conditions.}
\end{figure}

As the last example, we consider a gripper with four hands as shown in Fig. \ref{fig-sm-gripper-geom}. Similar to the previous example of the cantilever beam, the gripper comprises an active layer consisting of magneto-active polymer and a passive elastomeric layer. The loading is purely magnetic, and it is such that the specified value of magnetic potential is zero on the diagonal faces passing through the dashed lines and a time-dependent value of $\bar{\phi}$ is applied on all the end faces of the gripper fingers, see Fig. \ref{fig-sm-gripper-geom}. Due to the symmetry of geometry and loading conditions, only a quarter portion of the model is considered for the analysis. The finite element mesh of the quarter portion shown in Fig. \ref{fig-sm-gripper-mesh} consists of 6561 nodes and 640 BQ2/BQ1 elements. Similar to the hard-magnetic gripper example, the effect of the viscoelastic parameters on the response characteristics of the gripper under an applied magnetic load is assessed. The material properties for the hyperelastic and magnetic strain energy functions are the same as the previous example. The maximum value of the applied magnetic potential is $1200 \, \Ampere$ which is ramped up linearly for the first ten seconds and then kept constant at $1200 \Ampere$ for $t >= 10$ s.

\begin{figure}[H]
 \centering
  \includegraphics[clip, scale=0.9]{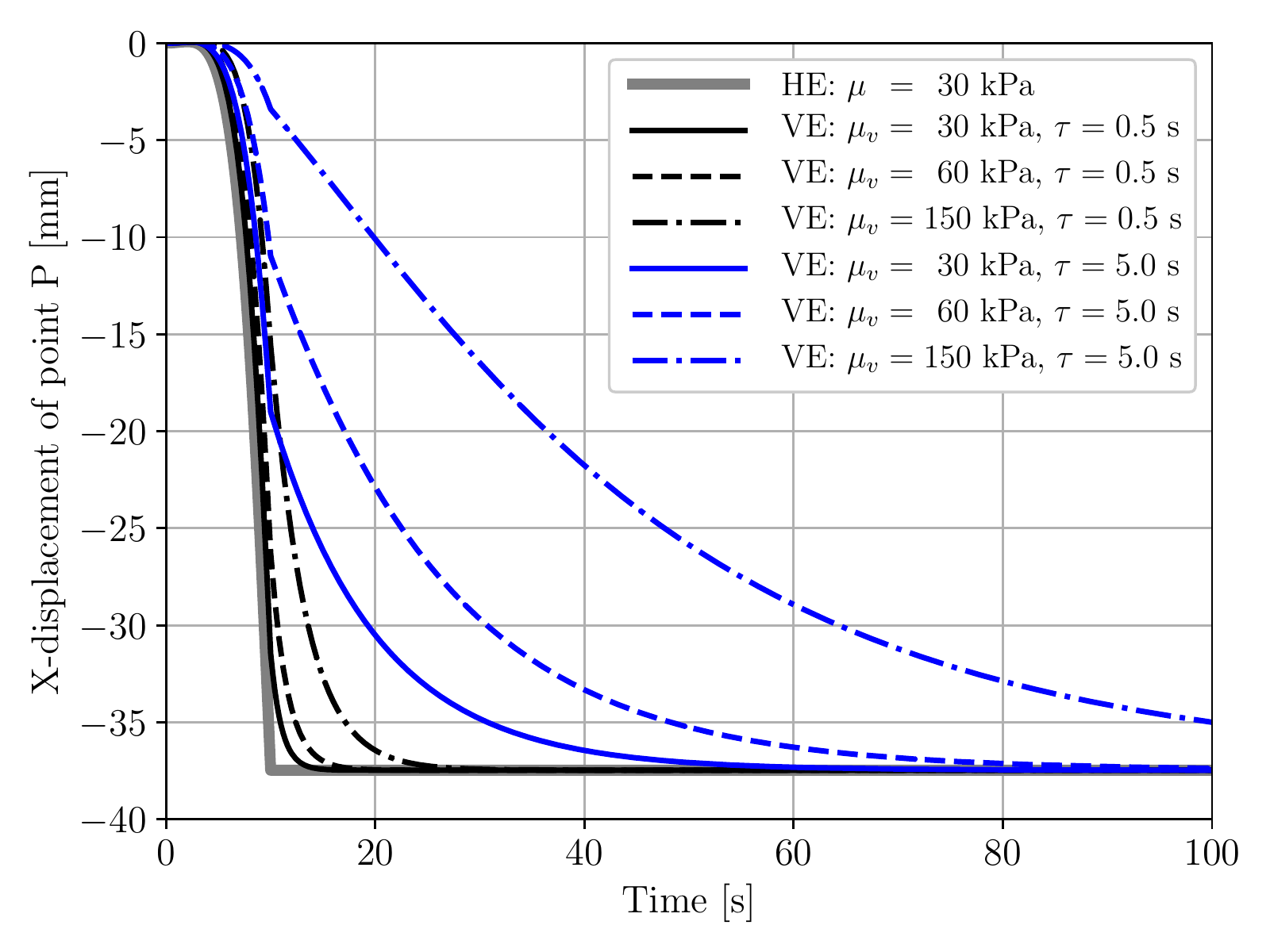}
 \caption{Soft-magnetic gripper: time evolution of X-displacement of point P obtained with the hyperelastic (HE) model and viscoelastic (VE) model for $\mu_{v} = \{ 30, 60, 150 \}$ kPa and $\tau = \{0.5,5.0 \}$ s.}
 \label{fig-sm-gripper-graph}
\end{figure}

\begin{figure}[H]
\centering
 \subfloat[$\mu_{v} = 30$ kPa and $\tau=0.5$ s]{\includegraphics[trim=10mm 0mm 10mm 0mm, clip, scale=0.4]{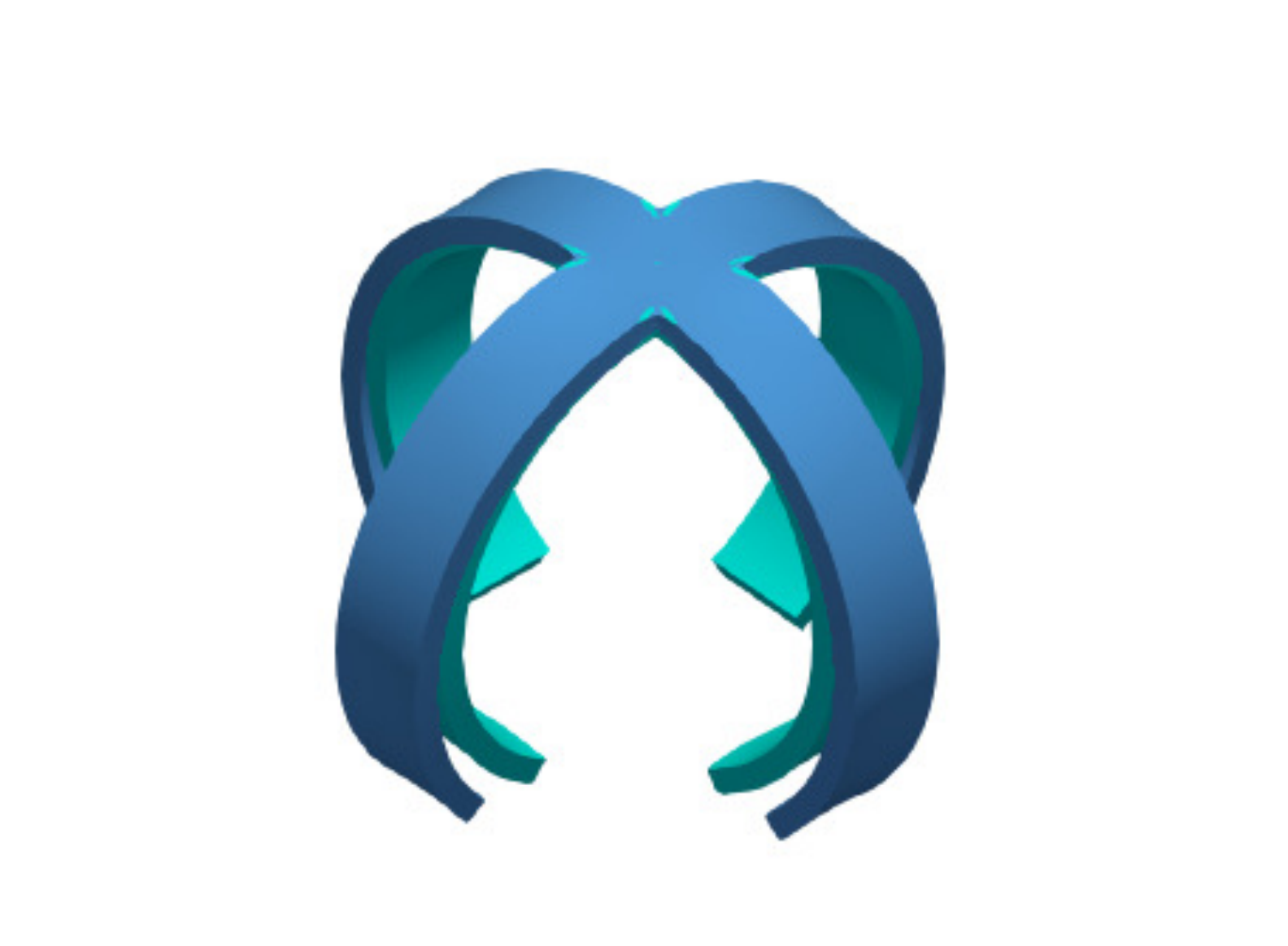}}
 \subfloat[$\mu_{v} = 60$ kPa and $\tau=0.5$ s]{\includegraphics[trim=10mm 0mm 10mm 0mm, clip, scale=0.4]{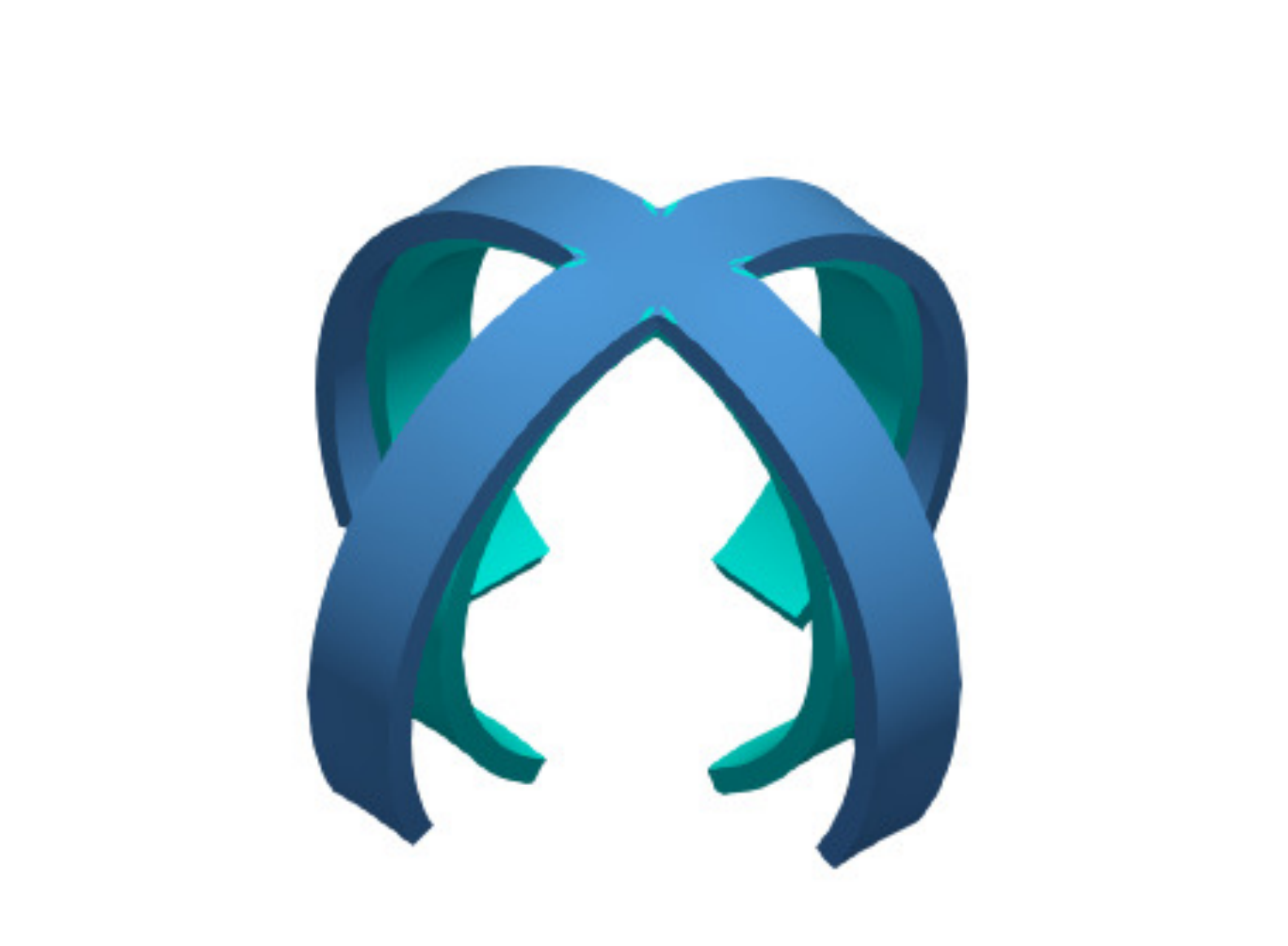}}
 \subfloat[$\mu_{v} = 150$ kPa and $\tau=0.5$ s]{\includegraphics[trim=10mm 0mm 10mm 0mm, clip, scale=0.4]{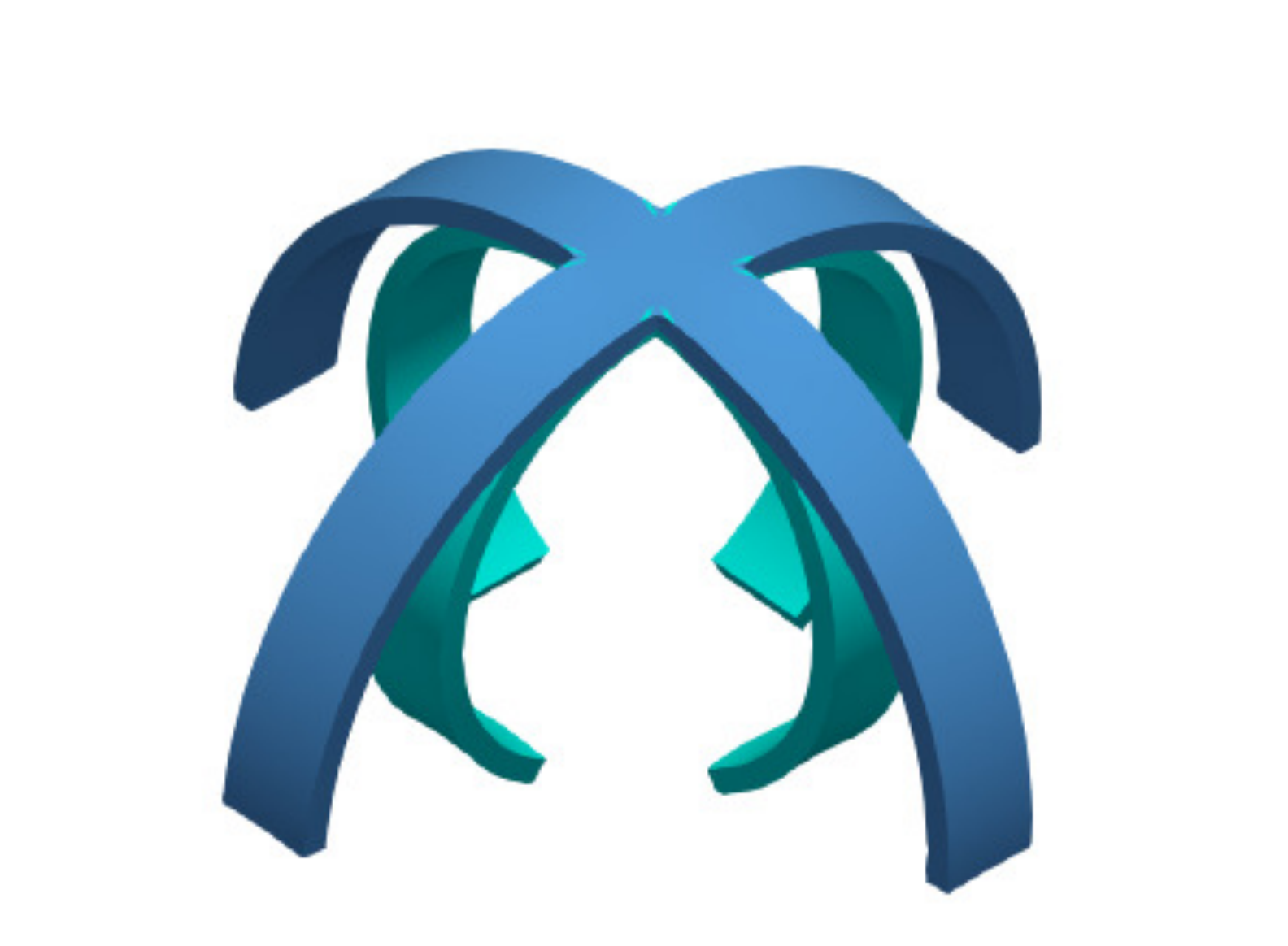}} \\
 \subfloat[$\mu_{v} = 30$ kPa and $\tau=5.0$ s]{\includegraphics[trim=10mm 0mm 10mm 0mm, clip, scale=0.4]{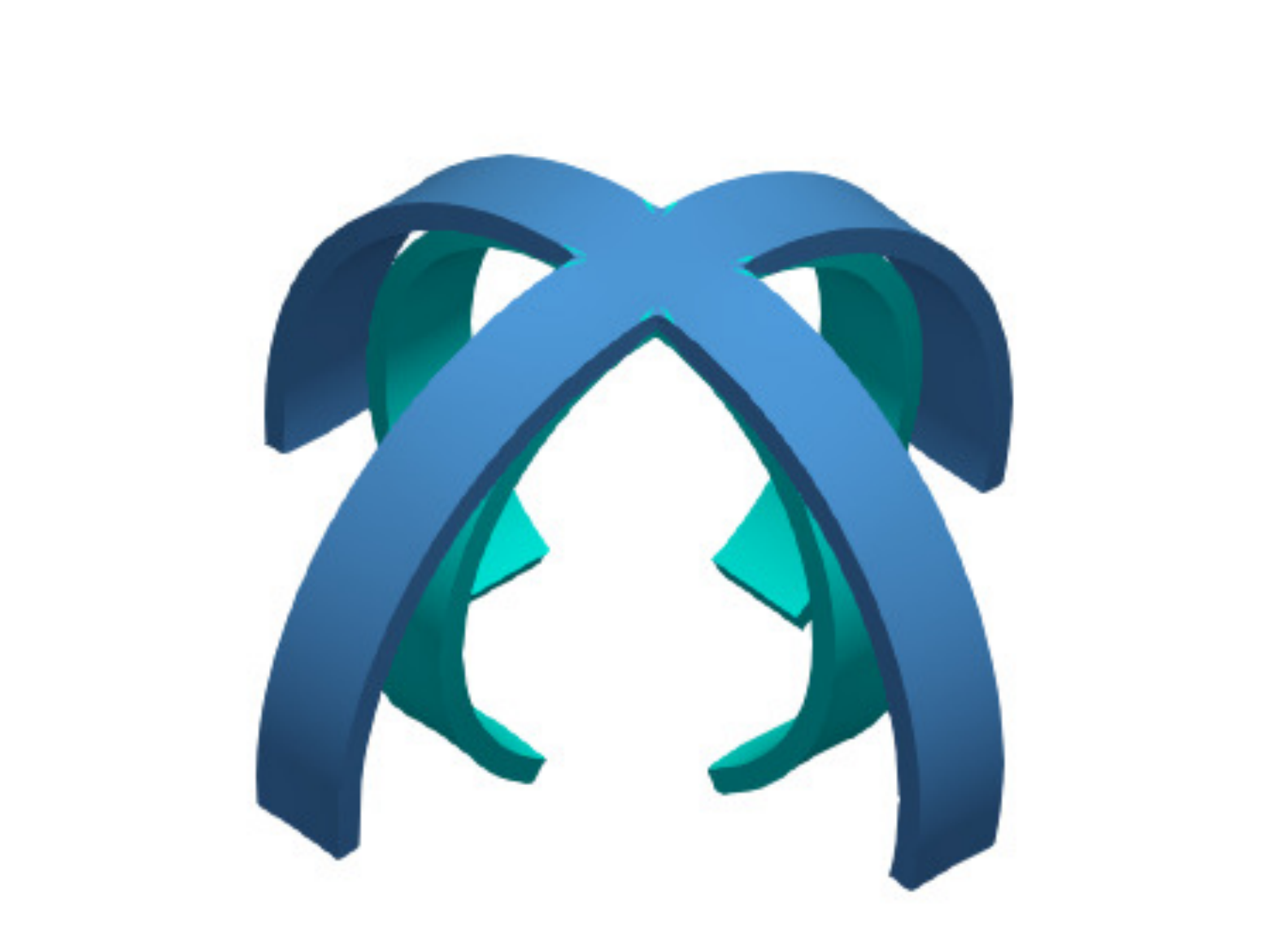}}
 \subfloat[$\mu_{v} = 60$ kPa and $\tau=5.0$ s]{\includegraphics[trim=10mm 0mm 10mm 0mm, clip, scale=0.4]{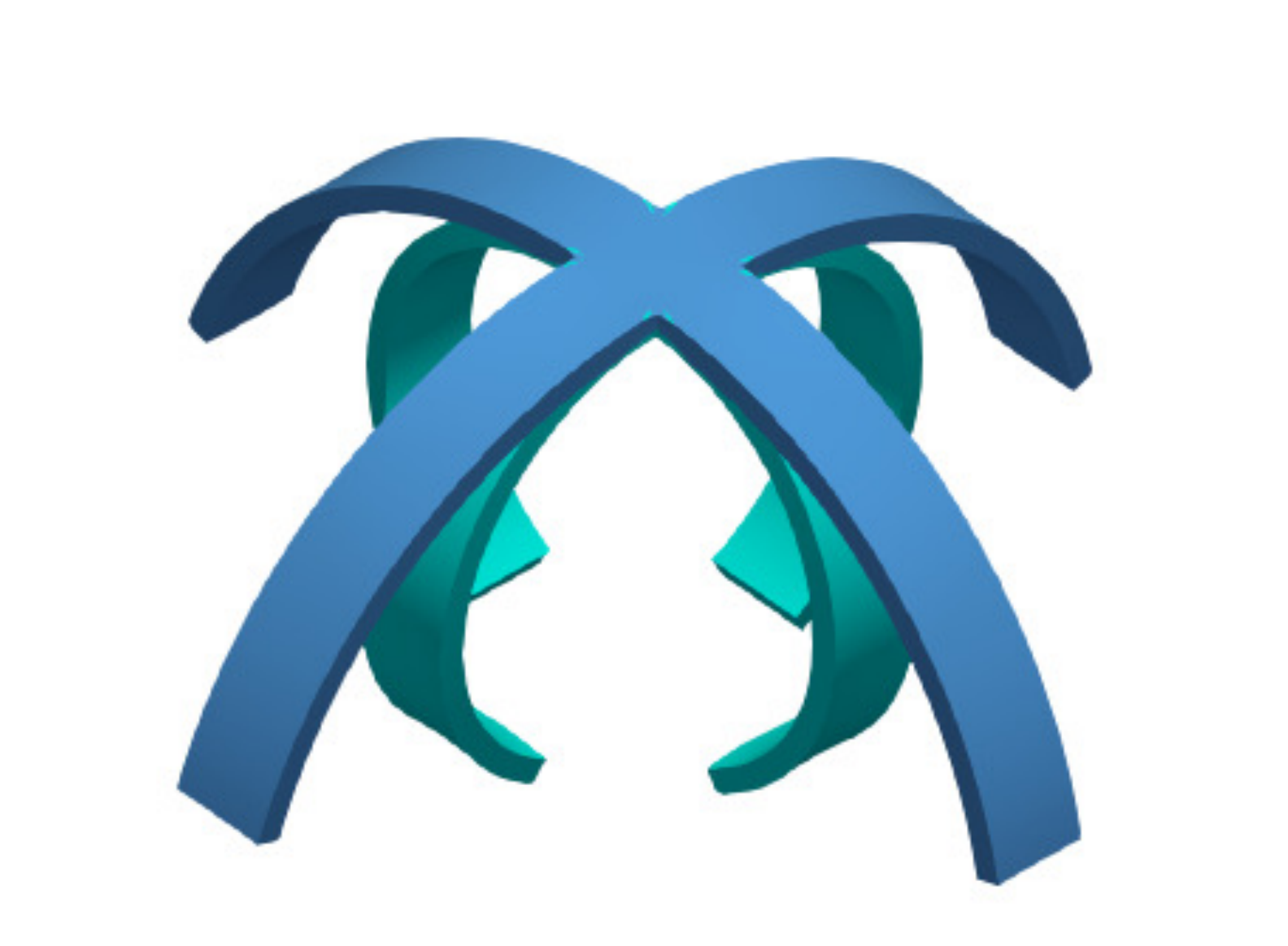}}
 \subfloat[$\mu_{v} = 150$ kPa and $\tau=5.0$ s]{\includegraphics[trim=10mm 0mm 10mm 0mm, clip, scale=0.4]{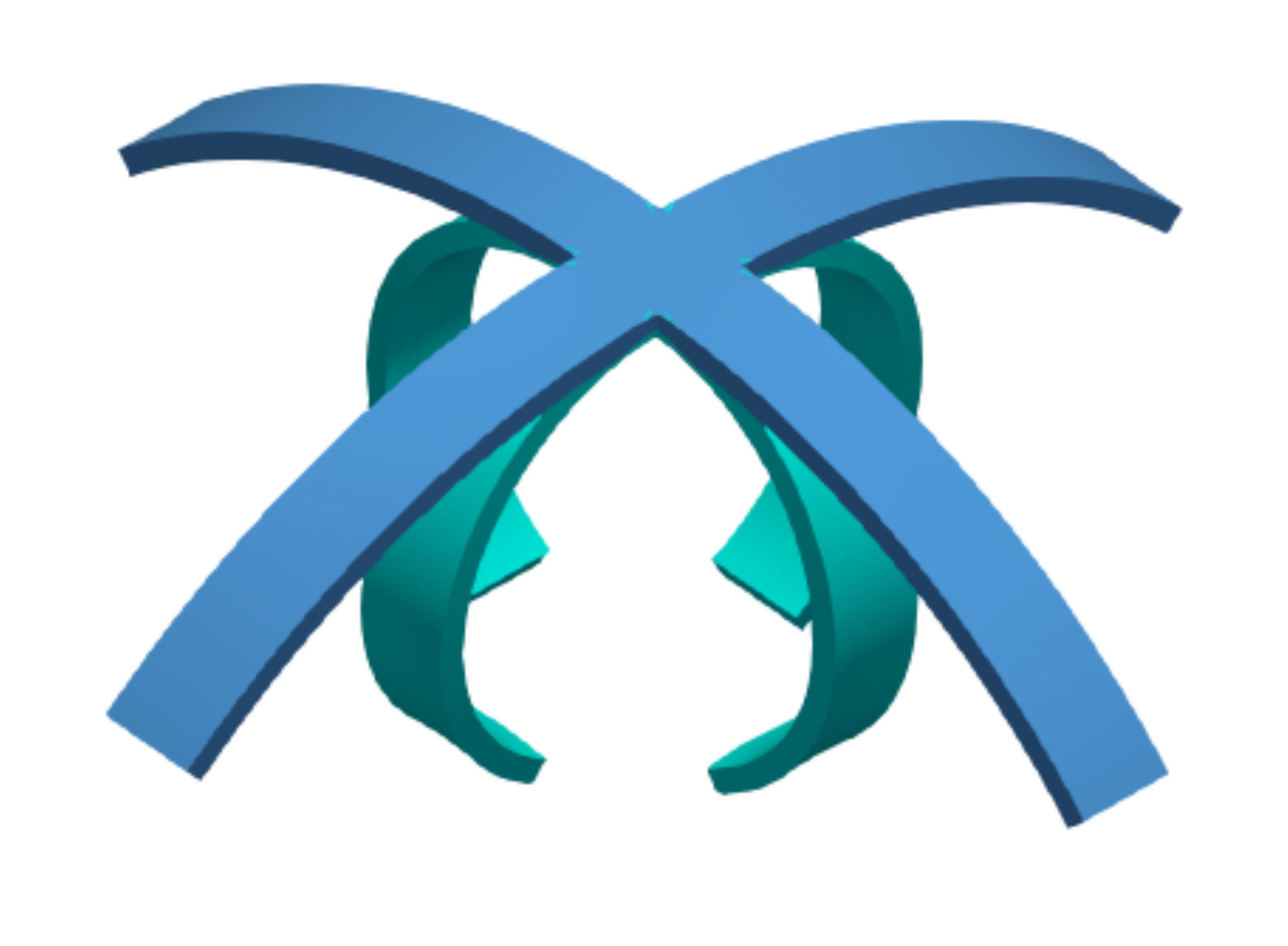}}
 \caption{Soft-magnetic gripper: deformed shapes at $t=10$ s obtained with hyperelastic model (cyan) and viscoelastic model (blue).}
 \label{fig-sm-gripper-defshape-1}
\end{figure}

\begin{figure}[H]
\centering
 \subfloat[$\mu_{v} = 30$ kPa and $\tau=0.5$ s]{\includegraphics[trim=10mm 0mm 10mm 0mm, clip, scale=0.4]{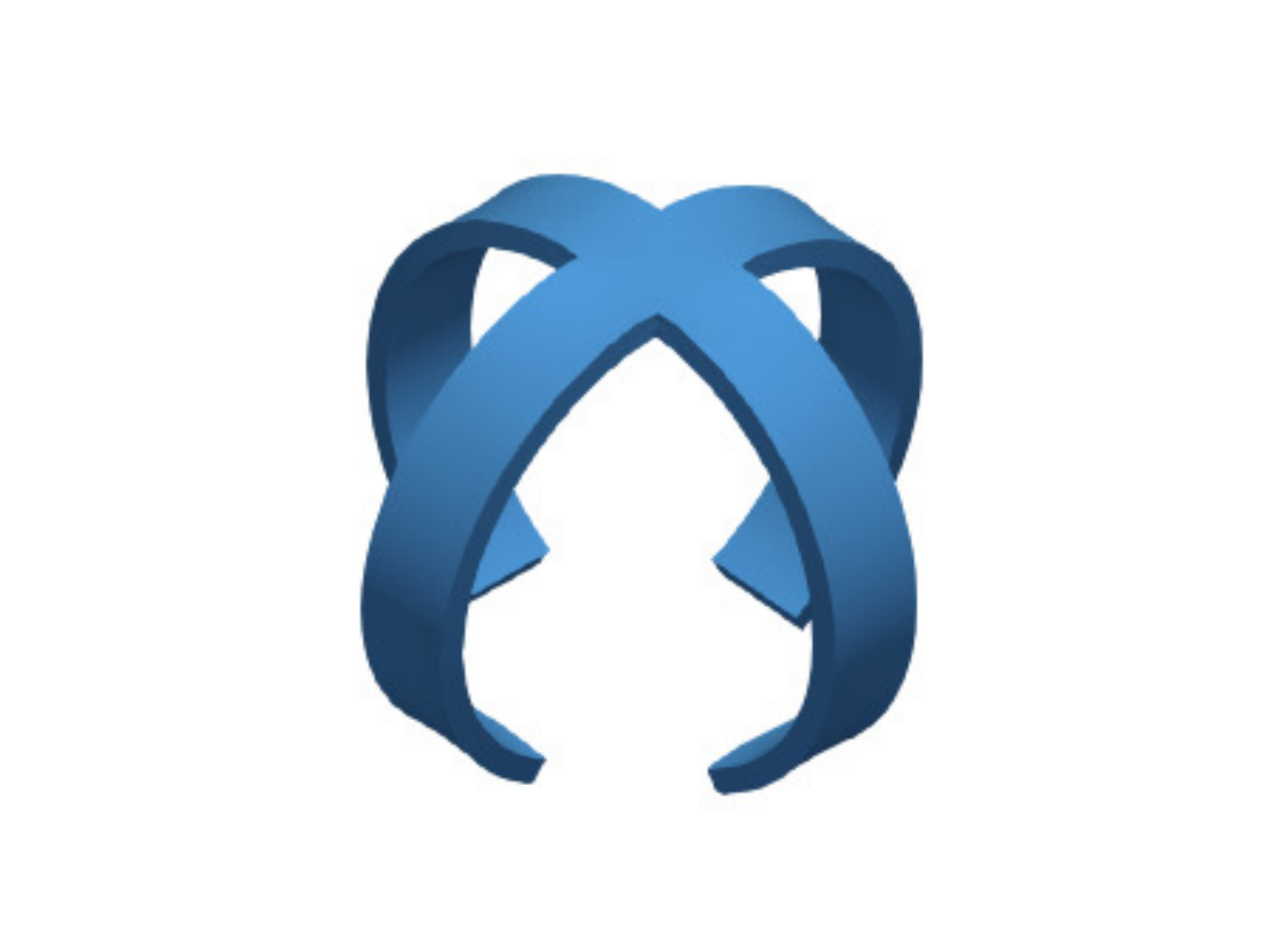}}
 \subfloat[$\mu_{v} = 60$ kPa and $\tau=0.5$ s]{\includegraphics[trim=10mm 0mm 10mm 0mm, clip, scale=0.4]{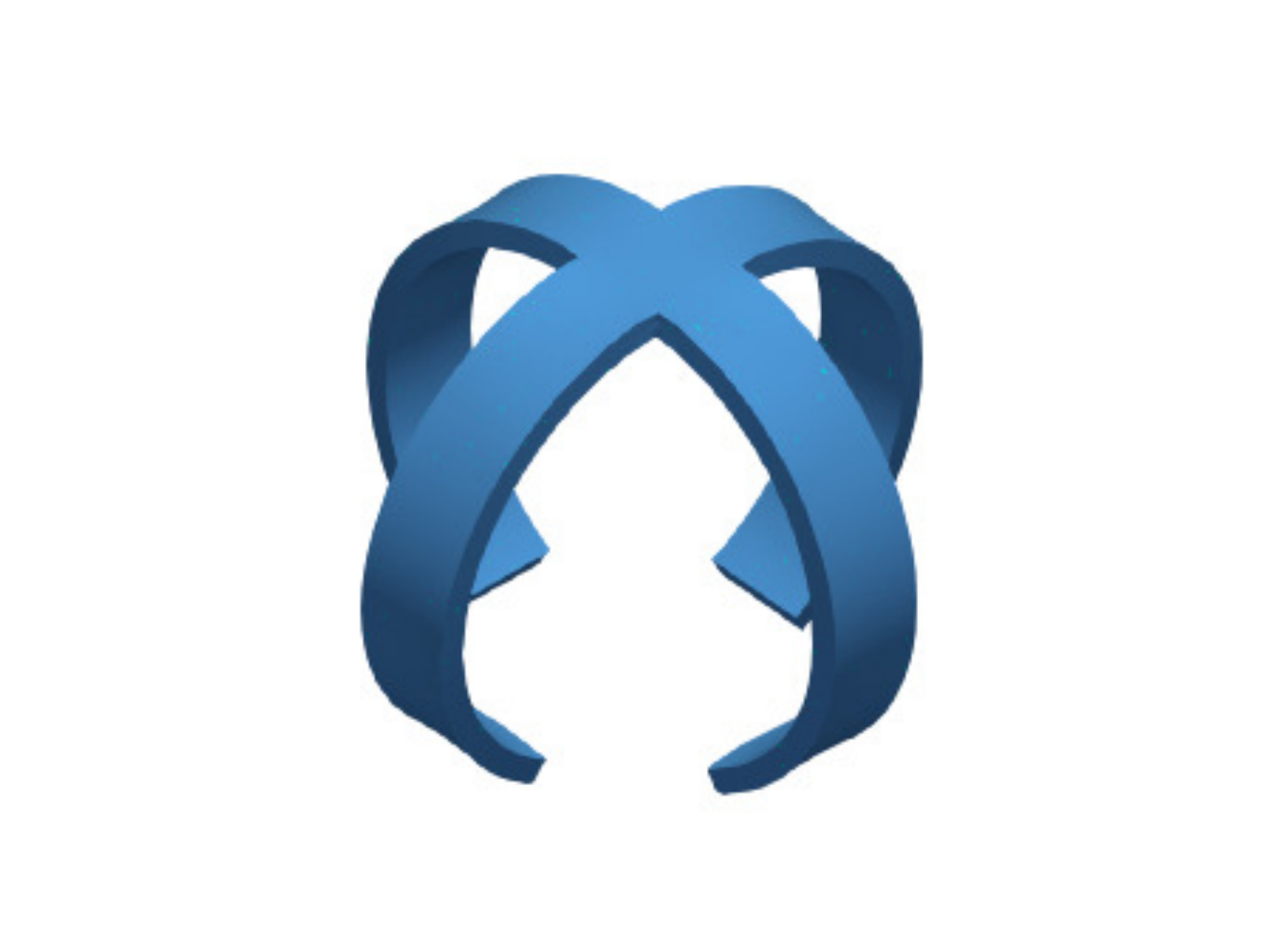}}
 \subfloat[$\mu_{v} = 150$ kPa and $\tau=0.5$ s]{\includegraphics[trim=10mm 0mm 10mm 0mm, clip, scale=0.4]{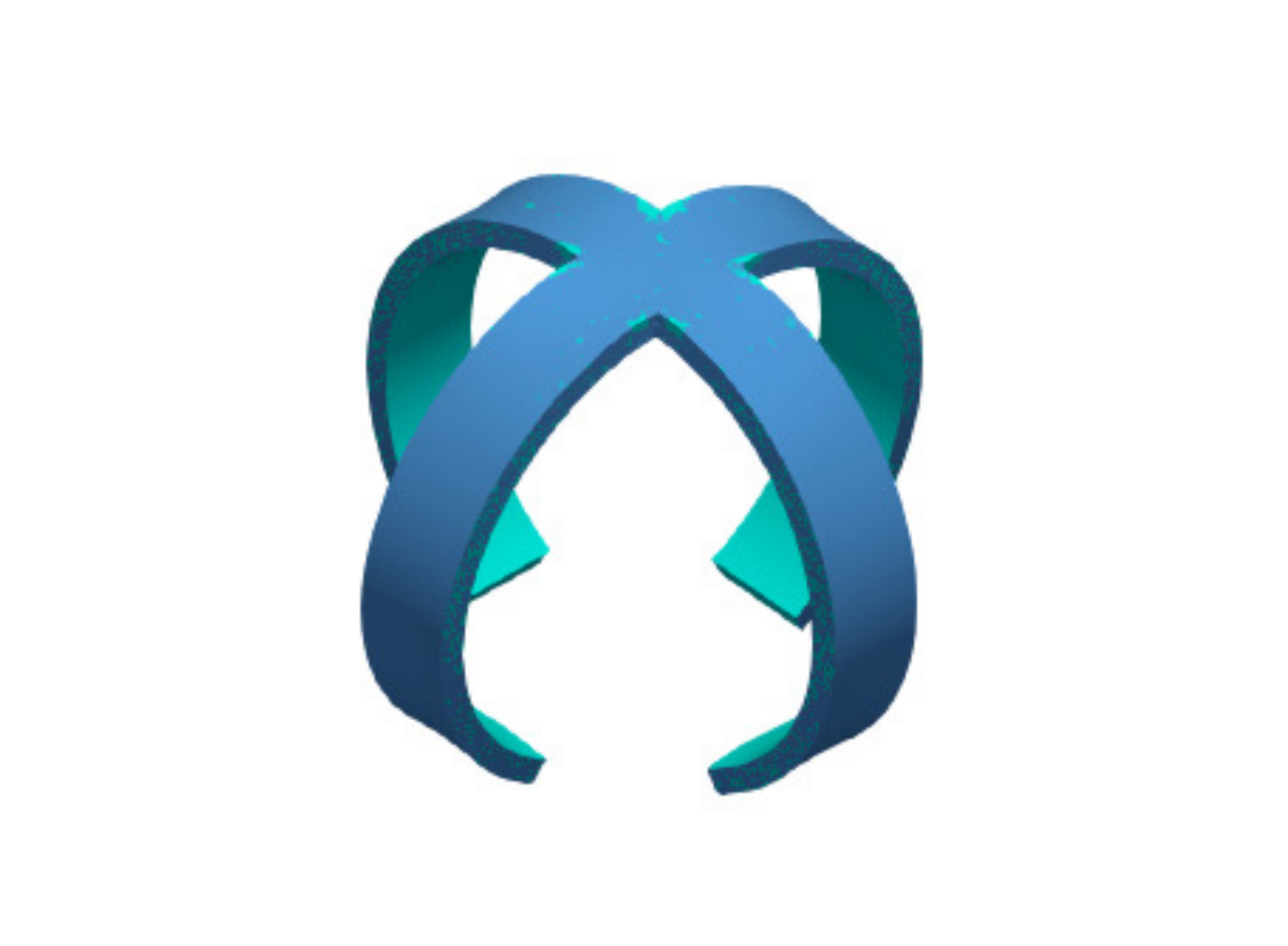}} \\
 \subfloat[$\mu_{v} = 30$ kPa and $\tau=5.0$ s]{\includegraphics[trim=10mm 0mm 10mm 0mm, clip, scale=0.4]{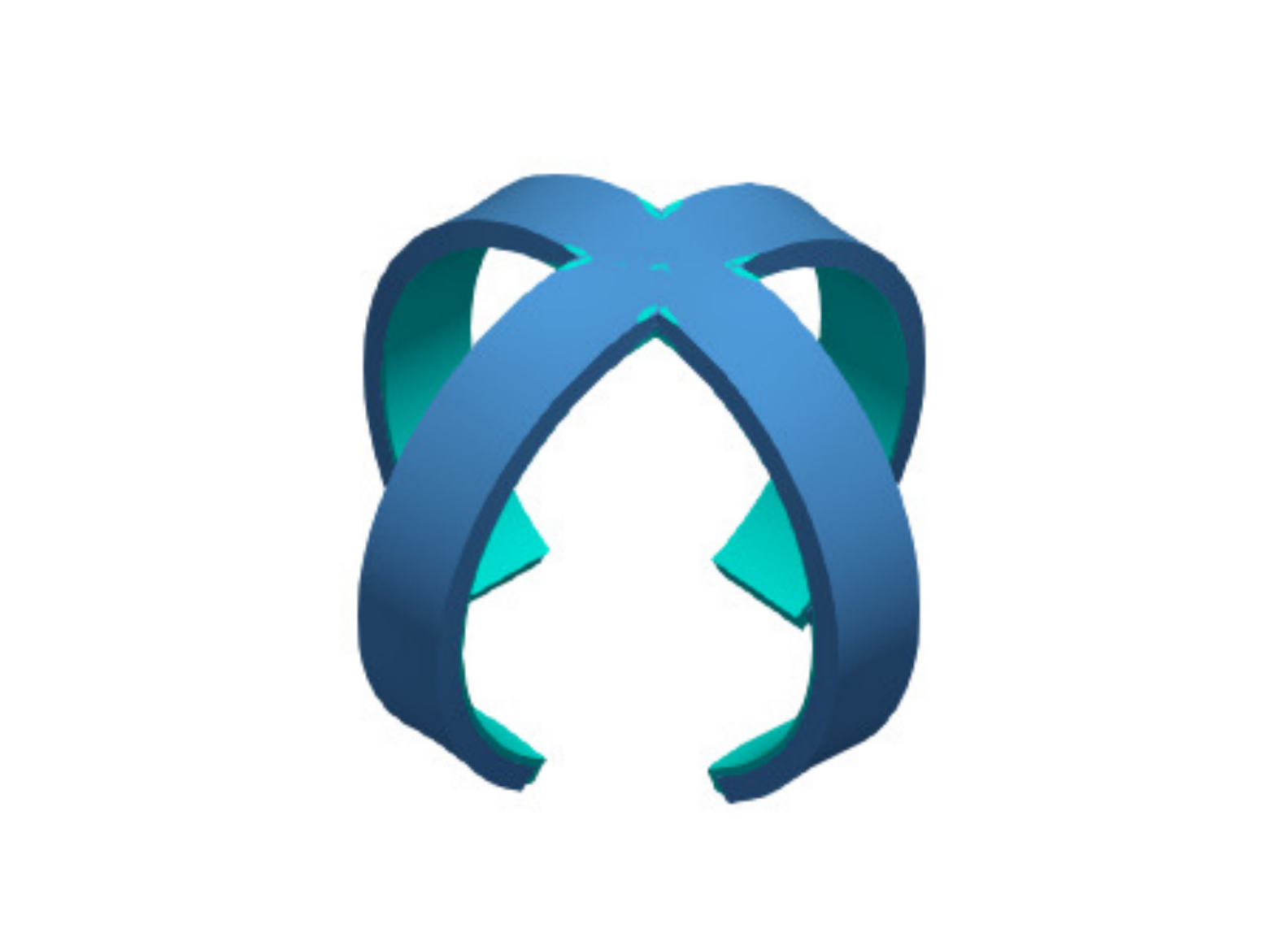}}
 \subfloat[$\mu_{v} = 60$ kPa and $\tau=5.0$ s]{\includegraphics[trim=10mm 0mm 10mm 0mm, clip, scale=0.4]{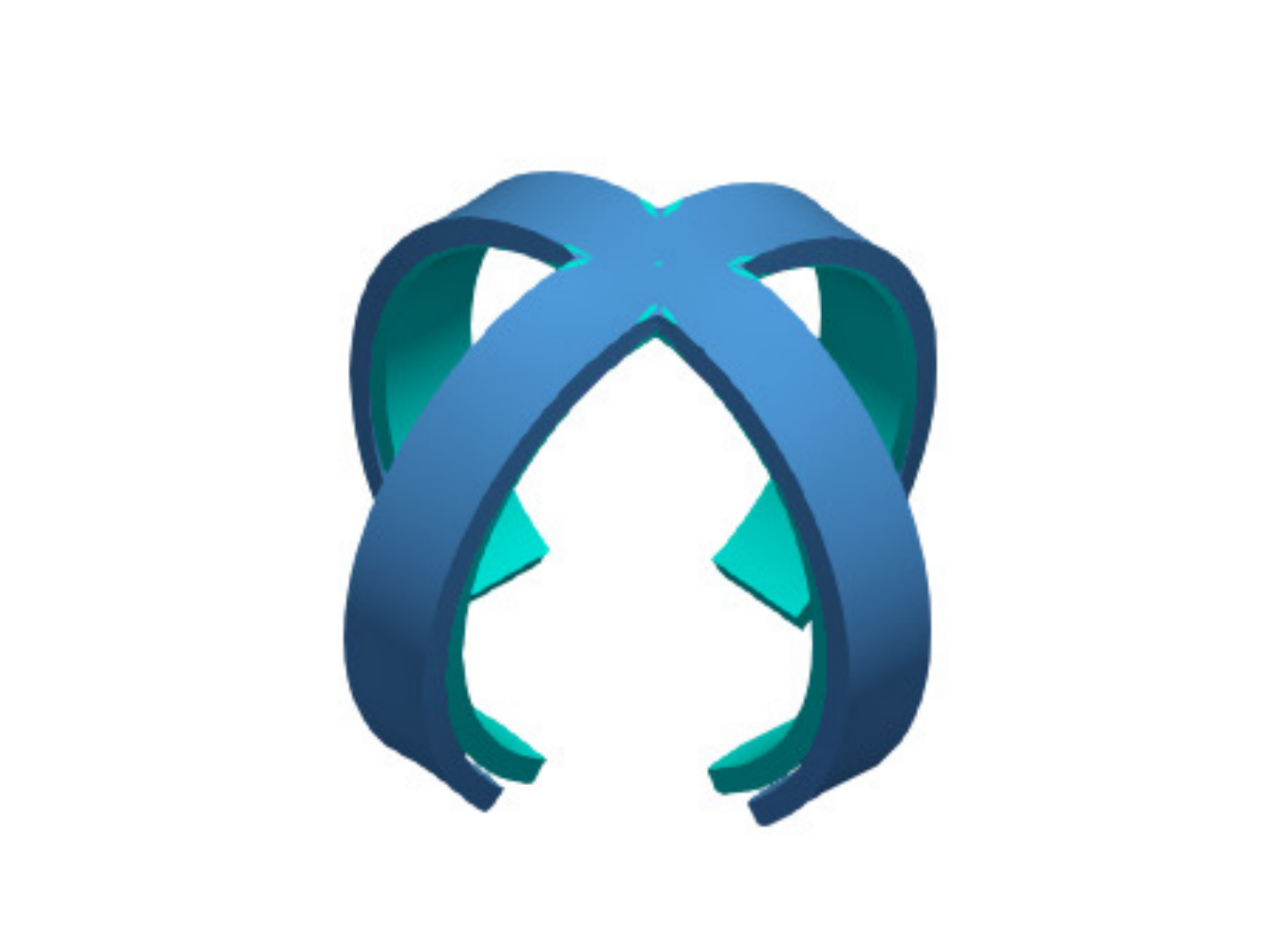}}
 \subfloat[$\mu_{v} = 150$ kPa and $\tau=5.0$ s]{\includegraphics[trim=10mm 0mm 10mm 0mm, clip, scale=0.4]{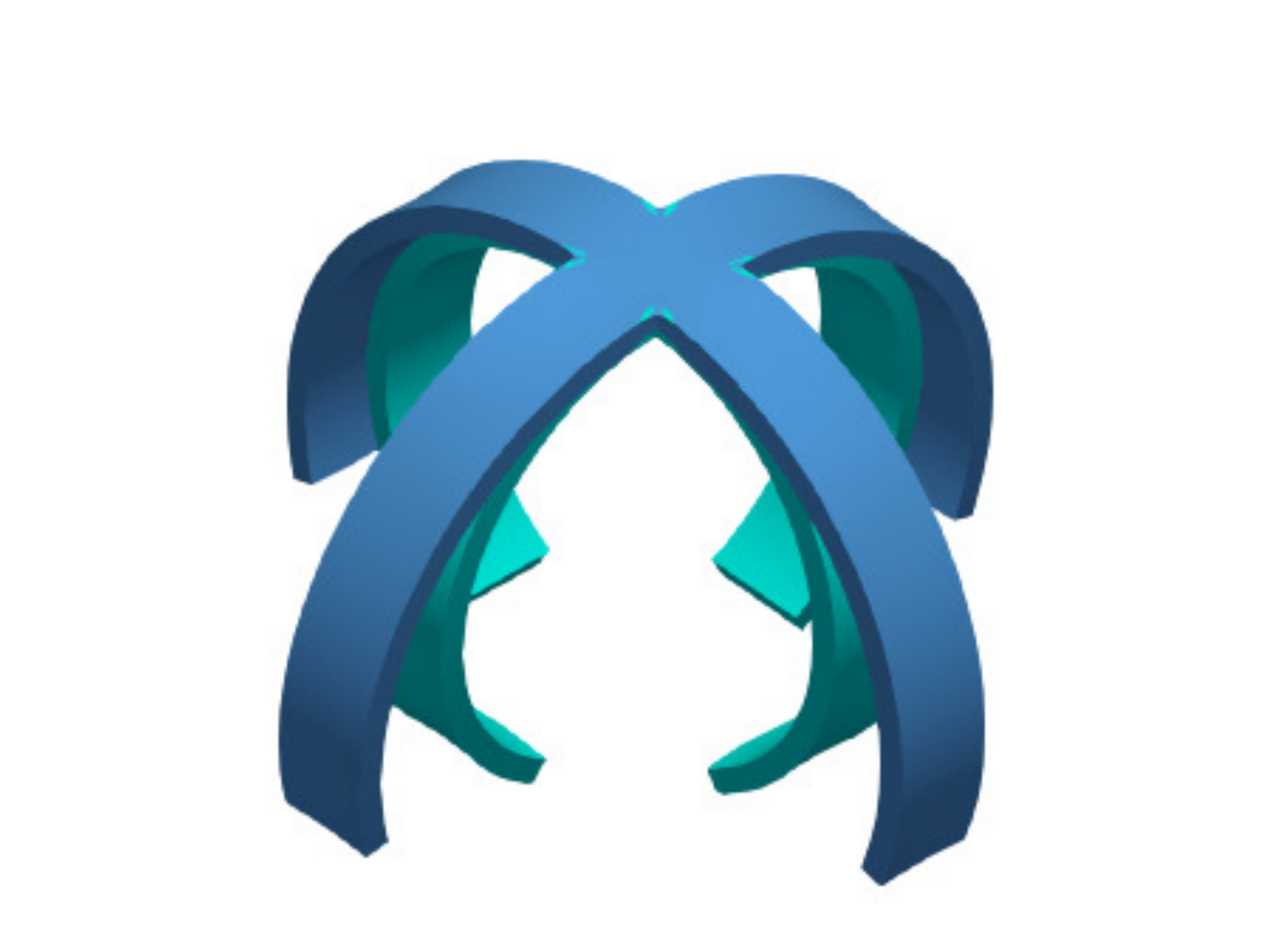}}
 \caption{Soft-magnetic gripper: deformed shapes at $t=40$ s obtained with hyperelastic model (cyan) and viscoelastic model (blue).}
 \label{fig-sm-gripper-defshape-2}
\end{figure}

Simulations are carried out for different combinations of the viscous shear modulus, $\mu_{v} = \{ 30, 60, 150 \}$ kPa, and the relaxation time, $\tau = \{ 0.5, 5.0 \}$ s. The response of the gripper is presented as the time evolution of X-displacement of point P (see Fig. \ref{fig-sm-gripper-mesh}) for different combinations of the viscoelastic parameters is shown in Fig. \ref{fig-sm-gripper-graph}. Deformed shapes of the gripper obtained with the viscoelastic model are shown in Figs. \ref{fig-sm-gripper-defshape-1} and \ref{fig-sm-gripper-defshape-2}, respectively, at time instants 10 s and 40 s. From these graphs, we can observe clear creep behaviour in the response of the gripper, similar to the one observed in the hard magnetic case. As expected, for a constant $\mu_{v}$, the higher the relaxation time, the longer it takes for the gripper to achieve the desired deformed shape. Moreover, for a constant $\tau$, the higher the value of the modulus, the longer the gripper takes to reach the reference deformed shape.

\section{Summary and conclusions}  \label{sec-conclusion}
In this manuscript, we proposed a novel unified finite element framework for performing robust and computationally efficient simulations of hard as well as soft MAPs. In particular, we developed displacement-pressure and displacement-pressure-potential formulations, respectively, for the hard magnetic and soft magnetic cases for modelling the finite strain deformation behaviour, including the time/strain rate-dependent viscoelastic effects. The novel aspects of the present work are:
\begin{itemize}
\item Adaptation of the mixed displacement-pressure formulation to successfully capture truly incompressible deformation behaviour in MAPs without having to approximate it by adjusting the bulk modulus.
\item Use of the generalised-alpha time integration scheme, which is second-order accurate in time, for solving the evolution equations of internal variables in the viscoelastic models.
\item Simulation of finite strain viscoelastic deformation behaviour of carefully devised experimentally-driven examples made of hard and soft MAPs.
\end{itemize}

The accuracy and efficiency of the framework are demonstrated using several numerical examples of beams and grippers made of MAPs. Using the model of a cantilever beam in the hard-magnetic case, the computational benefits of second-order elements over first-order elements are demonstrated. The ability of the proposed framework to capture complex deformation behaviour using coarse meshes is illustrated using the example of a pattern devised with hard magnetic fillers. The effect of viscoelasticity on the response characteristics of soft robotic grippers made of MAPs is demonstrated using comprehensive studies of parameters such as viscoelastic shear moduli and relaxation times.

In conclusion, the present work provides a versatile finite element analysis framework for the simulation of coupled as well as decoupled interactions in soft magneto-active polymeric materials. The use of higher-order elements for spatial discretisation and a second-order time integration scheme for the viscoelastic evolution equations facilitate computationally efficient simulations of magneto-mechanics problems in 3D. In a forthcoming contribution, an extension of the proposed work will be considered taking acceleration effects into account for the simulation of MAPs subjected to high-strain rates and impact loads. The use of B\'ezier elements for the spatial discretisation provides straightforward extensions to such problems using either implicit or explicit time integration schemes \cite{KadapaIJNME2019bbar,KadapaIJNME2019mixed,KadapaAMech2021}.

\section*{ACKNOWLEDGEMENTS}
\noindent  M. H. acknowledges the funding through an Engineering and Physical Sciences Research Council (EPSRC) Impact Acceleration Award (EP/R511614/1).


\begin{appendices}


\setcounter{equation}{0}

\appendix
\section{Generalised-alpha scheme for the evolution equation}  \label{section-appndx-evol}

Using generalised-alpha scheme for first-order system of equations \cite{JansenCMAME2000}, equation (\ref{eqn-iv-evol}) is written as
\begin{align} \label{eqn-evol-galpha}
\dot{\bm{A}}^{(k)}_{n+\alpha_m} = \frac{1}{\tau^{(k)}} \left[ \thickbar{C}^{-1}_{n+\alpha_f} - \bm{A}^{(k)}_{n+\alpha_f} \right]
\end{align}
where
\begin{subequations}  \label{eqn-evol-galpha-subs}
\begin{align}
\dot{\bm{A}}^{(k)}_{n+\alpha_m} &= \alpha_m \, \dot{\bm{A}}^{(k)}_{n+1} + (1-\alpha_m) \, \dot{\bm{A}}^{(k)}_{n}, \\
\bm{A}^{(k)}_{n+\alpha_f} &= \alpha_f \, \bm{A}^{(k)}_{n+1} + (1-\alpha_f) \, \bm{A}^{(k)}_{n}, \\
\dot{\bm{A}}^{(k)}_{n+1} &= \frac{1}{\gamma \, \Delta t}  \left[ \bm{A}^{(k)}_{n+1}-\bm{A}^{(k)}_{n} \right] + \frac{\gamma-1}{\gamma} \, \dot{\bm{A}}^{(k)}_{n}.
\end{align}
\end{subequations}

Here, $n+1$ and $n$ denote current and previous load steps, and $\Delta t$ is the size of the load step. For the scheme to unconditionally stable and second-order, the time integration parameters $\alpha_m$, $\alpha_f$ and $\gamma$ are chosen to be such that
\begin{align}
\alpha_f = \frac{1}{1+\rho_{\infty}}; \qquad
\alpha_m = \frac{3-\rho_{\infty}}{2 \, [ 1+\rho_{\infty} ]}; \qquad
\gamma = \frac{1}{2} + \alpha_m - \alpha_f, \qquad \text{for} \qquad
0 \leq \rho_{\infty} \leq 1,
\end{align}
where $\rho_{\infty}$ is the spectral radius, and it controls the amount of numerical damping. See \cite{JansenCMAME2000,KadapaCS2017} for comprehensive details of the scheme. In this work, we use $\rho_{\infty}=0$ for all the numerical examples.

Using (\ref{eqn-evol-galpha}) and (\ref{eqn-evol-galpha-subs}), the internal variable $\bm{A}^{(k)}$ at the current load step is evaluated using

\begin{align}
\left[ 1 + \frac{\alpha_f \, \gamma \, \Delta t}{\alpha_m \, \tau^{(k)}} \right] \, \bm{A}^{(k)}_{n+1} = \frac{\gamma \, \Delta t}{\alpha_m \, \tau^{(k)}} \, \thickbar{C}^{-1}_{n+\alpha_f} + \left[ 1 - \frac{(1-\alpha_f) \, \gamma \, \Delta t}{\alpha_m \, \tau^{(k)}} \right] \, \bm{A}^{(k)}_{n} - \frac{(\gamma - \alpha_m) \, \Delta t}{\alpha_m \, \gamma} \, \dot{\bm{A}}^{(k)}_{n}
\end{align}

The contribution from the viscoelastic free energy function to the deviatoric part of Cauchy stress is
\begin{align}
\sigma_{ij} = \sum_{k=1}^{s} \Bigg[ \mu^{(k)} \, J^{-5/3} \, \left[ \widehat{A}_{ij}^{(k)} - \frac{1}{3} \, I_{AC}^{(k)} \, \delta_{ij} \right]  \Bigg]
\end{align}
and to the elasticity tensor is
\begin{align}
\mathsf{e}_{ijkl} 
&= \sum_{k=1}^{s} \mu^{(k)} \, J^{-5/3} \, \left[ \delta_{ik} \, \widehat{A}_{jl}^{(k)} - \frac{2}{3} \, \delta_{kl}  \, \widehat{A}_{ij}^{(k)} - \frac{2}{3} \, \delta_{ij}  \, \widehat{A}_{kl}^{(k)} + \frac{2}{9} \, I_{AC}^{(k)} \, \delta_{ij} \, \delta_{kl}  + \frac{1}{3} \, I_{AC}^{(k)} \, \delta_{il} \, \delta_{jk} \right] \nonumber \\
&- \sum_{k=1}^{s}  \mu^{(k)} \, J^{-1} \, \Lambda \, \left[ \delta_{ik} \, \delta_{jl} +  \delta_{il} \, \delta_{jk} - \frac{2}{3} \, \delta_{ij} \, \delta_{kl} \right]
\end{align}
where
\begin{align}
\widehat{A}_{ij}^{(k)} &= \frac{1}{2} \left[ F_{iM} \, A_{MN}^{(k)} \, F_{jN} + F_{jM} \, A_{MN}^{(k)} \, F_{iN} \right], \\
I_{AC}^{(k)}  &= A_{MN}^{(k)} \, C_{MN}, \\
\Lambda &= \frac{\gamma \, \Delta t}{\alpha_f \, \gamma \, \Delta t + \alpha_m \, \tau^{(k)}}.
\end{align}

\end{appendices}

\end{document}